\documentclass[aoas]{imsart}

\RequirePackage{amsthm,amsmath,amsfonts,amssymb}
\RequirePackage[authoryear]{natbib}
\RequirePackage[colorlinks,citecolor=blue,urlcolor=blue]{hyperref}
\RequirePackage{graphicx}
\usepackage{booktabs}
\usepackage{subcaption}
\usepackage{float}
\usepackage{comment}
\usepackage{placeins}
\startlocaldefs
\theoremstyle{plain}

\theoremstyle{definition}

\newtheorem{assumption}{Assumption}
\endlocaldefs

\begin{document}

\begin{frontmatter}
\title{Generalized propensity score weighting for functional causal inference framework}
\runtitle{Generalized propensity score weighting for functional causal inference}
%\thankstext{T1}{A sample additional note to the title.}

\begin{aug}
\thankstext{t1}{These authors contributed equally to this work.}
%%%%%%%%%%%%%%%%%%%%%%%%%%%%%%%%%%%%%%%%%%%%%%%
%% Only one address is permitted per author. %%
%% Only division, organization and e-mail is %%
%% included in the address.                  %%
%% Additional information such as            %%
%% identifying the corresponding author must %%
%% be included in in the Acknowledgments     %%
%% section if necessary.                     %%
%% ORCID can be inserted by command:         %%
%% \orcid{0000-0000-0000-0000}               %%
%%%%%%%%%%%%%%%%%%%%%%%%%%%%%%%%%%%%%%%%%%%%%%%
\author[A]
{\fnms{Simone}~\snm{Ciardulli} %\thanks{[\textbf{Corresponding author indication should be put in the Acknowledgment section if necessary.}]}
\ead[label=e1]{simone.ciardulli@polimi.it}\thanksref{t1}},
\author[A,B]{\fnms{Nicole}~\snm{Fontana}\ead[label=e2]{nicole.fontana@polimi.it}\orcid{0009-0007-4416-1921}\thanksref{t1}},
\author[A]{\fnms{Simone}~\snm{Vantini}\ead[label=e3]{simone.vantini@polimi.it}\orcid{0000-0001-8255-5306}}
\and
\author[A,B]{\fnms{Francesca}~\snm{Ieva}\ead[label=e4]{francesca.ieva@polimi.it}\orcid{0000-0003-0165-1983}}

\address[A]{MOX, Department of Mathematics, Politecnico di Milano, Milan, Italy
%\textbf{[Additional affiliations should be put in the Acknowledgments section]}
\printead[presep={ ,\ }]{e1,e2,e3,e4}}
\address[B]{Health Data Science Research Centre, Human Technopole, Milan, Italy
%\textbf{[Additional affiliations should be put in the Acknowledgments section]}
\printead[presep={ \ }]{}}
\end{aug}

\begin{abstract}
Estimating causal effects in observational studies requires adjustment for confounding, a task that becomes challenging when the exposure is a function observed over a continuous domain rather than a scalar variable. We develop a functional propensity score weighting framework that achieves covariate balance by removing dependence between time-varying treatments and observed confounders, thereby enabling estimation of marginal causal effects in settings with functional treatments, covariates, and outcomes.
%We develop a functional propensity score weighting framework that enables covariate balancing and causal effect estimation when treatments, covariates, and outcomes are functional. 
We propose a dual formulation of the weight estimation problem that yields a smooth unconstrained optimization and improves computational scalability. The proposed framework extends naturally to settings with time-varying covariates and to longitudinal outcomes via a function-on-function marginal structural model, allowing estimation of causal effect surfaces. The proposed method improves covariate balance, estimation accuracy, and computational efficiency compared to the existing approach and retains these properties when extended to functional covariates and outcomes.  We apply the method to data from the UK Biobank to estimate the causal effect of body mass index trajectories on the risk of Type~2 Diabetes and on subsequent glycated hemoglobin trajectories, a functional measure of metabolic status.
\end{abstract}

%\begin{keyword}
%\kwd{Causal inference}
%\kwd{functional data analysis}
%\kwd{functional regression}
%\kwd{functional propensity score}
%\end{keyword}

\end{frontmatter}
%%%%%%%%%%%%%%%%%%%%%%%%%%%%%%%%%%%%%%%%%%%%%%
%% Please use \tableofcontents for articles %%
%% with 50 pages and more                   %%
%%%%%%%%%%%%%%%%%%%%%%%%%%%%%%%%%%%%%%%%%%%%%%
%\tableofcontents

\section{Introduction}
\label{sec:introduction}%
Propensity score methods are a central tool for causal inference in observational studies, enabling the estimation of treatment effects by adjusting for systematic differences in pre-treatment covariates between exposure groups~\citep{rosenbaum, rubin, Austin}. Existing methodology, however, has been developed for settings in which treatments, covariates, and outcomes are represented as scalar or low-dimensional vectors. This assumption is increasingly restrictive in modern applications, where variables are observed repeatedly over time and can be represented more naturally as functions. In this setting, standard propensity score methods are not directly applicable, as they rely on balancing finite-dimensional covariate representations. 

Functional Data Analysis~\citep{Ramsay} provides the statistical tools to model longitudinal information, typically via Functional Principal Component Analysis (FPCA), which approximates infinite-dimensional functional objects through a finite set of principal component scores~\citep{Yao}. Building on the extension of propensity score theory to continuous treatments through the Generalized Propensity Score (GPS)~\citep{Hirano_Imbens}, recent work has begun to integrate FDA with causal inference.~\citet{Zhang} proposed the Functional Propensity Score (FPS), i.e., a covariate-balancing approach that represents a functional treatment through its Karhunen--Lo\`eve expansion and then estimates inverse-probability weights that remove the correlation between the FPC scores and observed confounders, extending the empirical-likelihood balancing framework of~\citet{Owen} and~\citet{Fong}. Despite these advances, the FPS framework has two important limitations. First, weight estimation relies on constrained optimization with sequential tuning over a dimensionality-reduction parameter, imposing a proportionality constraint on the balancing target and yielding computational burdens at sample sizes typical of modern observational studies. Second, the current FPS formulation is restricted to settings with scalar covariates and scalar outcomes, whereas many applications involve confounders and outcomes that are observed repeatedly over time and for which a functional representation may be more appropriate. Related work has addressed causal inference with functional covariates under a scalar treatment \citep{miao2020} and with functional outcome under a binary treatment \citep{ecker2024, fontana2025}, but a general framework that accommodates functional treatments, functional covariates, and functional outcomes simultaneously under a covariate-balancing strategy has not been developed. The absence of such a framework reflects the difficulty of simultaneously representing infinite-dimensional objects while maintaining covariate balance to control for confounding.

In this article, we develop a unified propensity score weighting framework for causal inference with functional data. We show that the proposed formulation yields a general covariate-balancing representation that extends propensity score methods to functional treatments, covariates, and outcomes. We develop a dual formulation of the functional propensity score weight estimation problem, yielding a smooth, unconstrained convex optimization that avoids sequential tuning or proportionality constraints. This formulation leads to improved computational scalability and covariate balance compared to existing methods. We then extend the FPS framework to accommodate functional confounders by representing them via their own FPCA scores and augmenting the balancing constraint vector accordingly, without altering the structure of the dual optimization. Finally, we develop a function-on-function marginal structural model for the case where both treatment and outcome are functional, and estimate a causal effect surface that characterizes how the treatment trajectory influences the outcome trajectory over the domain. The proposed framework applies to a broad class of observational studies with complex data structures.

We conduct simulation studies to evaluate the finite-sample performance of the proposed methodology under both the original functional propensity score setting and its extensions to longitudinal covariates and outcomes. The simulations assess covariate balance, bias reduction, estimation accuracy, and computational scalability under varying degrees of confounding, model complexity and sample sizes. The results show that the proposed weighting procedure consistently improves covariate balance and substantially reduces computational burden relative to the existing approach, while providing effective bias reduction and accurate estimation of the causal effect.

Our methodological innovation is motivated by the study of longitudinal biomarkers and their impact on long-term health outcomes. In such settings, scalar summaries of exposure fail to capture critical features such as timing, duration, and cumulative burden. In particular, we apply the proposed framework to identify how body mass index (BMI) during midlife causally affects subsequent metabolic health, leveraging data from the UK Biobank \citep{sudlow2015}. These data provide longitudinal BMI measurements from linked primary care records alongside rich clinical covariates and multiple metabolic outcomes. First, we estimate the time-varying causal effect of BMI trajectories observed during individuals' midlife on the risk of incident Type 2 Diabetes. We then extend the analysis by treating both the BMI exposure and the glycaemic control profile measured by glycated hemoglobin (HbA1c) as functional, and estimate the causal effect surface that describes how BMI at each exposure age shapes glycaemic control at each subsequent outcome age. 

The remainder of the article is organized as follows. Section~\ref{sec:method} introduces the methodology, covering the FPS weight estimation (Section~\ref{sec:fps}), the extension to functional covariates (Section~\ref{sec:func_cov}), and the functional outcome model (Section~\ref{sec:func_out}). Section~\ref{sec:sim} presents simulation studies evaluating the proposed weight estimator, including comparisons with unweighted analysis and with \citet{Zhang}, as well as an assessment of the extended framework with functional covariates and outcomes. In Section~\ref{sec:realdata}, we apply the proposed framework to estimate the causal effect of BMI trajectories on Type 2 Diabetes risk and on longitudinal HbA1c trajectories. Section~\ref{sec:discussion} concludes with a discussion and directions for future work.
The proposed method and the code used to reproduce the simulation studies are implemented in the R package \texttt{FPScausal}, available on GitHub at \url{https://github.com/NicoleFontana/FPScausal}.
%\textcolor{red}{The proposed method is implemented in the R package \texttt{FPScausal}, available from CRAN~\cite{mvfmr_package}.}

\section{Methodology}
\label{sec:method}%
\subsection{Setup and identification}
\label{sec:setup}

We consider $n$ i.i.d.\ observations $\{(Y_i, X_i(t), \boldsymbol{C}_i)\}_{i=1}^n$, where $Y_i \in \mathbb{R}$ is a scalar outcome, $X_i(t)$ is a zero-mean square-integrable functional treatment observed over a compact domain $\mathcal{T} \subset \mathbb{R}$, and $\boldsymbol{C}_i \in \mathbb{R}^p$ is a vector of observed confounders with $\mathbb{E}[\boldsymbol{C}^\top \boldsymbol{C}] < \infty$. Under the potential outcome framework \citep{rubin}, let $Y_i(x)$ denote the potential outcome of unit $i$ under functional treatment $x \in L^2(\mathcal{T})$ and let $Y_i = Y_i(X_i)$ denote the observed outcome. The causal estimand of interest is the effect function $\mu(t)$, defined through the marginal structural model \citep{Ramsay}:
\begin{equation}
\label{eq:outcome_model}
\mathbb{E}[Y(x)] = \mu_0 + \int_{\mathcal{T}} \mu(t) \cdot x(t) \, dt,
\end{equation}
where $\mu_0 = \mathbb{E}[Y]$ by the zero-mean assumption on $X(t)$. The function $\mu(t)$ characterizes how the treatment at each domain point $t$ contributes to the expected outcome at the population level. Identification of $\mu(t)$ from the observed data rests on two standard assumptions.

\begin{assumption}[Strong Ignorability]
\label{ass:ignorability}
Let $Y(x)$ denote the potential outcome under treatment $X = x$. Conditional on the observed confounders $\boldsymbol{C}$, the treatment $X$ actually received by a unit is independent of that unit's potential outcome under any fixed hypothetical treatment curve $x$:
%the functional treatment is independent of the potential outcomes:
$$X \perp\!\!\!\perp Y(x) \mid \boldsymbol{C} \qquad \forall\, x \in L^2(\mathcal{T}).$$
\end{assumption}

Assumption~\ref{ass:ignorability} requires that all confounding between the functional treatment and the outcome is captured by $\boldsymbol{C}$, so that adjusting for $\boldsymbol{C}$ is sufficient to remove confounding bias. Under this condition, identification of $\mu(t)$ reduces to the construction of weights that render the distribution of $X_i(t)$ independent of $\boldsymbol{C}_i$ in the weighted sample. For scalar continuous treatments, \citet{Hirano_Imbens} showed that such weights can be derived for a continuous treatment $T$ from the generalized propensity score (GPS) $r(t, \boldsymbol{c}) = f_{T \mid \boldsymbol{C}}(t \mid \boldsymbol{c})$. When the treatment is functional, however, no probability density is well-defined on $L^2(\mathcal{T})$ \citep{Delaigle}, and the GPS cannot be directly extended to this setting. We therefore represent the functional treatment $X(t)$ with a finite-dimensional representation via its Karhunen--Lo\`eve expansion, truncated at rank $L$:
\begin{equation}
\label{eq:kl}
X(t) \approx \sum_{k=1}^{L} A_k \cdot \phi_k(t),
\end{equation}
where $\phi_k(t)$ are the orthonormal eigenfunctions of $\mathrm{Cov}(X(s), X(t))$ and $A_k = \int_{\mathcal{T}} X(t) \cdot \phi_k(t) \, dt$ are the corresponding FPC scores satisfying $\mathbb{E}[A_k] = 0$ and $\mathrm{Var}(A_k) = \lambda_k$. The truncation level $L$ is chosen to explain a pre-specified proportion of the total variance of $X(t)$. The vector $\boldsymbol{A} = (A_1, \ldots, A_L)^\top$ serves as the finite-dimensional surrogate for $X(t)$ throughout the analysis.
The GPS naturally extends to this setting as the conditional density of $\boldsymbol{A}^*$ given $\boldsymbol{C}^*$, $r^*(\boldsymbol{a}, \boldsymbol{c}) = f_{\boldsymbol{A}^* \mid \boldsymbol{C}^*}(\boldsymbol{a} \mid \boldsymbol{c})$, with corresponding standardized functional propensity score (SFPS) weight $w_i^* = f_{\boldsymbol{A}^*}(\boldsymbol{A}_i^*) / r^*(\boldsymbol{A}^*_i, \boldsymbol{C}_i^*)$, where $\boldsymbol{A}^*$ and $\boldsymbol{C}^*$ denote the standardized scores $\boldsymbol{A}$ and confounders $\boldsymbol{C}$. 

\begin{assumption}[Positivity]
\label{ass:positivity}
For any value $\boldsymbol{c}^*$ of the observed confounders with positive density, the conditional distribution of the treatment scores $\boldsymbol{A}^*$ given $\boldsymbol{C}^*=\boldsymbol{c}^*$ has support containing the marginal support of $\boldsymbol{A}^*$. Equivalently:
\[
f_{\boldsymbol{A}^* \mid \boldsymbol{C}^*}(\boldsymbol{a}^* \mid \boldsymbol{c}^*) > 0
\]
for all $\boldsymbol{a}^*$ in the support of $\boldsymbol{A}^*$ and all $\boldsymbol{c}^*$ in the support of $\boldsymbol{C}^*$.
\end{assumption}

Assumption~\ref{ass:positivity} ensures that every treatment profile, represented by its FPC scores, has a nonzero probability of being observed at any covariate value, so that the reweighting scheme is well-defined over the entire support. Under assumptions~\ref{ass:ignorability} and~\ref{ass:positivity}, identification of $\mu(t)$ reduces to constructing weights $\{w_i\}_{i=1}^n$ that make $\boldsymbol{A}_i$ independent of $\boldsymbol{C}_i$ in the weighted sample, thereby recovering the marginal causal effect \citep{Hirano_Imbens}. In order to further simplify the expressions, we drop the asterisks in what follows.

\subsection{Functional propensity score estimation}
\label{sec:fps}

Building on the covariate-balancing framework for functional treatments~\citep{Zhang}, to avoid possible parametric misspecification for the SFPS weights, we rewrite $\{w_i\}_{i=1}^n$ as the solution to the following balancing conditions:
\begin{equation}
\label{eq:balance}
\frac{1}{n}\sum_{i=1}^n w_i \boldsymbol{A}_i \boldsymbol{C}_i^\top = \boldsymbol{0}, \qquad
\frac{1}{n}\sum_{i=1}^n w_i \boldsymbol{A}_i = \boldsymbol{0}, \qquad
\frac{1}{n}\sum_{i=1}^n w_i \boldsymbol{C}_i = \boldsymbol{0}, \qquad
\sum_{i=1}^n w_i = 1.
\end{equation}
Together, the constraints in~\eqref{eq:balance} require that, in the reweighted sample, treatment scores and confounders are mean-zero and mutually uncorrelated, thereby mimicking a population in which treatment assignment is independent of the confounders. Following \citet{Owen} and \citet{Fong}, the weights are obtained by maximizing the empirical likelihood subject to the balancing constraints, which is equivalent to solving the primal problem:
\begin{equation}
\label{eq:primal}
\min_{\boldsymbol{w}}\sum_{i=1}^n \log(w_i)
\quad \text{subject to} \quad
\sum_{i=1}^n w_i \boldsymbol{g}_i = \boldsymbol{0}, \quad
\sum_{i=1}^n w_i = 1, \quad w_i \geq 0,
\end{equation}
where $\boldsymbol{g}_i = \bigl[\boldsymbol{A}_i^\top,\; \boldsymbol{C}_i^\top,\; \mathrm{vec}(\boldsymbol{A}_i \boldsymbol{C}_i^\top)^\top \bigr]^\top \in \mathbb{R}^{L + p + Lp}$ stacks the balancing moments for unit $i$.
Problem~\eqref{eq:primal} is a constrained optimization whose dimension grows as $O(Lp)$, making it computationally demanding when $L$ and $p$ increase. We propose to solve it via its dual formulation (full derivation in Supplementary Material~A). Introducing a multiplier $\lambda \in \mathbb{R}$ for the normalization constraint and $\boldsymbol{\gamma} \in \mathbb{R}^{L+p+Lp}$ for the balancing constraints, the dual problem consists of maximizing the Lagrangian dual function:
\begin{equation}
\max_{\boldsymbol{\gamma},\lambda}\,\mathcal{D}(\lambda, \boldsymbol{\gamma})
= \max_{\boldsymbol{\gamma},\lambda}\,\inf_{\boldsymbol{w} \geq 0}
\mathcal{L}(\boldsymbol{w}, \boldsymbol{\gamma}, \lambda).
\end{equation}
Imposing the stationarity conditions and the primal constraints in~\eqref{eq:primal}, the dual function reduces to:
\begin{equation}
\mathcal{D}(\boldsymbol{\gamma}) = -\sum_{i=1}^n \log\!\bigl(n - \boldsymbol{\gamma}^\top \boldsymbol{g}_i\bigr).
\end{equation}
To guarantee strict positivity of the weights and remove the implicit constraint $n - \boldsymbol{\gamma}^\top \boldsymbol{g}_i > 0$, we introduce the reparametrization $n - \boldsymbol{\gamma}^\top \boldsymbol{g}_i = K\,e^{\boldsymbol{\theta}^\top \boldsymbol{g}_i}$ for $\boldsymbol{\theta} \in \mathbb{R}^{L+p+Lp}$ and $K > 0$. The normalization constraint uniquely determines $K = \sum_{i=1}^n e^{-\boldsymbol{\theta}^\top \boldsymbol{g}_i}$, and the weights are recovered as the softmax transformation of $-\boldsymbol{\theta}^\top \boldsymbol{g}_i$:
\begin{equation}
\label{eq:weights}
w_i = \frac{e^{-\boldsymbol{\theta}^\top \boldsymbol{g}_i}}{\sum_{j=1}^n e^{-\boldsymbol{\theta}^\top \boldsymbol{g}_j}}, \qquad i = 1, \ldots, n.
\end{equation}
Substituting into $\mathcal{D}$ and exploiting $\bar{\boldsymbol{g}} = \boldsymbol{0}$ (since FPC scores and confounders are standardized), the dual function becomes:
\begin{equation}
    \mathcal{D}(\boldsymbol{\theta}) = -n\log\!\left(\sum_{i=1}^n e^{-\boldsymbol{\theta}^\top \boldsymbol{g}_i}\right).
\end{equation}
The dual problem reduces then to the following smooth, unconstrained convex minimization in $\boldsymbol{\theta} \in \mathbb{R}^{L + p + Lp}$:
\begin{equation}
\label{eq:dual}
\min_{\boldsymbol{\theta}} \; \log\!\left(\sum_{i=1}^n e^{-\boldsymbol{\theta}^\top \boldsymbol{g}_i}\right).
\end{equation}
A finite minimizer \(\boldsymbol{\theta}^\ast\) exists whenever the empirical balancing constraints in~\eqref{eq:primal} are feasible with positive weights. 
Under this condition, $\boldsymbol{\theta}^\ast$ can be obtained via the BFGS quasi-Newton algorithm, and the weights $w_i$ are then recovered via~\eqref{eq:weights}.
%The dual problem reduces to the following smooth, unconstrained minimization in $\boldsymbol{\theta} \in \mathbb{R}^{L + p + Lp}$:
%\begin{equation}
%\label{eq:dual}
%\min_{\boldsymbol{\theta}} \; \log\!\left(\sum_{i=1}^n e^{-\boldsymbol{\theta}^\top \boldsymbol{g}_i}\right).
%\end{equation}
%Problem~\eqref{eq:dual} involves no constraints and is strictly convex and therefore the optimal $\boldsymbol{\theta}^*$ exist and it is obtained via the BFGS quasi-Newton algorithm, and the weights are recovered as
%\begin{equation}
%\label{eq:weights}
%w_i^* = \frac{e^{-\boldsymbol{\theta}^{*\top} \boldsymbol{g}_i}}{\sum_{j=1}^n e^{-\boldsymbol{\theta}^{*\top} \boldsymbol{g}_j}}, \qquad i = 1, \ldots, n.
%\end{equation}
%For numerical stability, the log-sum-exp trick is employed when evaluating the objective in~\eqref{eq:dual}.
Once the weights $\{w_i\}_{i=1}^n$ are obtained, the effect function $\mu(t)$ can be estimated by approximating it, without loss of generality, over the same FPCA basis $\{\phi_k(t)\}$ used to represent $X(t)$: $$\mu(t) \approx \sum_{k=1}^{L^*} \mu_k \cdot \phi_k(t),$$ where the truncation level $L^*$ is chosen by a second proportion-of-variance-explained threshold and may differ from $L$. The coefficients $\boldsymbol{\mu} = (\mu_0, \mu_1, \ldots, \mu_{L^*})^\top$ are estimated by weighted least squares using the SFPS weights:
$$\hat{\boldsymbol{\mu}} = (\boldsymbol{A}^\top \boldsymbol{W} \boldsymbol{A})^{-1} \boldsymbol{A}^\top \boldsymbol{W} \boldsymbol{y},$$
where $\boldsymbol{y} = (Y_1, \ldots, Y_n)^\top$, $\boldsymbol{W} = \mathrm{diag}(w_1, \ldots, w_n)$, and $\boldsymbol{A} \in \mathbb{R}^{n \times (1 + L^*)}$ is the design matrix of treatment FPC scores augmented with an intercept column. The estimated effect function is then $\hat{\mu}(t) = \sum_{k=1}^{L^*} \hat{\mu}_k \cdot \phi_k(t)$. Although we use the FPCA basis to represent the effect function, the framework is not tied to this choice, and other fixed basis expansions could be used in the outcome model. We adopt the FPCA basis because {\color{black}it provides a data-adaptive and low-dimensional representation of the functional treatment, retaining the leading sources of variation with a limited number of components.} Its orthogonality yields a simple representation of the functional linear model in terms of FPC scores, reducing estimation to standard weighted linear regression. Full details of the estimation procedure are given in Supplementary Material~B.

\subsection{Extension to Functional Covariates}
\label{sec:func_cov}

The framework of Section~\ref{sec:fps} accommodates scalar confounders $\boldsymbol{C} \in \mathbb{R}^p$. We now extend it to settings where one or more covariates are themselves functional. Without loss of generality, consider a single functional covariate $D(t)$, $t \in \mathcal{T}$, observed alongside $\boldsymbol{C}$; the extension to multiple functional covariates follows by applying the same argument to each in turn. We apply FPCA to $D(t)$, representing it as:
$$D(t) \approx \sum_{k=1}^{K} b_k \cdot \eta_k(t),$$
where $\eta_k(t)$ are the orthonormal eigenfunctions of $\mathrm{Cov}(D(s), D(t))$, with corresponding eigenvalues $\nu_k$, and $b_k = \int_{\mathcal{T}} D(t) \cdot \eta_k(t) \, dt$ are the FPC scores satisfying $\mathbb{E}[b_k] = 0$ and $\mathrm{Var}(b_k) = \nu_k$. The expansion is truncated at $K$ components, chosen to explain a pre-specified proportion of the total variance in $D(t)$. Denote the $n \times K$ matrix of FPC scores as $\boldsymbol{B}$, with $[\boldsymbol{B}]_{ik} = b_{ik}$. Functional covariates are incorporated into the balancing procedure by augmenting the scalar covariate matrix $\boldsymbol{C}$ with $\boldsymbol{B}$:
$$\bar{\boldsymbol{C}} = \begin{bmatrix} \boldsymbol{C} \mid \boldsymbol{B} \end{bmatrix} \in \mathbb{R}^{n \times (p+K)}.$$
The augmented matrix $\bar{\boldsymbol{C}}$ replaces $\boldsymbol{C}$ throughout the weight estimation procedure of Section~\ref{sec:fps}. This leaves the structure of the dual optimization~\eqref{eq:dual} unchanged: the functional covariate contributes to balancing through its finite-dimensional FPC representation, and the number of optimization variables increases only by $K$, which remains small in practice.

\subsection{Extension to Functional Outcomes}
\label{sec:func_out}

We now extend the framework to settings where the outcome is itself a function. Given a functional treatment $X(s)$, $s \in \mathcal{S}$, and a functional outcome $Y(t)$, $t \in \mathcal{T}$, the causal effect is no longer a univariate function $\mu(t)$ but a bivariate causal effect surface $\mu(s,t)$, characterizing how the treatment at domain point $s$ causally influences the outcome at domain point $t$. The corresponding function-on-function marginal structural model \citep{Ramsay} is:
\begin{equation}
\label{eq:fof}
\mathbb{E}[Y(t) \mid X] = \mu_0(t) + \int_{\mathcal{S}} \mu(s, t) \cdot X(s) \, ds, \qquad t \in \mathcal{T},
\end{equation}
where $\mu_0(t)$ is a functional intercept capturing the baseline mean response trajectory.

To make estimation of~\eqref{eq:fof} feasible, we represent $X(s)$ and $Y(t)$ through their respective Karhunen--Lo\`eve expansions, truncated respectively at $L^*_{\mathcal{S}}$ and $L^*_{\mathcal{T}}$ components, with orthonormal eigenfunctions $\{\phi_k(s)\}$ and $\{\psi_j(t)\}$ and corresponding FPC scores $A_{ik}$ and $c_{ij}$. The causal effect surface $\mu(s,t)$ and the functional intercept $\mu_0(t)$ are expanded in the same bases:
$$\mu(s, t) = \sum_{k=1}^{L^*_{\mathcal{S}}} \sum_{j=1}^{L^*_{\mathcal{T}}} \mu_{kj} \cdot \phi_k(s) \cdot \psi_j(t), \qquad \mu_0(t) = \sum_{j=1}^{L^*_{\mathcal{T}}} \mu_{0j} \cdot \psi_j(t).$$
This choice simplifies the integral in~\eqref{eq:fof}, which reduces to the FPC scores $A_{ik}$ of $X(s)$ and yields a linear system that decouples across output components. Specifically, substituting the expansions into~\eqref{eq:fof} and matching coefficients of $\psi_j(t)$ yields, for each $j = 1, \ldots, L^*_{\mathcal{T}}$, the linear model
$$c_{ij} = \mu_{0j} + \sum_{k=1}^{L^*_{\mathcal{S}}} \mu_{kj} \cdot A_{ik}, \qquad i = 1, \ldots, n.$$
This decouples estimation of $\mu(s,t)$ into $L^*_{\mathcal{T}}$ independent regression problems, one per output component $j$. For each $j$, the coefficient vector $\boldsymbol{\mu}_j = (\mu_{0j}, \mu_{1j}, \ldots, \mu_{L^*_{\mathcal{S}}j})^\top$ is estimated by weighted least squares using the FPS weights $\boldsymbol{W} = \mathrm{diag}(w_1, \ldots, w_n)$:
$$\hat{\boldsymbol{\mu}}_j = (\boldsymbol{A}^\top \boldsymbol{W} \boldsymbol{A})^{-1} \boldsymbol{A}^\top \boldsymbol{W} \boldsymbol{c}_j,$$
where $\boldsymbol{c}_j = (c_{1j}, \ldots, c_{nj})^\top$ and $\boldsymbol{A} \in \mathbb{R}^{n \times (1 + L^*_{\mathcal{S}})}$ is the design matrix of treatment FPC scores augmented with an intercept. The estimated causal effect surface and functional intercept are then reconstructed as:
\begin{equation}
    \hat{\mu}(s, t) = \sum_{k=1}^{L^*_{\mathcal{S}}} \sum_{j=1}^{L^*_{\mathcal{T}}} \hat{\mu}_{kj} \cdot \phi_k(s) \cdot \psi_j(t), \qquad \hat{\mu}_0(t) = \sum_{j=1}^{L^*_{\mathcal{T}}} \hat{\mu}_{0j} \cdot \psi_j(t).
\end{equation}
As in the case with a scalar outcome, the SFPS weights are used to construct a weighted pseudo-population in which the treatment FPC scores are balanced with respect to the observed confounders. 
Under the identification assumptions stated above, the resulting weighted least squares regressions target the marginal causal effect surface rather than a confounded regression surface. Full algebraic details of the basis expansion and the derivation of the linear system are provided in Supplementary Material~B.

\subsubsection{Overlapping Domains}
\label{sec:overlapping}
The estimation procedure above assumes that the domains $\mathcal{S}$ and $\mathcal{T}$ are either distinct or do not impose structural constraints on the support of $\mu(s,t)$. When the two domains coincide or substantially overlap, the interpretation of $\mu(s,t)$ must respect the temporal ordering required for a causal effect. In causal inference, a cause must precede its effect; hence, if both the treatment and the outcome are observed over a common time domain $\mathcal{S}=\mathcal{T}=[\tau_0,\tau_1]$, the value of the outcome at time $t$ can only be causally affected by treatment values observed at times $s \leq t$. This temporal-precedence restriction is consistent with the intuition underlying Granger-type notions of causality for time-indexed data, where past information may be used to explain future outcomes, but future information cannot play a causal role for the present~\citep{granger1969investigating}.
This yields the non-anticipativity constraint:
\begin{equation}
\label{eq:historical_constraint}
\mu(s, t) = 0 \qquad \text{for all } s > t,
\end{equation}
which restricts the support of the causal effect surface to the lower-triangular region $\{(s,t): \tau_0 \leq s \leq t \leq \tau_1\}$, as in the historical functional linear model of~\citet{malfait2003}. This specification corresponds to a historical, non-concurrent function-on-function model, in which the outcome at time $t$ depends only on treatment values observed up to time $t$. The non-anticipativity restriction is therefore a causal constraint induced by the temporal ordering of treatment and outcome, rather than solely a technical estimation consideration.
Constraint~\eqref{eq:historical_constraint} affects only the estimation of $\mu(s,t)$ and has no bearing on the weight estimation procedure of Section~\ref{sec:fps}.

Since the tensor-product FPCA basis is globally supported over $\mathcal{S} \times \mathcal{T}$ and cannot encode constraint~\eqref{eq:historical_constraint}, we follow \citet{malfait2003} and replace it with a triangular basis $\{B_m(s,t)\}_{m=1}^{M}$ of piecewise linear functions defined on a finite-element grid confined to $\{(s,t): s \leq t\}$. The causal effect surface is then expanded as:
$$\mu(s, t) = \sum_{m=1}^{M} \alpha_m \cdot B_m(s, t),$$
which satisfies~\eqref{eq:historical_constraint} exactly by construction. The coefficients $\boldsymbol{\alpha} = (\alpha_1, \ldots, \alpha_M)^\top$ are estimated via weighted least squares using the FPS weights, following the same procedure as in Section~\ref{sec:func_out} with the triangular basis replacing the FPCA tensor-product basis. 

\section{Simulations}
\label{sec:sim}%
\subsection{Functional Propensity Score Simulation}
\label{sec:sim_fps}
We study the finite-sample performance of the proposed estimator under varying degrees of confounding and model misspecification. The simulation design follows~\citet{Zhang} and allows controlled variation in both the treatment–covariate and covariate–outcome relationships. We generate $R = 200$ independent datasets, each with $n = 200$ subjects for each simulation. The functional treatment is constructed as a linear combination of six trigonometric eigenfunctions $\phi_{2j-1}(t) = \sqrt{2}\sin(2\pi j t)$ and $\phi_{2j}(t) = \sqrt{2}\cos(2\pi j t)$ for $j = 1,2,3$, with corresponding FPC scores defined as $A_{i1} = 4Z_{i1}$, $A_{i2} = 2\sqrt{3}Z_{i2}$, $A_{i3} = 2\sqrt{2}Z_{i3}$, $A_{i4} = 2Z_{i4}$, $A_{i5} = Z_{i5}$, and $A_{i6} = Z_{i6}/\sqrt{2}$, where $Z_{i1}, \ldots, Z_{i6} \stackrel{\mathrm{i.i.d.}}{\sim} \mathcal{N}(0,1)$. The scalar outcome is generated according to:
\begin{equation*}
Y_i = 1 + \int_0^1 \mu(t) \cdot  X_i(t) \, dt + g(\boldsymbol{C}_i) + e_i, \quad e_i \sim \mathcal{N}(0,25),
\end{equation*}
with true effect function:
\begin{equation*}
\mu(t) = 2\sqrt{2}\sin(2\pi t) + \sqrt{2}\cos(2\pi t) + \tfrac{\sqrt{2}}{2}\sin(4\pi t) + \tfrac{\sqrt{2}}{2}\cos(4\pi t),
\end{equation*}
and a three-dimensional vector of confounders $\boldsymbol{C}_i = (C_{i1}, C_{i2}, C_{i3})^\top$. We consider four settings that vary the functional form of the treatment-covariate and covariate-outcome relationships. In Setting~1, both relationships are linear, whereas Settings~2 and~3 introduce nonlinearities in the treatment–covariate and covariate–outcome relationships, respectively, and Setting~4 combines both nonlinear components. Details of the data-generating mechanisms are provided in Supplementary Material~C. For each setting, the truncation levels $L$ and $L^*$ are selected based on the percentage of variance explained (PVE), with $\mathrm{PVE}_L, \mathrm{PVE}_{L^*} \in \{0.95, 0.99\}$. Estimation accuracy is evaluated using the integrated squared error (ISE), with median (MISE), average (AISE), and integrated squared bias (ISB) computed across simulation runs. Results are compared to the unweighted estimator as a baseline.

Table~\ref{tab:fps_main} summarizes the results and additional detailed diagnostics are provided in Supplementary Material~C. Across all settings, the proposed weighting approach substantially reduces bias relative to the unweighted estimator, indicating effective control of confounding regardless of the functional form of the relationships. The performance of MISE and AISE depends on the specification of the outcome model. When the truncation level is correctly specified ($\mathrm{PVE}_{L^*} = 0.95$), the weighted estimator improves estimation accuracy across all settings, while under misspecification ($\mathrm{PVE}_{L^*} = 0.99$), performance deteriorates in settings with simpler dependence structures but remains comparable or improves in settings with nonlinear treatment-covariate relationships, where additional components are required to achieve adequate balance. Increasing $\mathrm{PVE}_L$ mitigates this effect, highlighting the role of the treatment representation in balancing performance. Overall, the results support the use of $\mathrm{PVE}_L = \mathrm{PVE}_{L^*} = 0.95$ as a practical default choice.

\begin{table}[H]
\centering
\caption{MISE, AISE and ISB  for the weighted and unweighted estimators across simulation settings and truncation levels $(\mathrm{PVE}_L, \mathrm{PVE}_{L^*})$.}
\label{tab:fps_main}
\begin{tabular}{llllcccccc}
\toprule
& & & &
\multicolumn{3}{c}{$\mathrm{PVE}_{L^*} = 0.95$} &
\multicolumn{3}{c}{$\mathrm{PVE}_{L^*} = 0.99$} \\
\cmidrule(lr){5-7} \cmidrule(lr){8-10}
& & & & MISE & AISE & ISB & MISE & AISE & ISB \\
\midrule

& & & Unweighted & 0.3135 & 0.3288 & 0.2397 & 0.5938 & 0.7270 & 0.2431 \\
Setting 1 & $\mathrm{PVE}_L = 0.95$ & & Weighted
                          & 0.1623 & 0.2119 & 0.0009 & 0.8544 & 1.2635 & 0.0066 \\
& $\mathrm{PVE}_L = 0.99$ & & Weighted
                          & 0.1642 & 0.2202 & 0.0013 & 0.7636 & 1.1919 & 0.0072 \\
\midrule

& & & Unweighted & 0.3489 & 0.3660 & 0.2461 & 0.6902 & 0.8789 & 0.2487 \\
Setting 2 & $\mathrm{PVE}_L = 0.95$ & & Weighted
                          & 0.1104 & 0.1428 & 0.0010 & 0.7102 & 1.0280 & 0.0057 \\
& $\mathrm{PVE}_L = 0.99$ & & Weighted
                          & 0.1172 & 0.1461 & 0.0011 & 0.5616 & 0.8493 & 0.0075 \\
\midrule

& & & Unweighted & 0.3116 & 0.3296 & 0.2394 & 0.5987 & 0.7334 & 0.2430 \\
Setting 3 & $\mathrm{PVE}_L = 0.95$ & & Weighted
                          & 0.1639 & 0.2131 & 0.0008 & 0.8308 & 1.2758 & 0.0068 \\
& $\mathrm{PVE}_L = 0.99$ & & Weighted
                          & 0.1723 & 0.2216 & 0.0012 & 0.7792 & 1.2034 & 0.0073 \\
\midrule

& & & Unweighted & 0.3490 & 0.3667 & 0.2458 & 0.7051 & 0.8866 & 0.2487 \\
Setting 4 & $\mathrm{PVE}_L = 0.95$ & & Weighted
                          & 0.1119 & 0.1438 & 0.0009 & 0.7292 & 1.0331 & 0.0054 \\
& $\mathrm{PVE}_L = 0.99$ & & Weighted
                          & 0.1187 & 0.1472 & 0.0011 & 0.5880 & 0.8548 & 0.0074 \\
\bottomrule
\end{tabular}
\end{table}

\begin{figure}[H]
\centering
\begin{minipage}{\textwidth}
    \centering
    \includegraphics[width=\textwidth]{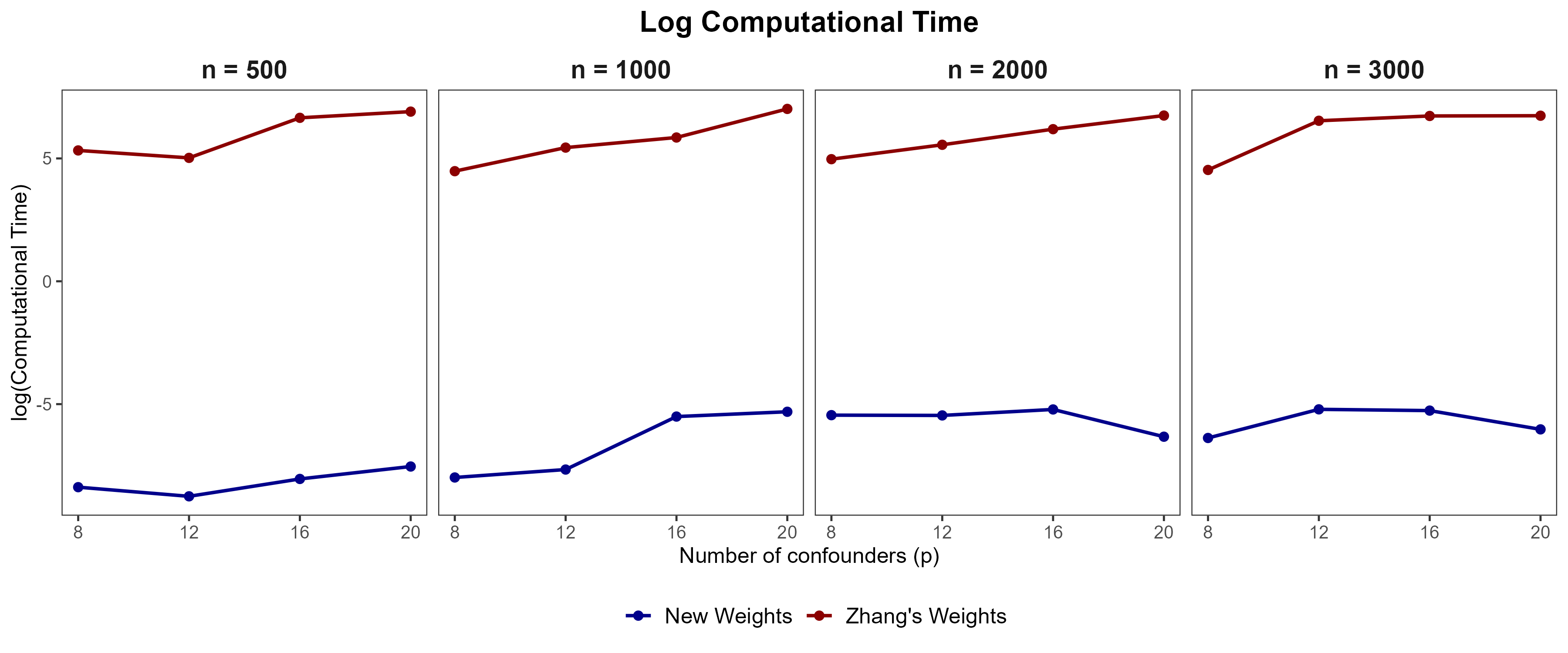}
    \subcaption{Log elapsed time by $n$}
\end{minipage}
\\[1em]
\begin{minipage}{\textwidth}
    \centering
    \includegraphics[width=\textwidth]{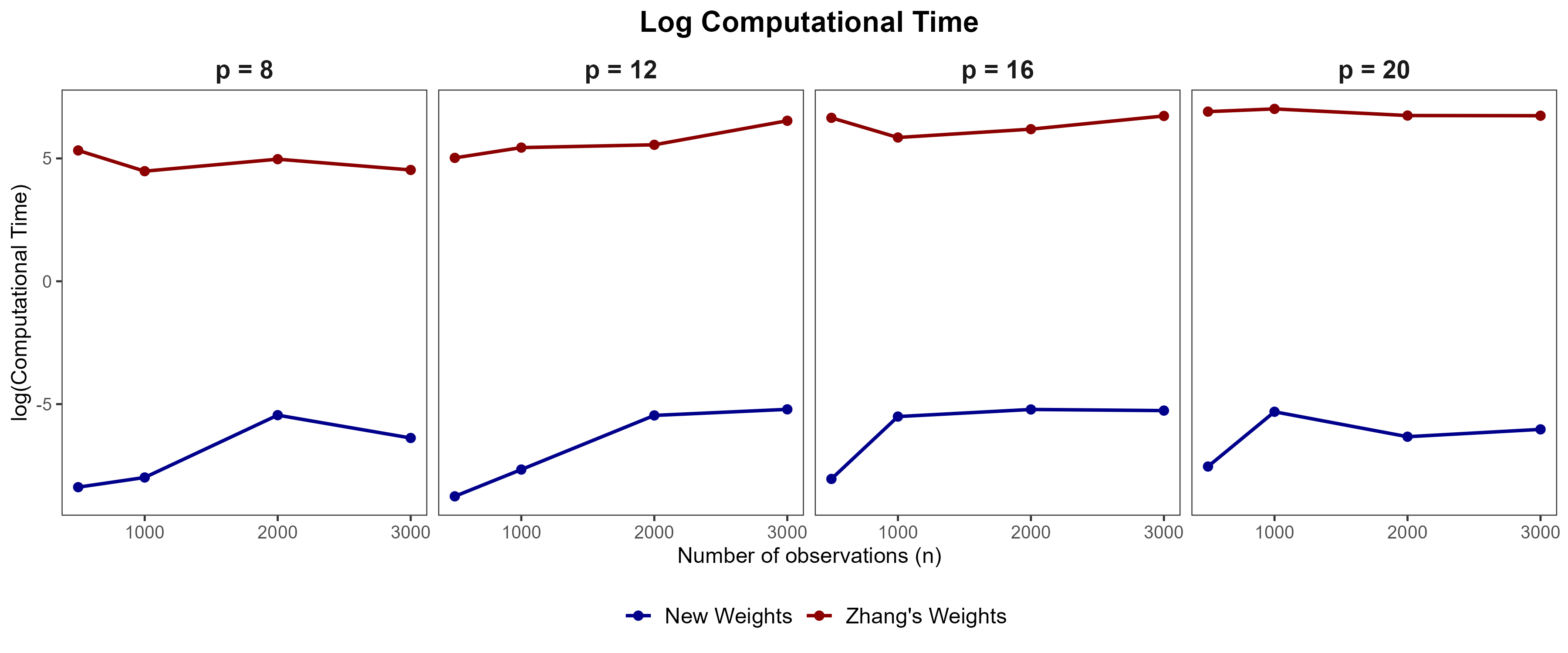}
    \subcaption{Log elapsed time by $p$}
\end{minipage}
\caption{Log wall-clock time for the proposed method and \citet{Zhang} as functions of $n$ (top) and $p$ (bottom), under Setting~1. Each curve corresponds to a fixed value of the other dimension.}
\label{fig:comptimes_log}
\end{figure}

We further compare the proposed method with \citet{Zhang} in terms of computational scalability and covariate balance. Under Setting~1, the proposed estimator achieves substantial reductions in computational time relative to the original approach, which requires approximately 1890 times longer running time. To study scaling behavior, we consider $n \in \{500, 1000, 2000, 3000\}$ and $p \in \{8, 12, 16, 20\}$. Figure~\ref{fig:comptimes_log} shows that the computational cost of \citet{Zhang} increases rapidly with $p$, whereas the proposed method remains stable across the range of $(n,p)$ considered. 

\begin{figure}[H]
\centering
\includegraphics[width=0.85\textwidth]{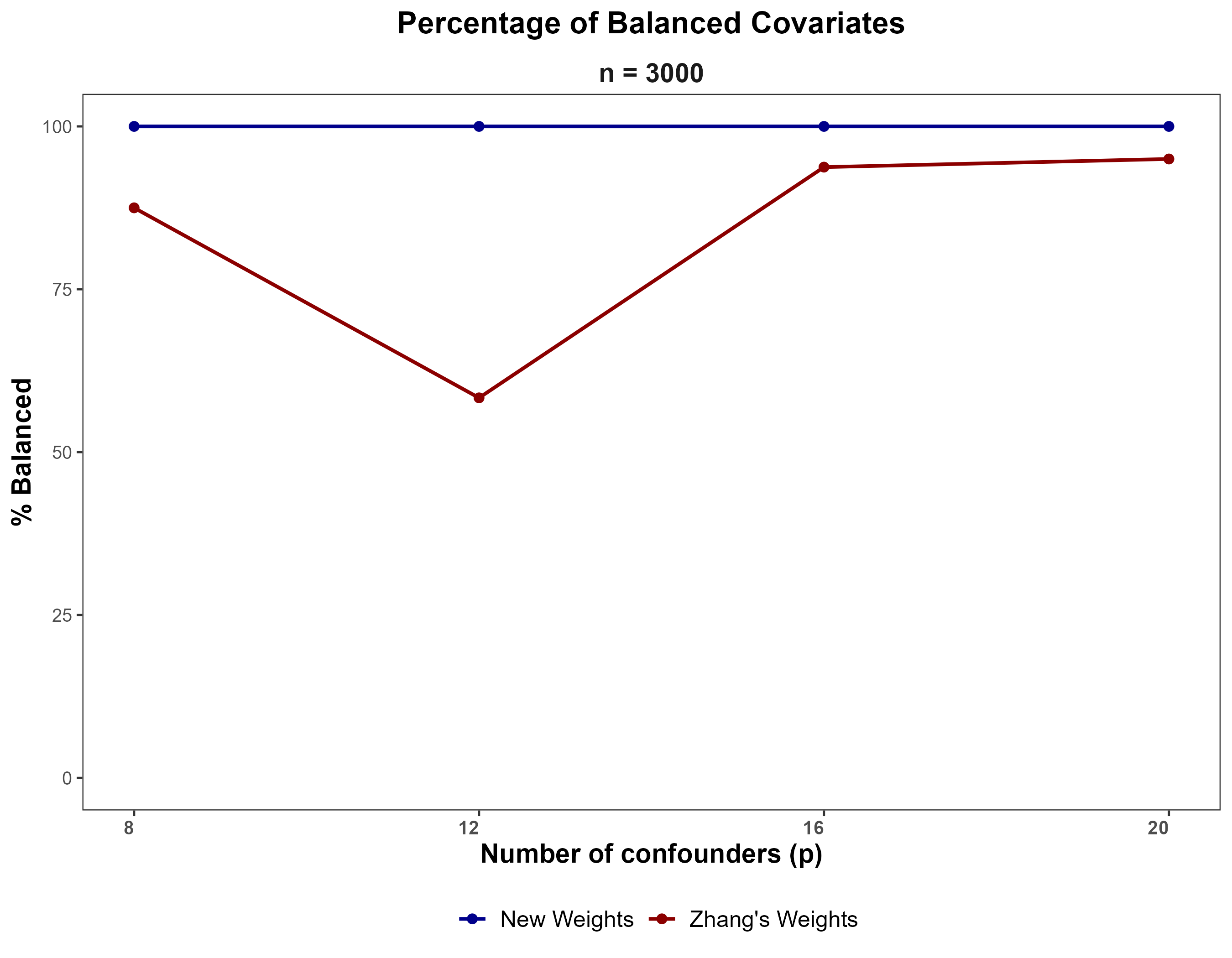}
\caption{Percentage of confounder-FPC pairs with absolute Pearson correlation below $0.1$, as a function of $p$, for $n = 3000$ under Setting~1.}
\label{fig:balance_pct}
\end{figure}

To assess whether the computational gains are accompanied by adequate covariate balance, we fix the largest sample size considered in the timing experiment ($n=3000$) and evaluate balance across the same values of $p$. For each value of $p$, we compute the absolute Pearson correlation between each confounder and each selected FPC score. A confounder-FPC pair is considered balanced if the absolute correlation is below $0.1$. Figure~\ref{fig:balance_pct} shows that the proposed method achieves stable covariate balance across all values of $p$, whereas the original FPS method exhibits erratic and unpredictable behavior, showing fluctuating performance across different values of $p$ and $n$, with no systematic pattern, occasionally dropping to very low values.

\subsection{Functional Covariates and Outcome Simulation}
\label{sec:sim_ext}
We evaluate the proposed method in the extended setting where both covariates and outcomes are functional. This simulation examines the ability of the framework to reduce confounding bias and accurately estimate causal effect surfaces when functional covariates and outcomes are present. We generate $R = 200$ datasets, each with $n = 1000$ subjects. The functional treatment is simulated as in Section~\ref{sec:sim_fps}. The functional covariate is defined as $D_i(t) = \sum_{k=1}^{4} \lambda_{ik} \cdot \eta_k(t)$, where $\eta_k = \phi_k$ and the FPC scores are given by $\lambda_{i1} = f(Z_{i1}) + 4\mathcal{U}_{i1}$, $\lambda_{i2} = Z_{i2} + 2\mathcal{U}_{i2}$, $\lambda_{i3} = Z_{i3} + \mathcal{U}_{i3}$, and $\lambda_{i4} = Z_{i4} + \mathcal{U}_{i4}/\sqrt{2}$, with $Z_{ik}, \mathcal{U}_{ik} \stackrel{\mathrm{i.i.d.}}{\sim} \mathcal{N}(0,1)$. The functional outcome is generated according to:
\begin{equation*}
Y_i(t) = \mu_0(t) + \int_0^1 \mu(s,t) \cdot X_i(s) \, ds + g(D_i) + \epsilon_i(t), \qquad \epsilon_i(t) \sim \mathrm{GP}(0, k_{\mathrm{SE}}),
\end{equation*}
where $\mu_0(t) = 1 + \cos(2\pi t)$ and $k_{\mathrm{SE}}(s,t) = \exp[-(s-t)^2/(2 \cdot 0.01)]$. The true causal surface $\mu(s,t)$ is defined as:
\begin{align*}
\mu(s,t) &= 2\sqrt{2}\sin(2\pi s)\cos(2\pi t) + 2\sqrt{2}\sin(2\pi t)\cos(2\pi s) \\
         &\quad + \sqrt{2}\cos(4\pi t)\sin(4\pi s) + \sqrt{2}\cos(4\pi s)\sin(4\pi t).
\end{align*}

As in the previous simulation, we consider four settings that vary the treatment-covariate and covariate-outcome relationships. Details of the data-generating process are provided in Supplementary Material~D. The FPC scores of the functional covariate are incorporated into the balancing constraints as described in Section~\ref{sec:func_cov}. Truncation levels $L_{\mathcal{T}}$ and $L^*_{\mathcal{T}} = L^*_{\mathcal{S}}$  are selected using $\mathrm{PVE}_{L_{\mathcal{T}}}, \mathrm{PVE}_{L^*_{\mathcal{T}}}, \mathrm{PVE}_{L^*_{\mathcal{S}}} \in \{0.95, 0.99\}$. Estimation accuracy is assessed using ISE over the bivariate domain, along with MISE, AISE and ISB across runs. 

Table~\ref{tab:ext_main} summarizes the results and more detailed diagnostics are provided in Supplementary Material~D. The proposed estimator consistently reduces bias across all settings and truncation levels, indicating effective adjustment for confounding in the functional setting. The behavior of MISE and AISE depends on the specification of the outcome model, mirroring the findings of Section~\ref{sec:sim_fps}. Under correct specification, the weighted estimator improves estimation accuracy across settings. Under misspecification, performance deteriorates, but this effect is mitigated by increasing the truncation level used for the treatment representation. Overall, the results confirm that the bias reduction and robustness properties observed in the scalar-outcome setting extend to the more general case with functional covariates and outcomes.

\begin{table}[H]
\centering
\caption{MISE, AISE and ISB for the weighted and unweighted estimators, functional covariate and outcome simulation ($R = 200$ runs, $n = 1000$).}
\label{tab:ext_main}
\begin{tabular}{llllcccccc}
\toprule
& & & & \multicolumn{3}{c}{$\mathrm{PVE}_{L^*_{\mathcal{T}}} = \mathrm{PVE}_{L^*_{\mathcal{S}}}= 0.95$} & \multicolumn{3}{c}{$\mathrm{PVE}_{L^*_{\mathcal{T}}} = \mathrm{PVE}_{L^*_{\mathcal{S}}} = 0.99$} \\
\cmidrule(lr){5-7} \cmidrule(lr){8-10}
& & & & MISE & AISE & ISB & MISE & AISE & ISB \\
\midrule
& & & Unweighted & 1.2750 & 1.2792 & 1.2395 & 1.4272 & 1.4593 & 1.2408 \\
Setting 1 & $\mathrm{PVE}_{L_{\mathcal{T}}} = 0.95$ & & Weighted & 1.0083 & 1.0094 & 0.9989 & 1.9844 & 2.6861 & 1.0085 \\
& $\mathrm{PVE}_{L_{\mathcal{T}}} = 0.99$ & & Weighted & 1.0090 & 1.0099 & 0.9990 & 1.0260 & 1.0291 & 0.9991 \\
\midrule
& & & Unweighted & 1.2760 & 1.3178 & 1.2415 & 1.4455 & 1.5154 & 1.2440 \\
Setting 2 & $\mathrm{PVE}_{L_{\mathcal{T}}} = 0.95$ & & Weighted & 1.0085 & 1.0393 & 0.9994 & 2.0180 & 2.7749 & 1.0049 \\
& $\mathrm{PVE}_{L_{\mathcal{T}}} = 0.99$ & & Weighted & 1.0089 & 1.0399 & 0.9995 & 1.0276 & 1.0594 & 0.9996 \\
\midrule
& & & Unweighted & 3.3053 & 3.3079 & 3.1589 & 3.5760 & 3.6798 & 3.1589 \\
Setting 3 & $\mathrm{PVE}_{L_{\mathcal{T}}} = 0.95$ & & Weighted & 3.2619 & 3.3839 & 2.9272 & 4.9133 & 5.8024 & 2.9354 \\
& $\mathrm{PVE}_{L_{\mathcal{T}}} = 0.99$ & & Weighted & 3.2761 & 3.4019 & 2.9265 & 3.9724 & 4.2508 & 2.9295 \\
\midrule
& & & Unweighted & 3.3076 & 3.3162 & 3.1620 & 3.6202 & 3.7087 & 3.1621 \\
Setting 4 & $\mathrm{PVE}_{L_{\mathcal{T}}} = 0.95$ & & Weighted & 3.2748 & 3.3876 & 2.9289 & 4.9398 & 5.7833 & 2.9338 \\
& $\mathrm{PVE}_{L_{\mathcal{T}}} = 0.99$ & & Weighted & 3.2861 & 3.4131 & 2.9278 & 3.9598 & 4.2760 & 2.9296 \\
\bottomrule
\end{tabular}
\end{table}

\section{Real Case Application}
\label{sec:realdata}%
\subsection{The data}
\label{sec:data}
Our application is based on data from the UK~Biobank \citep{sudlow2015}, a large-scale prospective cohort study that recruited approximately 500,000 participants aged 40-69 across the United Kingdom between 2006 and 2010.  Our objective is to evaluate the effect of midlife body mass index (BMI) trajectories on subsequent metabolic health, controlling for differences in patient health status. A study estimated the causal effect of BMI on metabolic outcomes in the UK Biobank~\citep{wainberg2019}, but it relies on a single summary measure of adiposity at baseline. Rather than focusing on a single adiposity measure, we treat BMI as a functional exposure and estimate its time-varying causal impact. We consider two complementary analyses that differ in the outcome considered: (1) incident Type 2 Diabetes (T2D), and (2) longitudinal HbA1c levels. UK Biobank is an ideal setting for this application due to its large sample size, rich longitudinal data, and comprehensive covariate information, guaranteed by the resource linkage of many data sources (baseline assessments, primary care records, hospital episode statistics). 

BMI and HbA1c trajectories are reconstructed from repeated measurements recorded at UK Biobank assessment visits and linked general practitioner records. To ensure data quality, BMI values are restricted to the clinically plausible range $[12,75]$ kg/m$^2$, while HbA1c values are restricted to the range $[4,15.7]$. Participants are required to satisfy common eligibility criteria: absence of T2D diagnosis (defined using the International Classification of Diseases and Related Health Problems system with the code E11) at age 50, survival throughout follow-up, complete baseline covariates information, and at least one BMI observation within each five-year interval of the exposure window to ensure adequate functional representation. We select a common set of clinical confounders measured at baseline: sex, ethnicity, smoking status, family history of diabetes, hypertension, dyslipidemia, ischaemic heart disease, chronic kidney disease, depression, anemia, and antihypertensive medication use. These variables are established determinants of both adiposity and metabolic outcomes and exhibit associations with BMI trajectories prior to adjustment. Analysis 1 additionally includes baseline HbA1c as a continuous covariate to control for pre-existing metabolic status. Descriptive summaries of the baseline covariates for both analyses are reported in Supplementary Material E.

\subsection{Analysis 1: BMI trajectories and incident T2D}
\label{sec:analysis1}
\begin{figure}[H]
    \centering
    \includegraphics[width=0.75\textwidth]{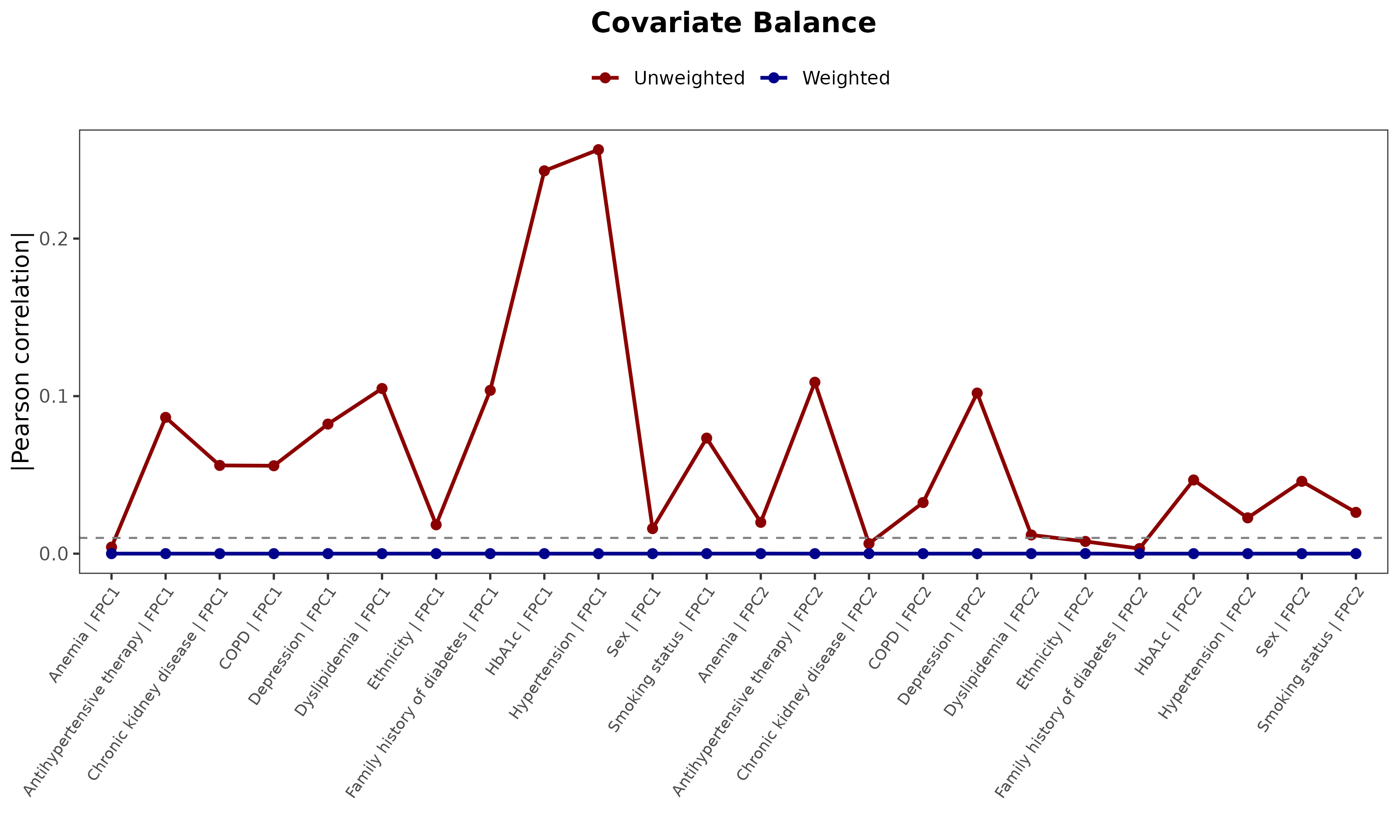}
    \caption{Absolute Pearson correlations between each FPC score of the BMI trajectory and each baseline covariate in Analysis 1, before (red) and after (blue) applying the FPS weights.}
    \label{fig:balance1}
\end{figure}

After applying the inclusion criteria, the cohort consists of $596$ individuals. The exposure is the BMI trajectory over the age range $\mathcal{T} = [50, 65]$, and the outcome is a binary indicator of incident T2D. We represent BMI trajectories using $2$ principal components, which explain 99.3\% of the total variability (97.7\% and 1.6\%, respectively). The leading component captures the overall BMI level, while the second reflects the temporal contrast between earlier and later exposure periods. The selected FPC components are shown in Supplementary Material~E. We employ the FPS to balance the distribution of baseline covariates across the two-dimensional FPCA score space. The FPS weighting procedure removes systematic associations between BMI trajectory and baseline health status, thus isolating the marginal effect of BMI on diabetes risk from confounding by health-related factors. After applying FPS weights, all correlations between FPS and baseline covariates are reduced to zero (Figure~\ref{fig:balance1}), demonstrating that the weighting scheme effectively induces covariate balance in the target population.

\begin{figure}[H]
    \centering
    \includegraphics[width=0.75\textwidth]{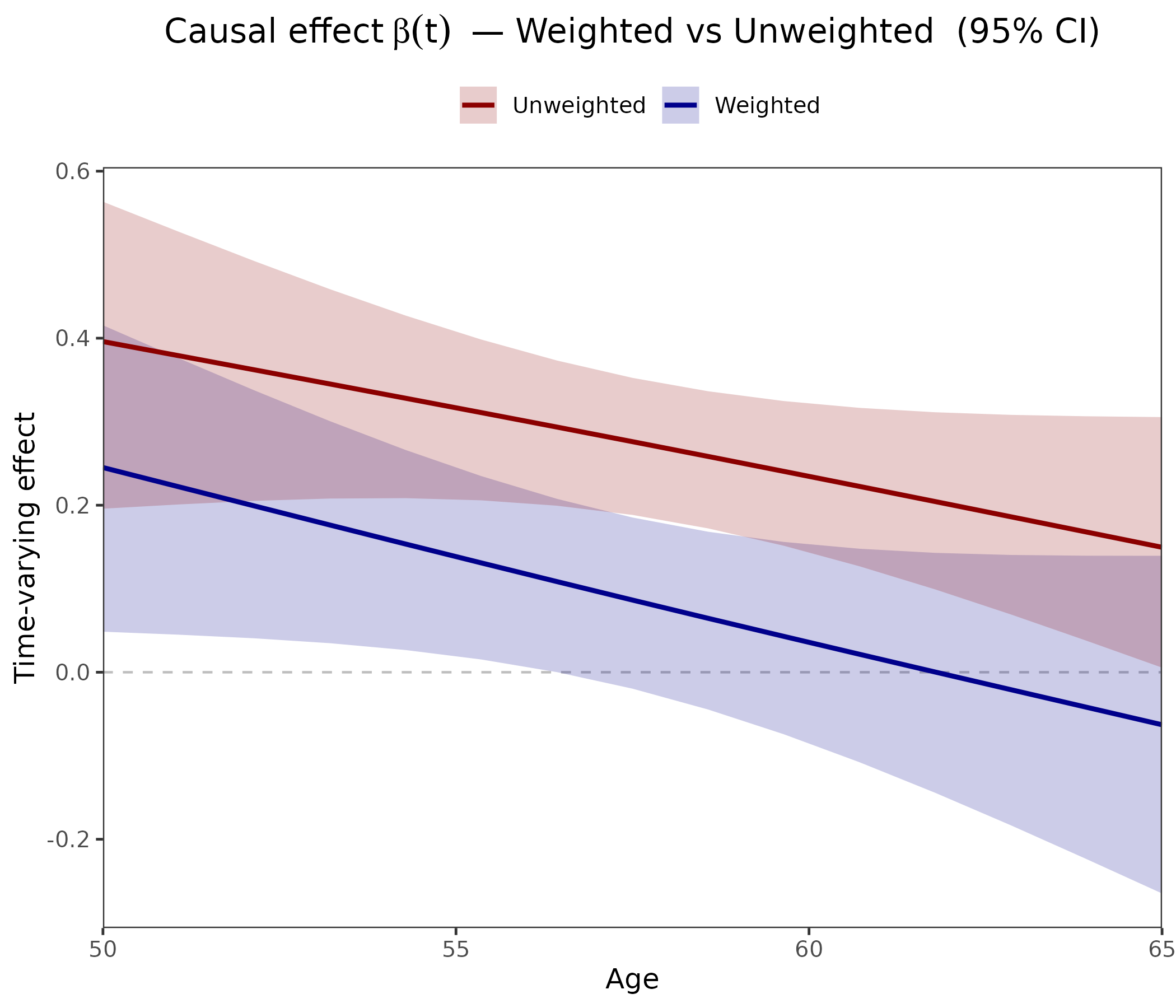}
    \caption{Estimated causal effect function $\hat{\mu}(t)$ for the BMI trajectory on incident T2D risk, with 95\% pointwise bootstrap confidence intervals (Analysis~1).}
    \label{fig:mu_t2d}
\end{figure}

We estimate the causal effect of the BMI trajectory on the risk of T2D using a functional marginal structural logistic model.Figure~\ref{fig:mu_t2d} reports the estimated weighted effect function $\hat{\mu}(t)$ and the corresponding unweighted (naive) estimate, together with 95\% pointwise reverse-percentile bootstrap confidence intervals based on subject-level resampling ($B=1000$). The estimated effect is positive across the observed period, indicating that higher BMI increases the log-odds of developing T2D. The approximately linear shape of the estimated effect function is consistent with the empirical structure of the BMI trajectories in this age range. Adult BMI is generally relatively stable over time~\citep{Heo2002}, and the trajectories in our data show gradual temporal variation with limited evidence of sharp oscillations. Additional plots of the smoothed BMI trajectories are reported in Supplementary Material~E.
Notably, the effect is strongest at the beginning of the exposure window and declines steadily with age, becoming non statistically significant from zero after approximately age 57. The unweighted analysis yields uniformly larger effects that remain significant throughout the exposure window, reflecting possible confounding by baseline health status, which is associated with both higher BMI and increased diabetes risk.  

\subsection{Analysis 2: BMI trajectories and HbA1c dynamics}
\label{sec:analysis2}
\begin{figure}[H]
    \centering
    \includegraphics[width=0.75\textwidth]{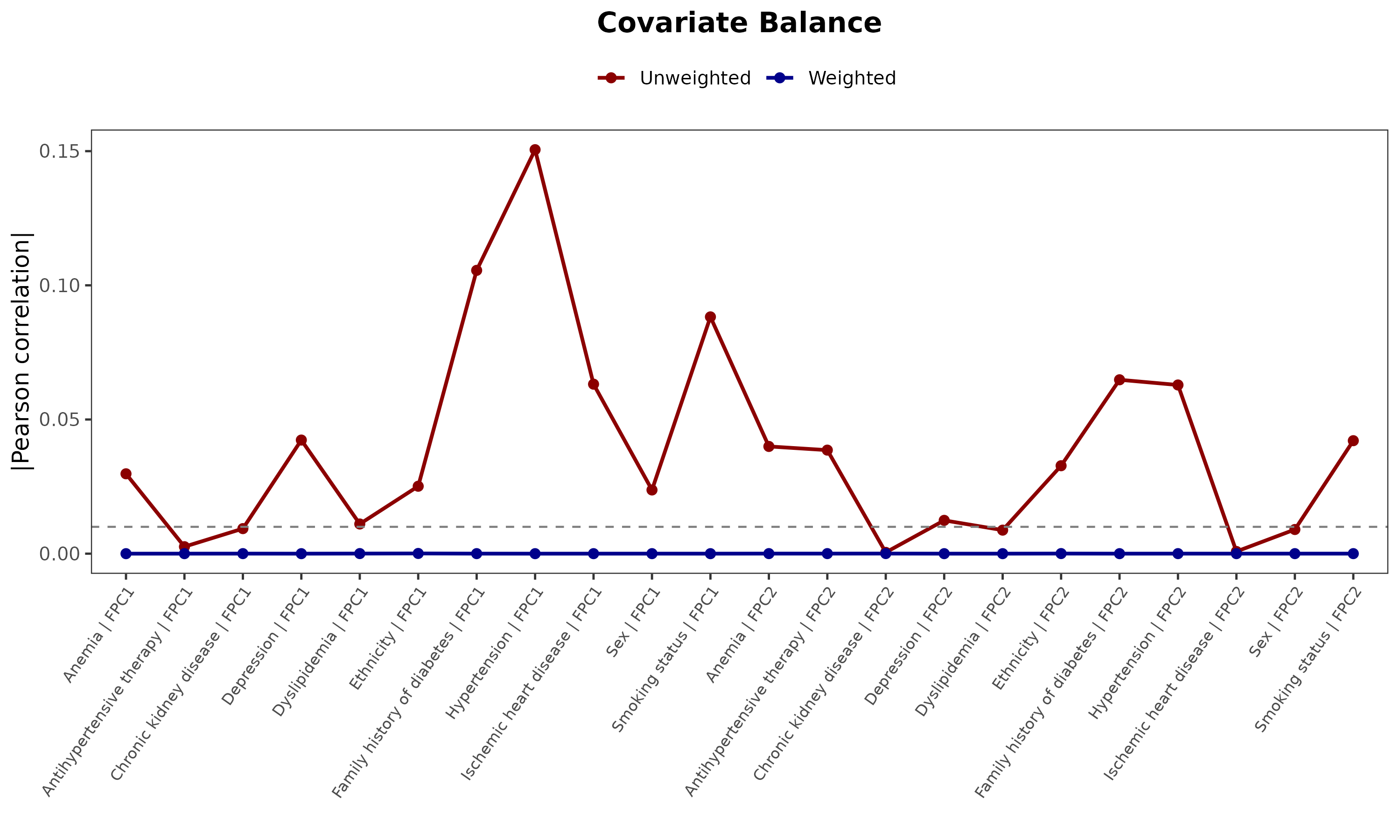}
    \caption{Absolute Pearson correlations between each FPC score of the BMI trajectory and each baseline covariate in Analysis 2, before (red) and after (blue) applying the FPS weights.}
    \label{fig:balance2}
\end{figure}

After applying the inclusion criteria, the cohort for the second analysis consists of $1713$ individuals. The exposure is the BMI trajectory over ages $\mathcal{S} = [50, 60]$, and the outcome is the HbA1c trajectory over ages $\mathcal{T} = [60, 70]$. Both exposure and outcome processes are represented via FPCA and two components are retained for each process, capturing the dominant modes of variation. The selected FPC components for both exposure and outcome are shown in Supplementary Material~E. FPS weights are estimated using the same baseline covariates as in Analysis 1, excluding baseline HbA1c. As before, weighting eliminates systematic associations between exposure features and covariates, achieving near-perfect balance (Figure~\ref{fig:balance2}).
%We apply FPCA to the BMI trajectories retaining $L = 2$ components explains ~1\% of the total trajectory variance (FPC1: 94.4\%; FPC2: 5.7\%). For the HbA1c outcome, two components retain ~1\% of the total variance (FPC1: 89.2\%; FPC2: 10.8\%)

We estimate the causal effect surface $\hat{\mu}(s,t)$, which quantifies how BMI at age $s$ affects HbA1c at age $t$. {\color{black} The estimated surface is shown in Figure~\ref{fig:surface_a2}.
\begin{figure}[H]
    \centering
    \includegraphics[width=0.50\textwidth]{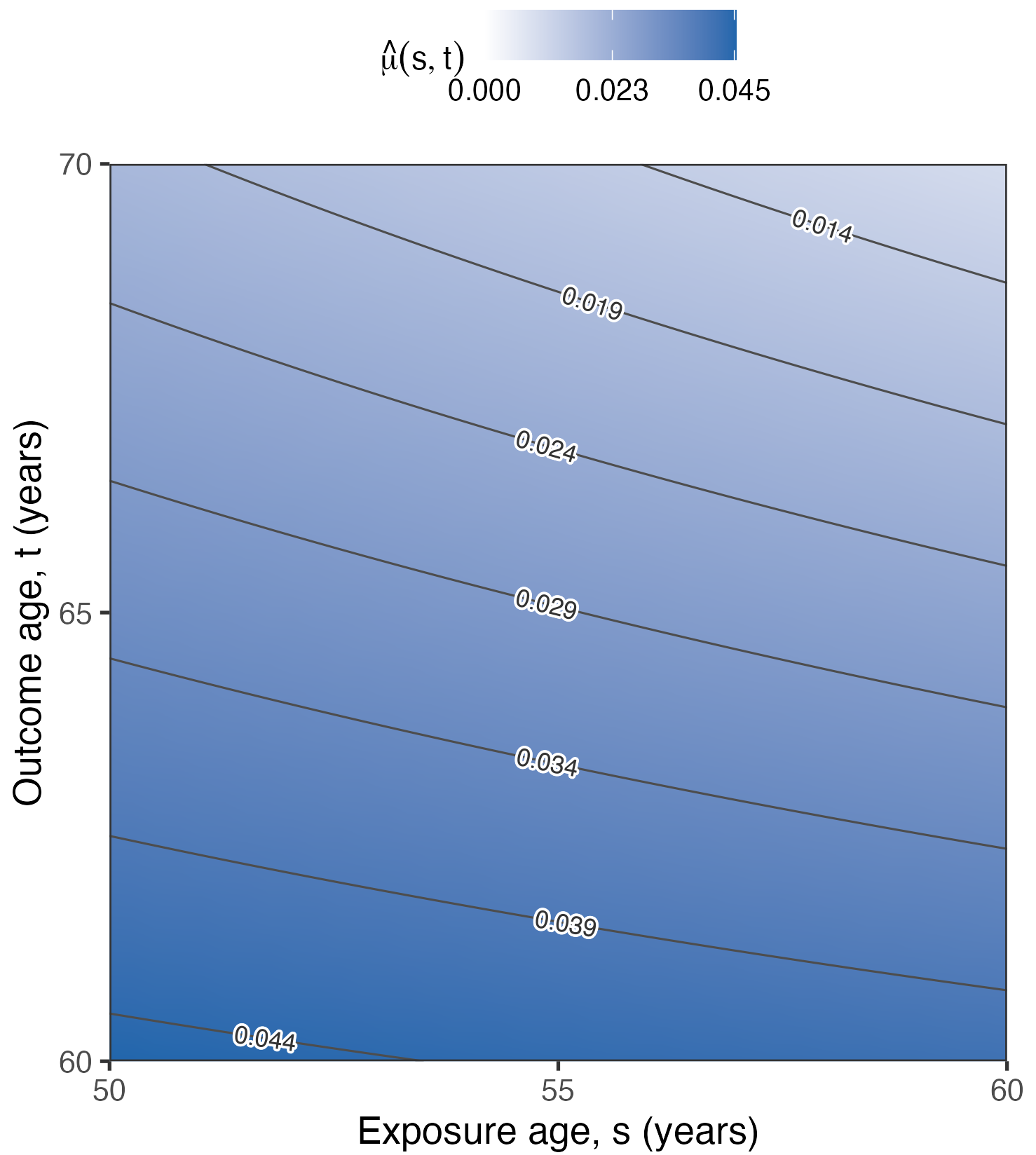}
    \caption{Estimated causal effect surface $\hat{\mu}(s,t)$ in Analysis~2, where $s \in [50,60]$ denotes BMI exposure age and $t \in [60,70]$ denotes HbA1c outcome age. Warmer colors indicate larger positive estimated effects. Contour lines indicate levels of equal estimated effect and are added to facilitate visual comparison.}
    \label{fig:surface_a2}
\end{figure}
}

\begin{figure}[H]
    \centering
    \includegraphics[width=\textwidth]{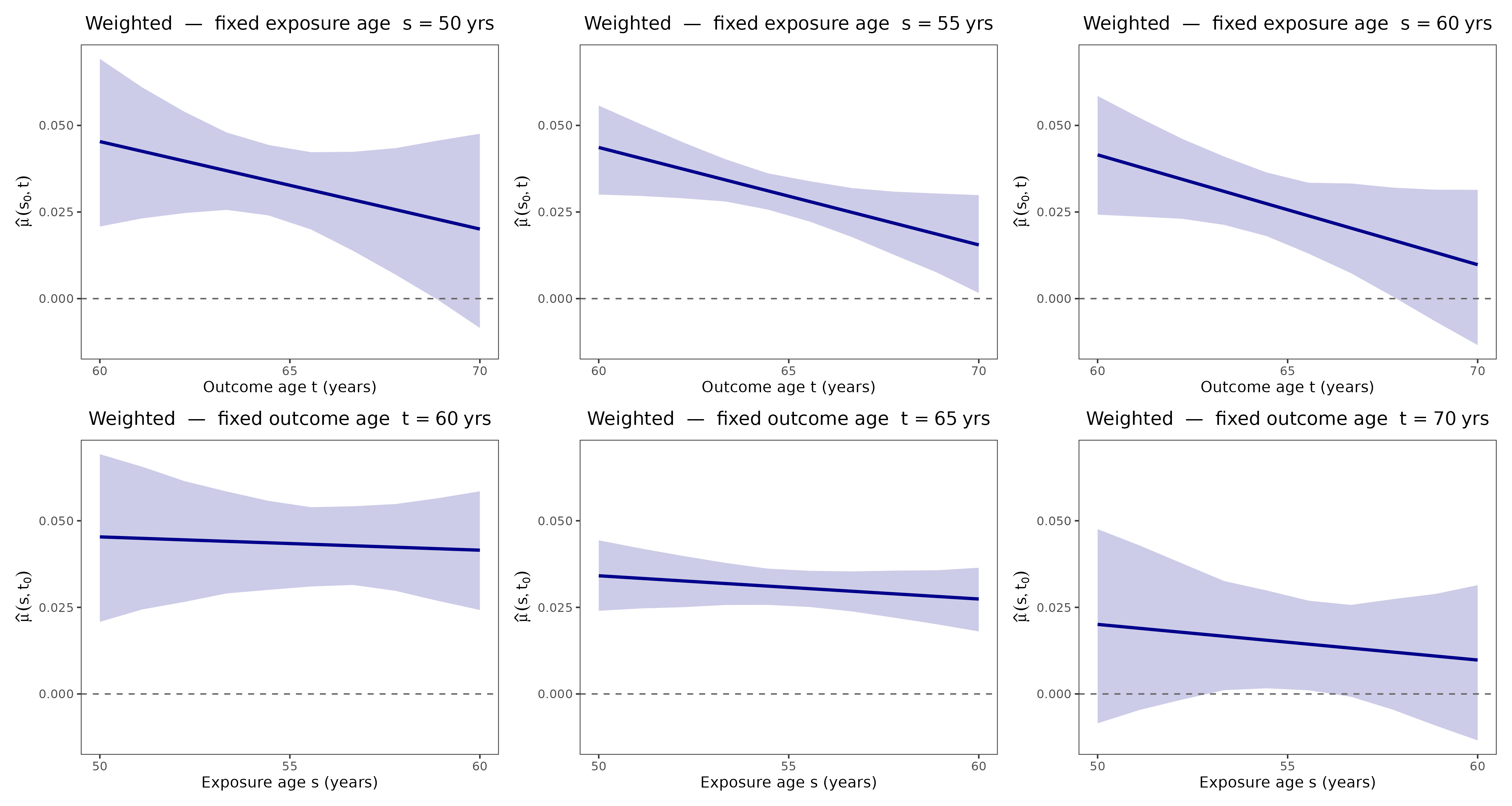}
    \caption{Slices of the estimated causal effect surface $\hat\mu(s, t)$ with 95\% pointwise bootstrap confidence bands. Stars mark pointwise significance ($\alpha = 0.05$). \textit{Top:} $\hat\mu(\cdot,\, t_0)$ as a function of exposure age $s$, for fixed outcome ages $t_0 \in \{60, 65, 70\}$. \textit{Bottom:} $\hat\mu(s_0,\, \cdot)$ as a function of outcome age $t$, for fixed exposure ages $s_0 \in \{50, 55, 60\}$.}
    \label{fig:slices}
\end{figure}

Figure~\ref{fig:slices} presents one-dimensional slices of the estimated effect surface $\hat{\mu}(s,t)$ for selected values of $s \in \{50,55,60\}$ and $t \in \{60,65,70\}$, together with 95\% pointwise reverse-percentile bootstrap confidence intervals based on subject-level resampling ($B=1000$). The corresponding unweighted estimates are reported in Supplementary Material~E. The estimated surface is uniformly positive, indicating that higher BMI at any exposure age increases subsequent HbA1c levels, consistent with the findings in Section~\ref{sec:analysis1}. As in Analysis~1, this smooth pattern is consistent with the gradual temporal evolution expected for adult BMI trajectories~\citep{Heo2002} and with the slowly varying BMI and HbA1c trajectories observed in our data, which are reported in Supplementary Material~E.
The effect exhibits a clear dependence on the timing of exposure, showing a stronger effect in the early part of the exposure window and attenuates as $s$ approaches 60. For a fixed exposure age, the effect shows a mild declining trend over outcome ages $t$, although this variation is less pronounced than the change across exposure ages. Effects at later exposure ages are both weaker and less precisely estimated, with wider confidence intervals that, in some cases (notably for $t = 70$), include zero. These findings align with the pattern observed in Analysis~1: BMI in the early part of midlife exerts the strongest estimated causal effect, while the influence of BMI at later exposure ages weakens. This pattern may reflect the cumulative nature of metabolic risk, where prolonged exposure to elevated BMI in midlife has a more profound impact on long-term glycaemic control than BMI at older ages. 

\section{Discussion}
\label{sec:discussion}%
This article develops a unified propensity score weighting framework for causal inference with functional data. The central methodological contribution is a dual formulation of the functional propensity score weight estimation problem, yielding a smooth, unconstrained optimization that eliminates the proportionality constraint on the weighted covariance structure and avoids the sequential line search required by existing approaches. This formulation yields a scalable estimator whose dimension depends on $p$ and $L$ rather than on the sample size $n$, making it computationally feasible for large-scale observational datasets. Simulation results confirm that this reformulation achieves better covariate balance at substantially reduced computational cost. The framework is extended to accommodate functional confounders by augmenting the balancing constraint vector with FPCA scores for each functional covariate and extending it to functional outcomes through a function-on-function marginal structural model whose estimation reduces to a sequence of independent weighted least squares problems in the outcome FPC basis. Together, these extensions allow causal questions involving functional treatments, functional covariates, and functional outcomes to be addressed within a single coherent framework.

Applying the proposed framework to data from the UK Biobank, we find consistent evidence that midlife BMI has a positive causal effect on subsequent metabolic health. In Analysis 1, the estimated causal effect of the BMI trajectory on the log-odds of incident Type~2 Diabetes is strongest in the early part of the exposure window, approximately ages 50 to 57, and attenuates thereafter, becoming non-significant beyond age 57. The unweighted analysis yields uniformly larger estimates throughout the exposure window, consistent with positive confounding by baseline health status, which is associated with both higher BMI and increased diabetes risk. In Analysis 2, the estimated causal effect surface is uniformly positive and exhibits a clear dependence on the timing of exposure: higher BMI in the early part of midlife drives the greatest increase in subsequent HbA1c, while the causal effect of BMI at later exposure ages is both weaker and less precisely estimated. Taken together, these findings suggest that early midlife represents a critical window during which elevated BMI most substantially raises long-term metabolic risk, with implications for the targeting and timing of preventive interventions. The results of this observational study must be interpreted with caution, however, given the possibility of residual confounding by factors not included in the analysis, such as genetic predisposition, dietary patterns, or physical activity level, that may be associated with both BMI trajectories and metabolic outcomes.

Although the application in this article is motivated by questions in metabolic epidemiology, the proposed framework is broadly applicable wherever an exposure or treatment of scientific interest is more naturally represented as a trajectory or curve than as a scalar summary. A key feature supporting this broad applicability is the scalability of the proposed estimator, which allows the method to remain feasible even in observational datasets with very large sample sizes. Relevant applications span pharmacoepidemiology (drug dosing trajectories and clinical outcomes), environmental epidemiology (pollution or temperature curves and disease risk), and economics (income or consumption profiles and long-term outcomes), among others. Although the present applications index functions by age, the framework places no restriction on the nature of the domain, and the treatment and outcome need not share a common one, as illustrated in our analysis, where BMI and HbA1c are observed over non-overlapping age intervals.

This paper has several limitations that warrant further investigation. 
First, the FPCA implementation used here relies on sufficiently dense, regularly observed trajectories for individual FPC scores to be estimated reliably. In many real-world applications, however, longitudinal data may be sparse or irregularly observed. In such settings, extensions could incorporate sparse FPCA methods, such as PACE, to estimate individual scores from incomplete trajectories \citep{Yao}. Alternatively, the functional propensity score could be defined through small-ball probability methods \citep{ferraty_vieu}, which characterize the conditional distribution of a functional treatment without requiring a finite-dimensional FPC representation. Extending the proposed covariate-balancing approach to sparse and irregular functional data is a natural direction for future research. Second, the marginal structural models in Sections~\ref{sec:setup} and~\ref{sec:func_out} assume a linear functional effect relationship. Nonlinear extensions, for example, through kernel-based or nonparametric functional regression, would increase the power of the outcome model, though at the cost of additional assumptions and increased complexity. Third, as in most observational causal analyses, identification rests on the strong ignorability assumption, which requires that all relevant confounders are observed. This assumption is not testable from the observed data, and unmeasured confounding cannot be ruled out. Nevertheless, it may be more plausible in settings with rich baseline phenotyping, repeated clinical measurements, and linkage to electronic health records or administrative registries, such as the UK Biobank. The increasing availability of longitudinal health data and linked records may therefore broaden the settings in which functional causal methods can be applied with credible adjustment for confounders. Developing sensitivity analysis procedures adapted to functional treatments would be a valuable complement to the proposed methodology. Finally, we used FPCA decomposition to estimate the outcome model using Weighted Least Squares (WLS); however, an alternative approach would be to employ roughness regularization methods. Developing a weighted extension of the S-Lasso estimator~\citep{s-lasso} may offer better finite-sample performance and improved interpretability, particularly when the effect function is represented using many principal components.

\begin{acks}
This study used data from the UKB Resource under Application Number 102297. The present research has been supported by MUR, grant Dipartimento di Eccellenza 2023-2027. F. Ieva acknowledges the National Plan for NRRP Complementary Investments "Advanced Technologies for Human-centred Medicine" (PNC0000003). 
We thank Prof. Piercesare Secchi for insightful discussions and valuable suggestions.
\end{acks}

\begin{supplement}
\stitle{Supplementary Material A: Dual Derivation of the Weight Estimation Problem}
\sdescription{This appendix provides the full derivation of the unconstrained dual optimization problem introduced in Section~\ref{sec:fps}.}
\end{supplement}

\begin{supplement}
\stitle{Supplementary Material B: Causal Effect Estimation via Weighted Least Squares}
\sdescription{This appendix provides the full computational details of the causal effect estimation procedures summarized in Sections~\ref{sec:fps} and~\ref{sec:func_cov}.}
\end{supplement}
\begin{supplement}
\stitle{Supplementary Material C: Additional Analysis and Diagnostics for Functional Propensity Score Simulation}
\sdescription{This section provides additional covariate balance diagnostics supporting the results of Section~\ref{sec:sim_fps}. We summarize the simulation settings and present detailed evidence on the ability of the proposed weighting scheme to reduce dependence between FPC scores and covariates.}
\end{supplement}
\begin{supplement}
\stitle{Supplementary Material D: Additional Analysis and Diagnostics for Functional Covariates and Outcome Simulation}
\sdescription{This section provides additional diagnostics supporting the simulation study reported in Section~\ref{sec:sim_ext}. We summarize the simulation settings and present detailed evidence on covariate balance and inferential behavior in the setting with a functional covariate and a functional outcome.}
\end{supplement}
\begin{supplement}
\stitle{Supplementary Material E: Additional descriptive and FPCA diagnostics for the UK Biobank Application}
\sdescription{This section provides additional descriptive summaries and FPCA diagnostics for the UK Biobank application in Section~\ref{sec:realdata}.}
\end{supplement}

\newpage
\setcounter{page}{1}
\setcounter{section}{0}
\setcounter{subsection}{0}
\setcounter{figure}{0}
\setcounter{table}{0}
\setcounter{equation}{0}
\renewcommand{\thepage}{\arabic{page}}
\renewcommand{\thesection}{\arabic{section}}
\renewcommand{\thetable}{S\arabic{table}}
\renewcommand{\thefigure}{S\arabic{figure}}
\renewcommand{\theequation}{S\arabic{equation}}

\clearpage
\section*{Supplementary Material A: Dual Derivation of the Weight Estimation Problem}
\label{sec:weights_derivation}
This appendix derives the unconstrained dual optimization problem introduced in Section 2.2. Starting from the primal formulation, we obtain the corresponding Lagrangian dual, derive the softmax representation of the weights, and show how the resulting objective reduces to a smooth unconstrained minimization problem. We also describe the log-sum-exp stabilization used in the numerical implementation.

The primal problem is defined as:
$$\min_{\boldsymbol{w}^*} \sum_{i=1}^n \log(w_i)
    \quad \text{subject to} \quad
    \sum_{i=1}^n w_i \boldsymbol{g}_i = \boldsymbol{0}, \quad
    \sum_{i=1}^n w_i = 1, \quad w_i \geq 0,$$

where $$\boldsymbol{g}_i = 
    \bigl[(\boldsymbol{A}_i)^\top,\; (\boldsymbol{C}_i)^\top,\;
    \mathrm{vec}\bigl(\boldsymbol{A}_i(\boldsymbol{C}_i)^\top\bigr)^\top
    \bigr]^\top \in \mathbb{R}^{L + p + L \times p}, \quad i = 1, \ldots, n.$$
   
Introducing a multiplier $\lambda \in \mathbb{R}$ for the normalization constraint and $\boldsymbol{\gamma} \in \mathbb{R}^{L + p + L \times p}$ for the balancing constraints, the Lagrangian is:
$$\mathcal{L}(\boldsymbol{w}, \boldsymbol{\gamma}, \lambda)
    =
    \sum_{i=1}^n \log(w_i)
    + \lambda \left( \sum_{i=1}^n w_i - 1 \right)
    + \boldsymbol{\gamma}^\top \sum_{i=1}^n w_i \boldsymbol{g}_i.$$

%Dual
The dual function is therefore defined as:
$$\mathcal{D}(\lambda, \boldsymbol{\gamma})
    =
    \inf_{\boldsymbol{w} \geq 0} \mathcal{L}(\boldsymbol{w}, \boldsymbol{\gamma}, \lambda).$$
Since the $w_i$ enter the Lagrangian independently, the infimum can be rewritten as:
$$\mathcal{D}(\lambda, \boldsymbol{\gamma})
    =
    \sum_{i=1}^n
    \inf_{w_i > 0}
    \left(
    \log(w_i) + w_i\bigl(\lambda + \boldsymbol{\gamma}^\top \boldsymbol{g}_i\bigr)
    \right)
    - \lambda.$$

The dual problem is then: $$\max_{\boldsymbol{\gamma}, \lambda}\,
\mathcal{D}(\lambda, \boldsymbol{\gamma}).$$

Setting partial derivatives of $\mathcal{L}$ to zero yields the stationarity conditions:
\begin{align*}
 \frac{\partial \mathcal{L}}{\partial w_i}
 &= \frac{1}{w_i} + \boldsymbol{\gamma}^\top \boldsymbol{g}_i + \lambda = 0
 \quad \Rightarrow \quad
 w_i = \frac{1}{-\lambda - \boldsymbol{\gamma}^\top \boldsymbol{g}_i},
 \qquad i = 1, \ldots, n, \\[8pt]
 \frac{\partial \mathcal{L}}{\partial \boldsymbol{\gamma}}
 &= \sum_{i=1}^n w_i \boldsymbol{g}_i = \boldsymbol{0}
 \quad \Rightarrow \quad
 \sum_{i=1}^n w_i \boldsymbol{g}_i = \boldsymbol{0}, \\[8pt]
 \frac{\partial \mathcal{L}}{\partial \lambda}
 &= \sum_{i=1}^n w_i - 1 = 0
 \quad \Rightarrow \quad
 \sum_{i=1}^n w_i = 1.
\end{align*}

To eliminate $\lambda$, we impose the weighted stationarity condition
$\sum_{i=1}^n w_i \,\frac{\partial \mathcal{L}}{\partial w_i} = 0$, which gives:
\[
    \sum_{i=1}^n w_i
    \left(
    \frac{1}{w_i} + \boldsymbol{\gamma}^\top \boldsymbol{g}_i + \lambda
    \right)
    =
    n
    + \boldsymbol{\gamma}^\top \sum_{i=1}^n w_i \boldsymbol{g}_i
    + \lambda \sum_{i=1}^n w_i
    = 0.\]

Substituting the constraints $\sum_{i=1}^n w_i \boldsymbol{g}_i = \boldsymbol{0}$
and $\sum_{i=1}^n w_i = 1$, it follows immediately that:
$$\lambda = -n.$$

Substituting $\lambda = -n$ into the dual function gives:
$$\mathcal{D}(\boldsymbol{\gamma})
    =
    \sum_{i=1}^n
    \inf_{w_i > 0}
    \left(
    \log(w_i) + w_i\bigl(-n + \boldsymbol{\gamma}^\top \boldsymbol{g}_i\bigr)
    \right)
    + n.$$

For each $i$, define the inner function
$\ell_i(w_i) = \log(w_i) + w_i(-n + \boldsymbol{\gamma}^\top \boldsymbol{g}_i)$.
Setting $\ell_i'(w_i) = 0$:
$$\frac{1}{w_i} - n + \boldsymbol{\gamma}^\top \boldsymbol{g}_i = 0
    \quad \Rightarrow \quad
    w_i^* = \frac{1}{n - \boldsymbol{\gamma}^\top \boldsymbol{g}_i},
    \qquad i = 1, \ldots, n.$$

Substituting back:
$$\inf_{w_i > 0} \ell_i(w_i)
    = -\log\!\bigl(n - \boldsymbol{\gamma}^\top \boldsymbol{g}_i\bigr) - 1,
    \qquad i = 1, \ldots, n.$$

Therefore,
$$\mathcal{D}(\boldsymbol{\gamma})
    =
    \sum_{i=1}^n
    \Bigl(-\log\!\bigl(n - \boldsymbol{\gamma}^\top \boldsymbol{g}_i\bigr) - 1\Bigr)
    + n
    =
    -\sum_{i=1}^n \log\!\bigl(n - \boldsymbol{\gamma}^\top \boldsymbol{g}_i\bigr).$$

To guarantee strict positivity of the weights and reduce the dual to an unconstrained problem in a free parameter vector, we introduce the reparametrization

$$n - \boldsymbol{\gamma}^\top \boldsymbol{g}_i
    = K \, e^{\boldsymbol{\theta}^\top \boldsymbol{g}_i},
    \qquad i = 1, \ldots, n,$$

for some $K > 0$ and $\boldsymbol{\theta} \in \mathbb{R}^{L + p + L \times p}$.
Under this parametrization, the weights become

$$w_i = \frac{1}{K} \, e^{-\boldsymbol{\theta}^\top \boldsymbol{g}_i},
    \qquad i = 1, \ldots, n.$$

Imposing the normalization constraint $\sum_{i=1}^n w_i = 1$ determines $K$
uniquely:

\begin{equation}
\sum_{i=1}^n \frac{1}{K} \, e^{-\boldsymbol{\theta}^\top \boldsymbol{g}_i} = 1
    \quad \Rightarrow \quad
    K = \sum_{i=1}^n e^{-\boldsymbol{\theta}^\top \boldsymbol{g}_i}.
\end{equation}

The normalized weights are therefore:

$$w_i(\boldsymbol{\theta})
    =
    \frac{e^{-\boldsymbol{\theta}^\top \boldsymbol{g}_i}}
    {\displaystyle\sum_{j=1}^n e^{-\boldsymbol{\theta}^\top \boldsymbol{g}_j}},
    \qquad i = 1, \ldots, n,$$
which is a softmax transformation of $-\boldsymbol{\theta}^\top \boldsymbol{g}_i$. Substituting the reparametrization into the dual function:
\begin{align*}
    \mathcal{D}(\boldsymbol{\theta})
    &= -\sum_{i=1}^n
    \log\!\left[
    \Bigl(\sum_{j=1}^n e^{-\boldsymbol{\theta}^\top \boldsymbol{g}_j}\Bigr)
    e^{-\boldsymbol{\theta}^\top \boldsymbol{g}_i}
    \right] \\
    &= -n \log\!\left(\sum_{j=1}^n e^{-\boldsymbol{\theta}^\top \boldsymbol{g}_j}\right)
    - \boldsymbol{\theta}^\top \sum_{i=1}^n \boldsymbol{g}_i.
\end{align*}
Since the FPC scores $\boldsymbol{A}$ and confounders $\boldsymbol{C}$ have been standardized to zero mean, the sample mean of $\boldsymbol{g}_i$ satisfies $\bar{\boldsymbol{g}} = \frac{1}{n} \sum_{i=1}^n \boldsymbol{g}_i = \boldsymbol{0},$ so the linear term $\boldsymbol{\theta}^\top \sum_{i=1}^n \boldsymbol{g}_i$ vanishes identically. The dual function simplifies to:
$$\mathcal{D}(\boldsymbol{\theta})
    = -n \log\!\left(\sum_{i=1}^n e^{-\boldsymbol{\theta}^\top \boldsymbol{g}_i}\right).$$

Maximizing $\mathcal{D}(\boldsymbol{\theta})$ is equivalent to minimizing
$f(\boldsymbol{\theta}) = \log\!\left(\sum_{i=1}^n e^{-\boldsymbol{\theta}^\top \boldsymbol{g}_i}\right)$, yielding the final unconstrained problem. The gradient of the objective with respect to $\boldsymbol{\theta}$ is:
$$\nabla_{\boldsymbol{\theta}}
    f(\boldsymbol{\theta})
    =
    \frac{\sum_{i=1}^n -\boldsymbol{g}_i \, e^{-\boldsymbol{\theta}^\top \boldsymbol{g}_i}}
    {\sum_{j=1}^n e^{-\boldsymbol{\theta}^\top \boldsymbol{g}_j}}
    =
    -\sum_{i=1}^n w_i(\boldsymbol{\theta}) \, \boldsymbol{g}_i,$$
which admits a closed-form expression in terms of the current weights
$w_i(\boldsymbol{\theta})$, making it directly available to the BFGS optimizer. Direct evaluation of $\sum_{i=1}^n e^{-\boldsymbol{\theta}^\top \boldsymbol{g}_i}$ can suffer from numerical overflow or underflow when the exponents are large in magnitude. To ensure numerical stability, we apply the log-sum-exp (LSE) trick.
Define:
$$M = \max_{i = 1,\ldots,n} \bigl(-\boldsymbol{\theta}^\top \boldsymbol{g}_i\bigr).$$

Then:
\begin{align*}
    \log\!\left(\sum_{i=1}^n e^{-\boldsymbol{\theta}^\top \boldsymbol{g}_i}\right)
    &=
    \log\!\left(\sum_{i=1}^n e^{-\boldsymbol{\theta}^\top \boldsymbol{g}_i - M + M}\right) \\
    &=
    \log\!\left(e^{M} \sum_{i=1}^n e^{-\boldsymbol{\theta}^\top \boldsymbol{g}_i - M}\right) \\
    &=
    M + \log\!\left(\sum_{i=1}^n e^{-\boldsymbol{\theta}^\top \boldsymbol{g}_i - M}\right).
\end{align*}
Since $-\boldsymbol{\theta}^\top \boldsymbol{g}_i - M \leq 0$ for all $i$, all exponentiated terms lie in $(0, 1]$, eliminating the risk of overflow. The stabilized objective function used in the numerical implementation is therefore:
\begin{equation}
\label{eq:lse}
    \mathcal{F}(\boldsymbol{\theta})
    =
    M + \log\!\left(\sum_{i=1}^n e^{-\boldsymbol{\theta}^\top \boldsymbol{g}_i - M}\right),
    \qquad
    M = \max_{i}\, \bigl(-\boldsymbol{\theta}^\top \boldsymbol{g}_i\bigr).
\end{equation}
The gradient of~\eqref{eq:lse} retains the same closed form derived before, with exponents shifted by $M$, and is passed analytically to the BFGS
algorithm, ensuring both numerical stability and fast convergence.

\newpage
\section*{Supplementary Material B: Causal Effect Estimation via Weighted Least Squares}
\label{sec:outcome_estimation}

This appendix provides the full computational details of the causal effect estimation procedures summarized in Sections 2.2 and 2.4 of the main paper.

\subsection*{B.1\quad Scalar Outcome}
\label{sec:outcome_scalar}

Once the SFPS weights $w_i$ are obtained, the causal effect function $\mu(t)$ is estimated by fitting the outcome model:
$$\mathbb{E}[Y \mid X] = \mu_0 + \int_{\mathcal{T}} \mu(t) \cdot X(t) \, dt,$$
where $\mu_0 = \mathbb{E}[Y]$ by the zero-mean assumption on $X(t)$. To make
estimation feasible, $\mu(t)$ is approximated through a truncated basis expansion.
While any complete orthonormal basis would be admissible, we adopt, without loss of generality, the eigenfunctions $\{\phi_k(t)\}_{k=1}^{L^*}$ of $X(t)$ already computed during the FPCA decomposition:
$$\mu(t) \approx \sum_{k=1}^{L^*} \mu_k \cdot \phi_k(t),$$
where the coefficients $\mu_k = \int_{\mathcal{T}} \mu(t) \cdot \phi_k(t) \, dt$ are
the projections of $\mu(t)$ onto the $k$-th basis function, and $L^*$ is chosen by a second PVE threshold that may differ from $L$ used in the treatment decomposition.
Substituting this expansion into the outcome model gives:
$$\mathbb{E}[Y \mid X] \approx \mu_0 + \sum_{k=1}^{L^*} \mu_k \cdot
 \int_{\mathcal{T}} X(t) \cdot \phi_k(t) \, dt
 = \mu_0 + \sum_{k=1}^{L^*} \mu_k \cdot A_k,$$
where $A_k = \int_{\mathcal{T}} X(t) \cdot \phi_k(t) \, dt$ are the FPC scores of $X(t)$. The coefficients $(\mu_1, \ldots, \mu_{L^*})$ are estimated via weighted least squares using the SFPS weights:
$$(\hat{\mu}_1, \ldots, \hat{\mu}_{L^*})
 = \operatorname*{arg\,min}_{\mu_1, \ldots, \mu_{L^*}}
 \sum_{i=1}^n \hat{w}_i \cdot
 \left(Y_i - \bar{Y} - \sum_{k=1}^{L^*} \mu_k \cdot A_{ik}\right)^2,$$
where $\bar{Y} = \frac{1}{n}\sum_{i=1}^n Y_i$ is the sample mean of the outcomes, $A_{ik} = \int_{\mathcal{T}} X_i(t) \cdot \phi_k(t) \, dt$ are the individual FPC scores, and $\hat{w}_i$ are the estimated SFPS weights. The final estimate of the causal effect function is:
$$\hat{\mu}(t) = \sum_{k=1}^{L^*} \hat{\mu}_k \cdot \phi_k(t).$$

\subsection*{B.2\quad Functional Outcome}
\label{sec:outcome_functional}

When the outcome is itself a function, both $X(s)$ and $Y(t)$ are represented through their respective Karhunen--Lo\`eve expansions, truncated at $L_{\mathcal{S}}$ and $L_{\mathcal{T}}$ components:
$$X_i(s) = \sum_{k=1}^{L_{\mathcal{S}}} A_{ik} \cdot \phi_k(s), \qquad
 Y_i(t) = \sum_{j=1}^{L_{\mathcal{T}}} c_{ij} \cdot \psi_j(t), \qquad i = 1, \ldots, n,$$
where $\phi_k(s)$ and $\psi_j(t)$ are the orthonormal eigenfunctions of the covariance operators of $X(s)$ and $Y(t)$, respectively, and $A_{ik}$, $c_{ij}$ are the corresponding FPC scores. 

The causal effect surface and the functional intercept are expanded in the same bases, but using possibly different truncation levels, denoted by $L^*_{\mathcal{S}}$ and $L^*_{\mathcal{T}}$:
$$\mu(s, t) = \sum_{k=1}^{L^*_{\mathcal{S}}} \sum_{j=1}^{L^*_{\mathcal{T}}}
 \mu_{kj} \cdot \psi_j(t) \cdot \phi_k(s), \qquad
 \mu_0(t) = \sum_{j=1}^{L^*_{\mathcal{T}}} \mu_{0j} \cdot \psi_j(t).
 $$

Substituting these expansions into model~\eqref{eq:fof}:
$$\sum_{j=1}^{L^*_{\mathcal{T}}} c_{ij} \cdot \psi_j(t)
 =
 \sum_{j=1}^{L^*_{\mathcal{T}}} \mu_{0j} \cdot \psi_j(t)
 +
 \sum_{k=1}^{L^*_{\mathcal{S}}} \sum_{j=1}^{L^*_{\mathcal{T}}}
 \mu_{kj} \cdot \psi_j(t) \cdot
 \int_{\mathcal{S}} X_i(s) \cdot \phi_k(s) \, ds.$$

Since the integral term is precisely the $k$-th FPC score of $X_i(s)$, that is $A_{ik} = \int_{\mathcal{S}} X_i(s) \cdot \phi_k(s) \, ds$, the expression simplifies to:
$$\sum_{j=1}^{L^*_{\mathcal{T}}} c_{ij} \cdot \psi_j(t)
 =
 \sum_{j=1}^{L^*_{\mathcal{T}}} \mu_{0j} \cdot \psi_j(t)
 +
 \sum_{k=1}^{L^*_{\mathcal{S}}} \sum_{j=1}^{L^*_{\mathcal{T}}}
 \mu_{kj} \cdot \psi_j(t) \cdot A_{ik}.$$

Matching coefficients of $\psi_j(t)$ for each $j = 1, \ldots, L_{\mathcal{T}}$ yields the system:
$$c_{ij} = \mu_{0j} + \sum_{k=1}^{L^*_{\mathcal{S}}} \mu_{kj} \cdot A_{ik},
 \qquad i = 1, \ldots, n,$$
which in matrix form reads $\boldsymbol{c}_j = \boldsymbol{A} \cdot \boldsymbol{\mu}_j$,
where:
$$\boldsymbol{c}_j = (c_{1j}, \ldots, c_{nj})^\top, \quad
 \boldsymbol{A} =
 \begin{bmatrix}
 1 & A_{11} & \cdots & A_{1L^*_{\mathcal{S}}}\\
 \vdots & \vdots & \ddots & \vdots\\
 1 & A_{n1} & \cdots & A_{nL^*_{\mathcal{S}}}
 \end{bmatrix}, \quad
 \boldsymbol{\mu}_j = (\mu_{0j}, \mu_{1j}, \ldots, \mu_{L^*_{\mathcal{S}} j})^\top.$$

For each $j = 1, \ldots, L^*_{\mathcal{T}}$, the coefficient vector $\boldsymbol{\mu}_j$ is estimated via weighted least squares using the FPS weights $\boldsymbol{W} = \mathrm{diag}(w_1, \ldots, w_n)$:
$$\hat{\boldsymbol{\mu}}_j
 = \operatorname*{arg\,min}_{\boldsymbol{\mu}_j}
 \bigl\| \boldsymbol{W}^{1/2}
 \bigl(\boldsymbol{c}_j - \boldsymbol{A} \cdot \boldsymbol{\mu}_j\bigr)
 \bigr\|^2
 = (\boldsymbol{A}^\top \boldsymbol{W} \boldsymbol{A})^{-1}
 \boldsymbol{A}^\top \boldsymbol{W} \boldsymbol{c}_j.$$

The estimated causal effect surface and functional intercept are then reconstructed as:
$$\hat{\mu}(s, t) = \sum_{k=1}^{L^*_{\mathcal{S}}} \sum_{j=1}^{L^*_{\mathcal{T}}}
 \hat{\mu}_{kj} \cdot \psi_j(t) \cdot \phi_k(s), \qquad
 \hat{\mu}_0(t) = \sum_{j=1}^{L^*_{\mathcal{T}}} \hat{\mu}_{0j} \cdot \psi_j(t).$$

\newpage
\section*{Supplementary Material C: Additional Analysis and Diagnostics for Functional Propensity Score Simulation}
\label{sec:sm_balance_fps}

This section provides additional covariate balance diagnostics supporting the results of Section 3.1. We summarize the simulation settings and present detailed evidence on the ability of the proposed weighting scheme to reduce dependence between FPC scores and covariates.

We consider four simulation settings that vary the form of the treatment–covariate and covariate–outcome relationships:
\begin{itemize}
 \item Setting 1 (linear treatment–covariate and covariate–outcome relationships):\\
 $C_{i1} = Z_{i1} + W_{i1}$, $C_{i2} = 0.2Z_{i2} + W_{i2}$,
 $C_{i3} = 0.2Z_{i3} + W_{i3}$, with $W_{i1} \sim \mathcal{N}(0,1)$ and
 $W_{i2}, W_{i3} \sim \mathcal{N}(0, 0.5)$ independent of $Z_{ik}$;
 $g(\boldsymbol{C}_i) = 2C_{i1}$.

 \item Setting 2 (nonlinear treatment–covariate, linear outcome):\\
 $C_{i1} = (Z_{i1} + 0.5)^2 + W_{i1}$; $C_{i2}$, $C_{i3}$ and $g$ as in Setting~1.

 \item Setting 3 (linear treatment–covariate, nonlinear outcome):\\
 Covariates as in Setting~1; $g(\boldsymbol{C}_i) = 2C_{i1} + C_{i2}^2$.

 \item Setting 4 (nonlinear treatment–covariate and nonlinear outcome):\\
 Covariates as in Setting~2; $g$ as in Setting~3.
\end{itemize}
Table~\ref{tab:fps_improvement} reports the percentage change of the weighted estimator relative to the unweighted estimator for MISE, AISE, and ISB across all settings and $(\mathrm{PVE}_L, \mathrm{PVE}_{L^*})$ combinations. 
\begin{table}[h]
\centering
\caption{Percentage improvement of the weighted estimator over the unweighted in MISE, AISE and ISB. Negative values indicate deterioration.}
\label{tab:fps_improvement}
\begin{tabular}{lllcccccc}
\toprule
& & & \multicolumn{3}{c}{$\mathrm{PVE}_{L^*} = 0.95$} & \multicolumn{3}{c}{$\mathrm{PVE}_{L^*} = 0.99$} \\
\cmidrule(lr){4-6} \cmidrule(lr){7-9}
& & & MISE & AISE & ISB & MISE & AISE & ISB \\
\midrule
Setting 1
& $\mathrm{PVE}_L = 0.95$ & & $48.23\%$ & $35.55\%$ & $99.62\%$ & $-43.89\%$ & $-73.80\%$ & $97.29\%$ \\
& $\mathrm{PVE}_L = 0.99$ & & $47.62\%$ & $33.03\%$ & $99.46\%$ & $-28.60\%$ & $-63.95\%$ & $97.04\%$ \\
\midrule
Setting 2
& $\mathrm{PVE}_L = 0.95$ & & $68.36\%$ & $60.98\%$ & $99.59\%$ & $-2.90\%$ & $-16.96\%$ & $97.71\%$ \\
& $\mathrm{PVE}_L = 0.99$ & & $66.41\%$ & $60.08\%$ & $99.55\%$ & $18.63\%$ & $3.37\%$ & $96.98\%$ \\
\midrule
Setting 3
& $\mathrm{PVE}_L = 0.95$ & & $47.40\%$ & $35.35\%$ & $99.67\%$ & $-38.77\%$ & $-73.96\%$ & $97.20\%$ \\
& $\mathrm{PVE}_L = 0.99$ & & $44.70\%$ & $32.77\%$ & $99.50\%$ & $-30.15\%$ & $-64.09\%$ & $97.00\%$ \\
\midrule
Setting 4
& $\mathrm{PVE}_L = 0.95$ & & $67.94\%$ & $60.79\%$ & $99.63\%$ & $-3.42\%$ & $-16.52\%$ & $97.83\%$ \\
& $\mathrm{PVE}_L = 0.99$ & & $65.99\%$ & $59.86\%$ & $99.55\%$ & $16.61\%$ & $3.59\%$ & $97.02\%$ \\
\bottomrule
\end{tabular}
\end{table}

\begin{figure}[h]
 \centering
 \begin{minipage}{0.35\textwidth}
 \centering
 \includegraphics[width=\textwidth]{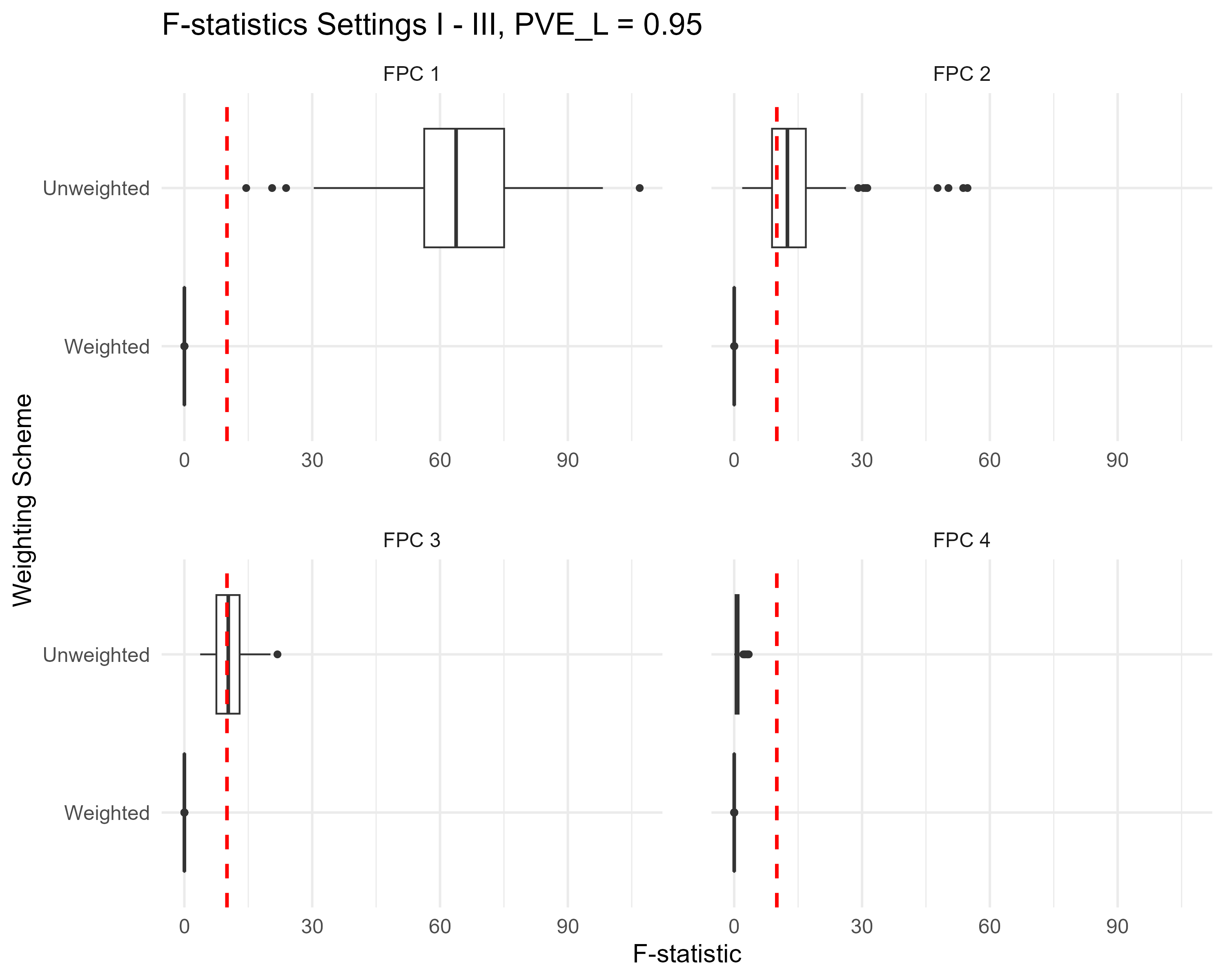}
 \subcaption{Settings~1 and~3, $\mathrm{PVE}_L = 0.95$}
 \end{minipage}
 \hfill
 \begin{minipage}{0.35\textwidth}
 \centering
 \includegraphics[width=\textwidth]{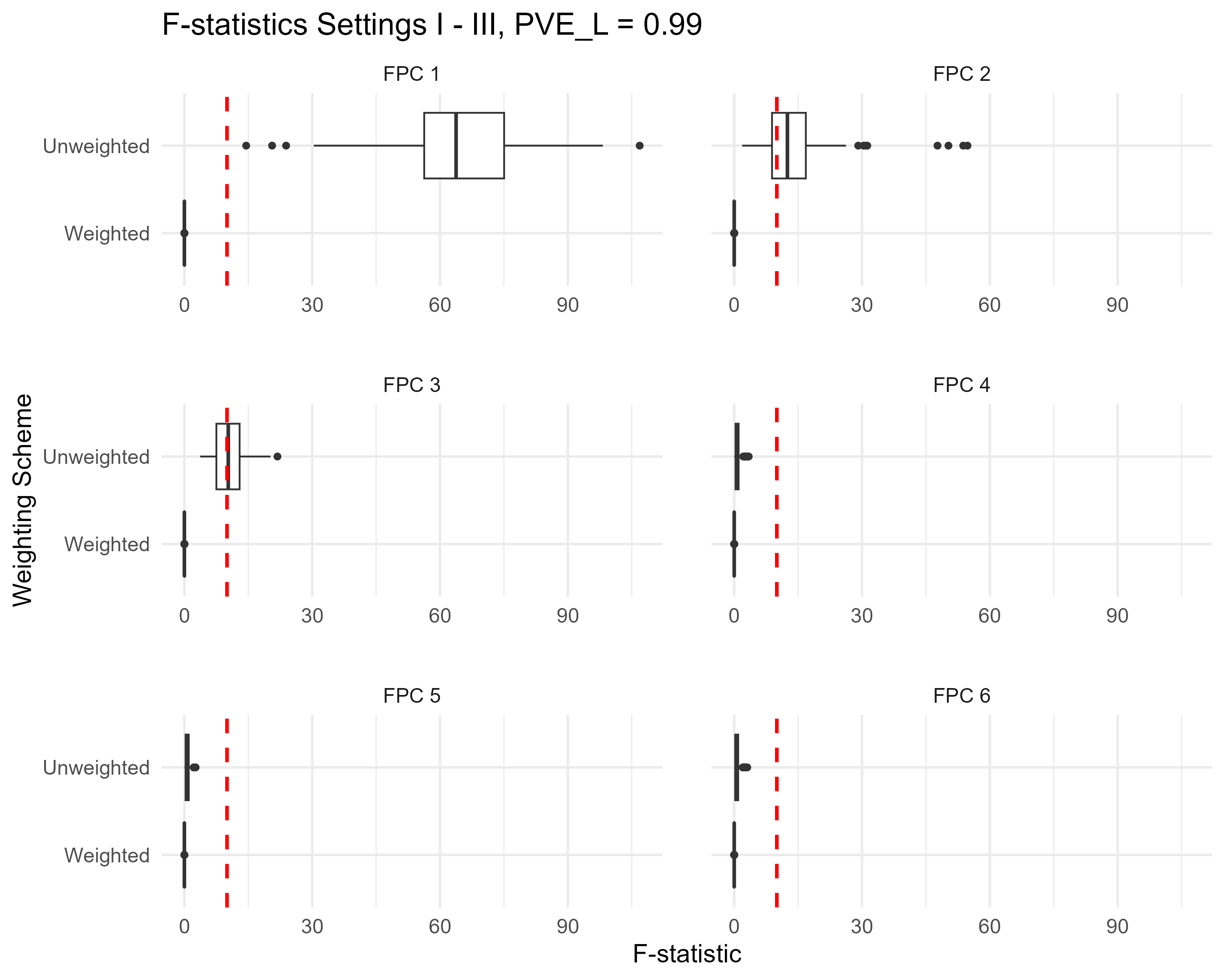}
 \subcaption{Settings~1 and~3, $\mathrm{PVE}_L = 0.99$}
 \end{minipage}

 \vspace{1em}

 \begin{minipage}{0.35\textwidth}
 \centering
 \includegraphics[width=\textwidth]{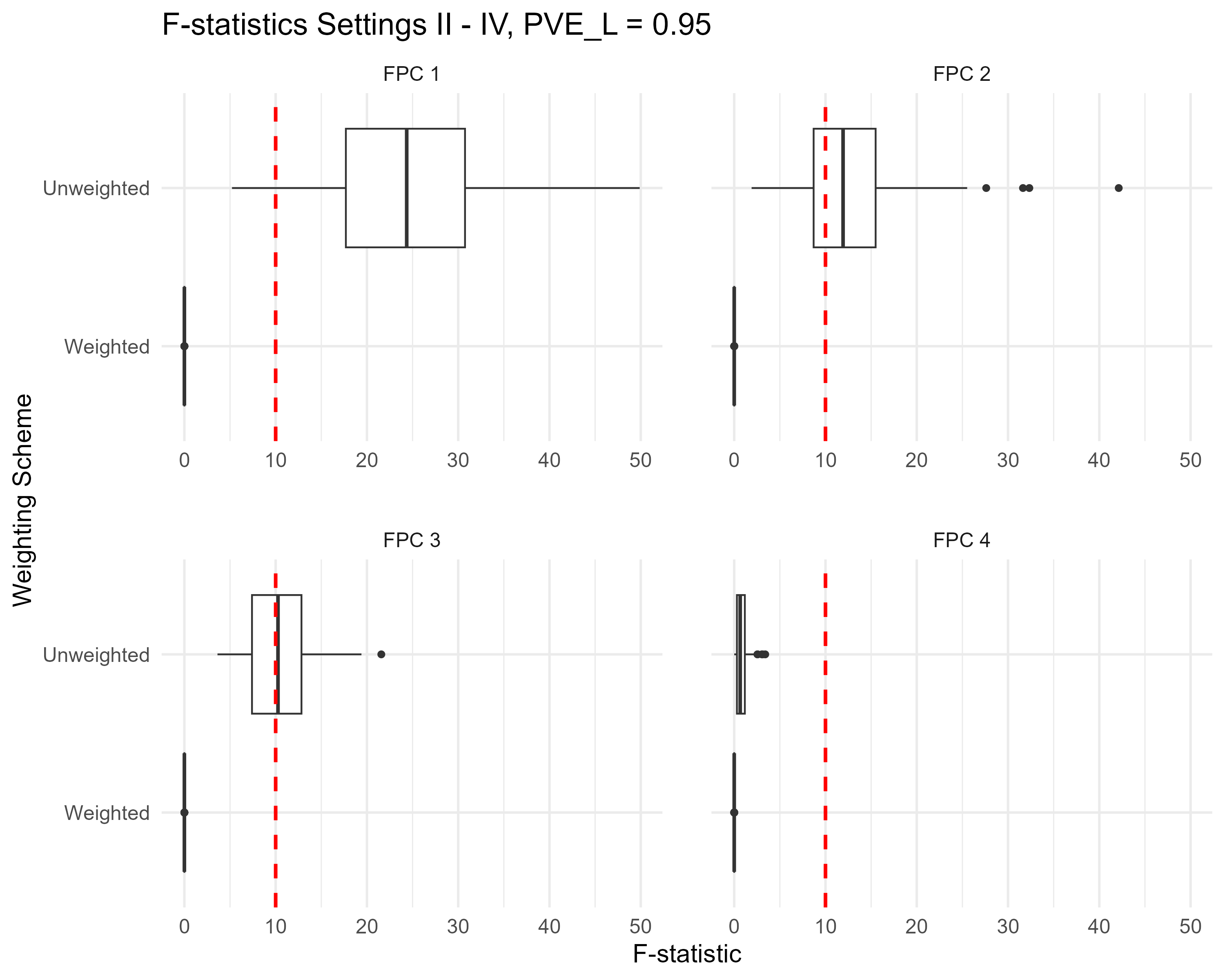}
 \subcaption{Settings~2 and~4, $\mathrm{PVE}_L = 0.95$}
 \end{minipage}
 \hfill
 \begin{minipage}{0.35\textwidth}
 \centering
 \includegraphics[width=\textwidth]{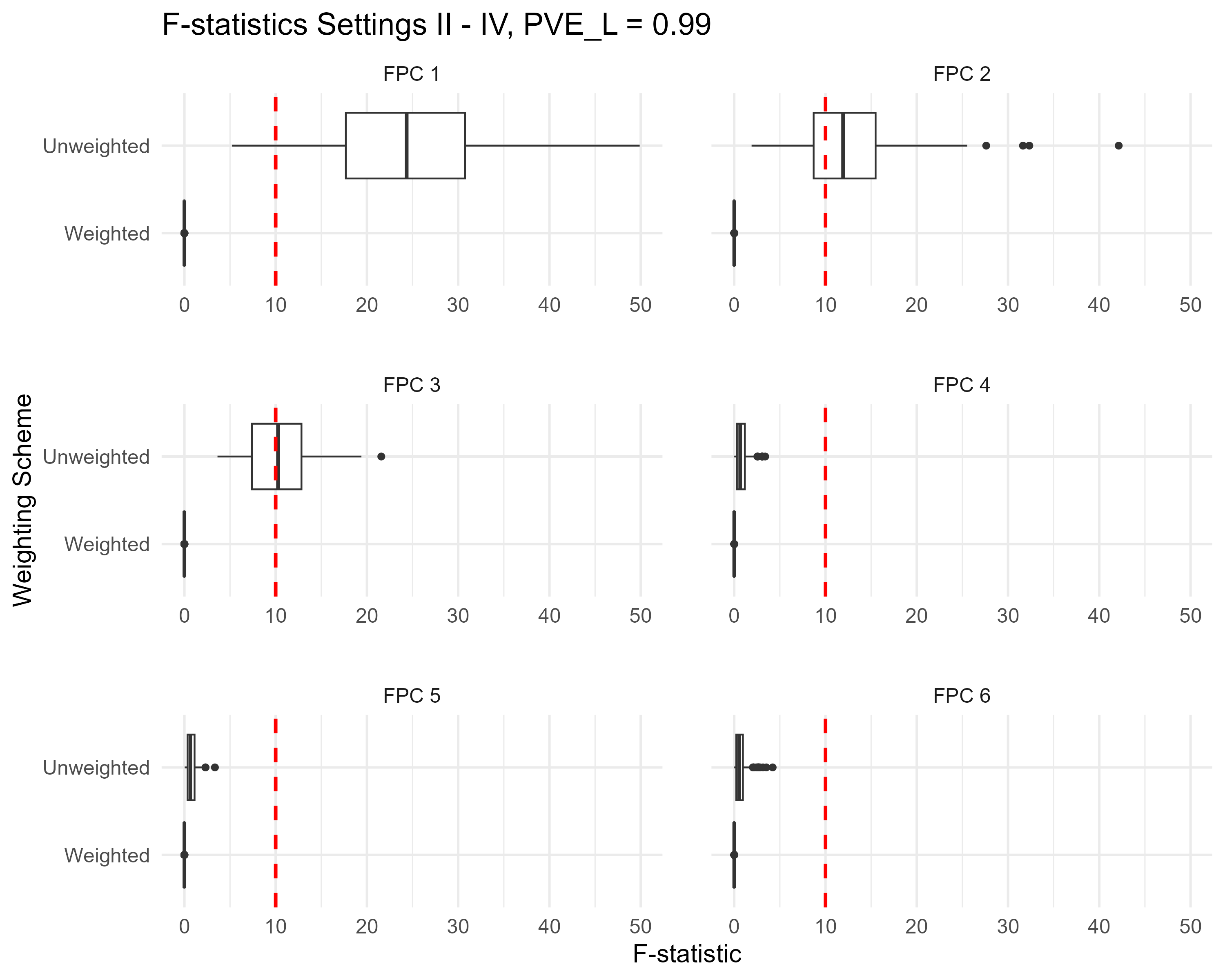}
 \subcaption{Settings~2 and~4, $\mathrm{PVE}_L = 0.99$}
 \end{minipage}

 \caption{Boxplots of $F$-statistics from weighted linear regressions of each
 selected FPC score on the covariate vector $\boldsymbol{C}$, across $R = 200$
 simulation runs. The dashed horizontal line indicates the reference threshold
 $F = 10$.}
 \label{fig:fstat}
\end{figure}

\begin{figure}[H]
 \centering
 \begin{minipage}{0.35\textwidth}
 \centering
 \includegraphics[width=\textwidth]{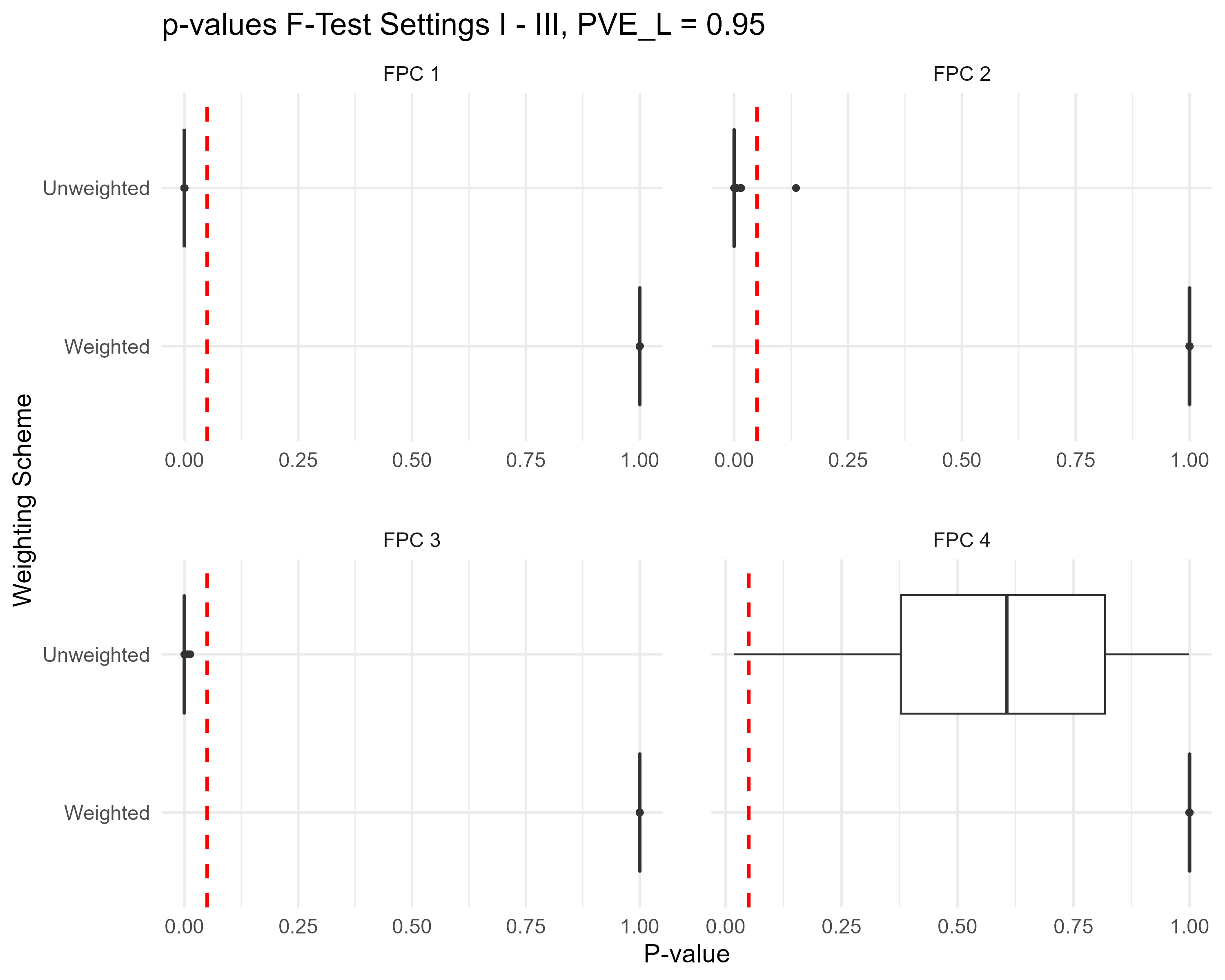}
 \subcaption{Settings~1 and~3, $\mathrm{PVE}_L = 0.95$}
 \end{minipage}
 \hfill
 \begin{minipage}{0.35\textwidth}
 \centering
 \includegraphics[width=\textwidth]{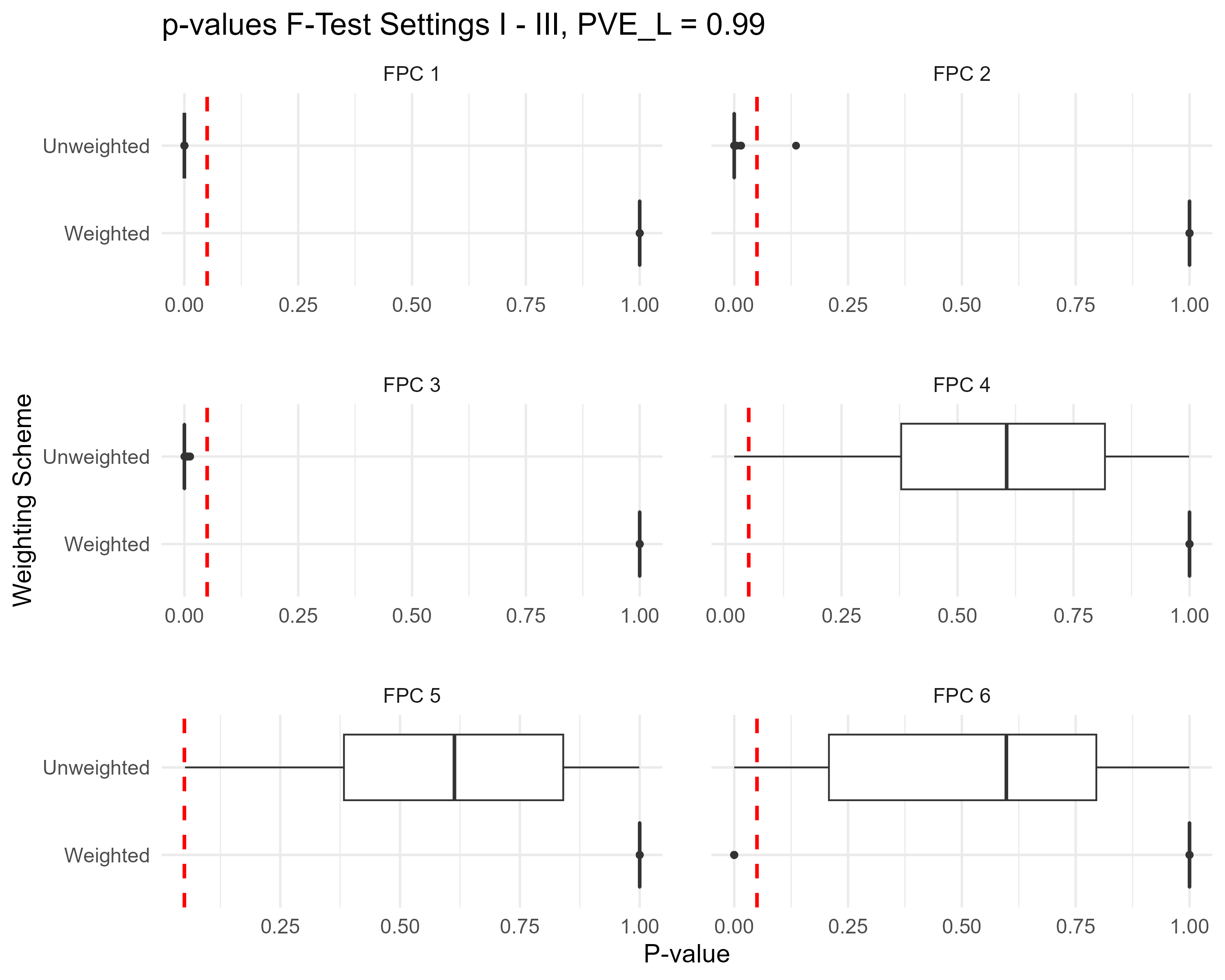}
 \subcaption{Settings~1 and~3, $\mathrm{PVE}_L = 0.99$}
 \end{minipage}

 \vspace{1em}

 \begin{minipage}{0.35\textwidth}
 \centering
 \includegraphics[width=\textwidth]{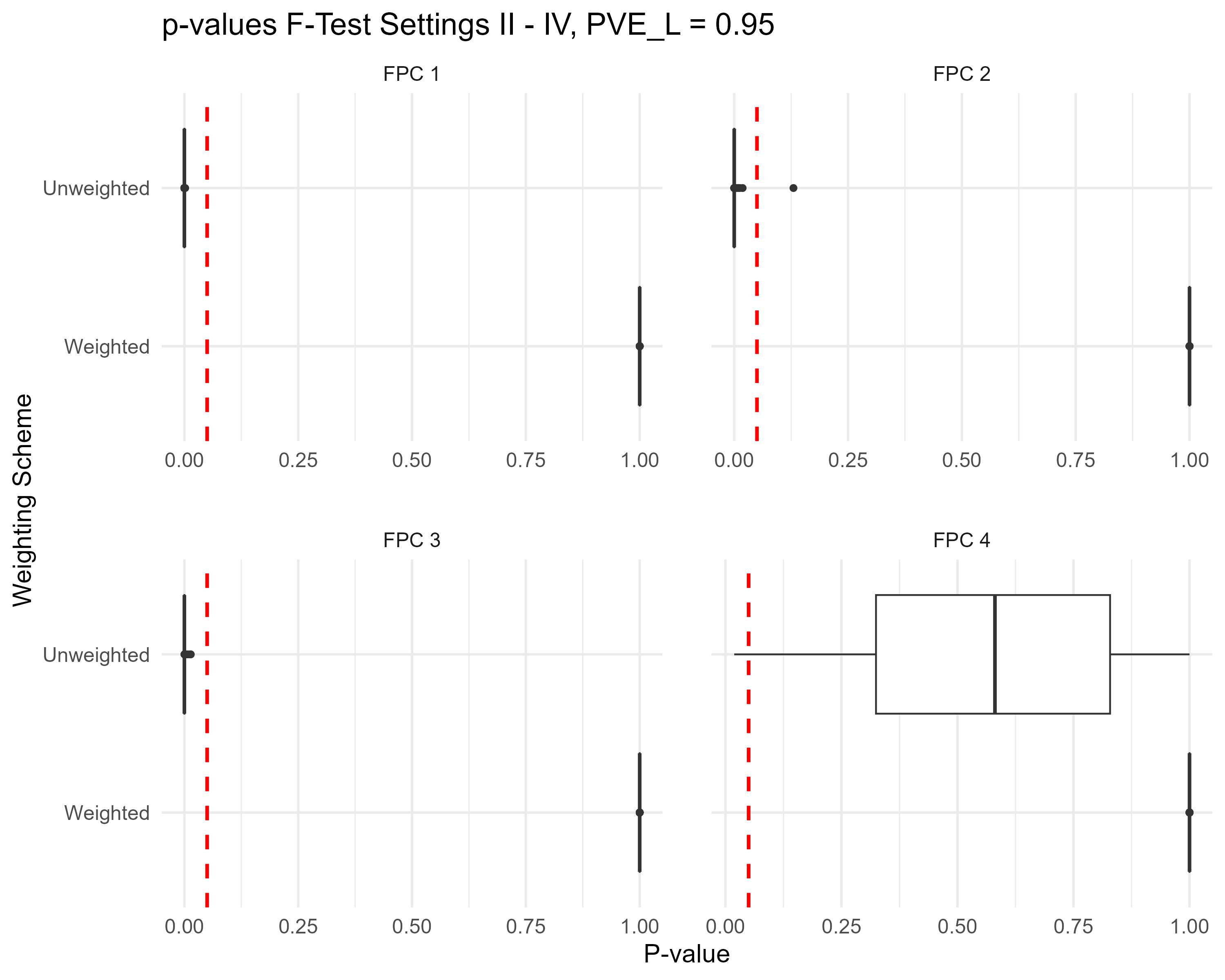}
 \subcaption{Settings~2 and~4, $\mathrm{PVE}_L = 0.95$}
 \end{minipage}
 \hfill
 \begin{minipage}{0.35\textwidth}
 \centering
 \includegraphics[width=\textwidth]{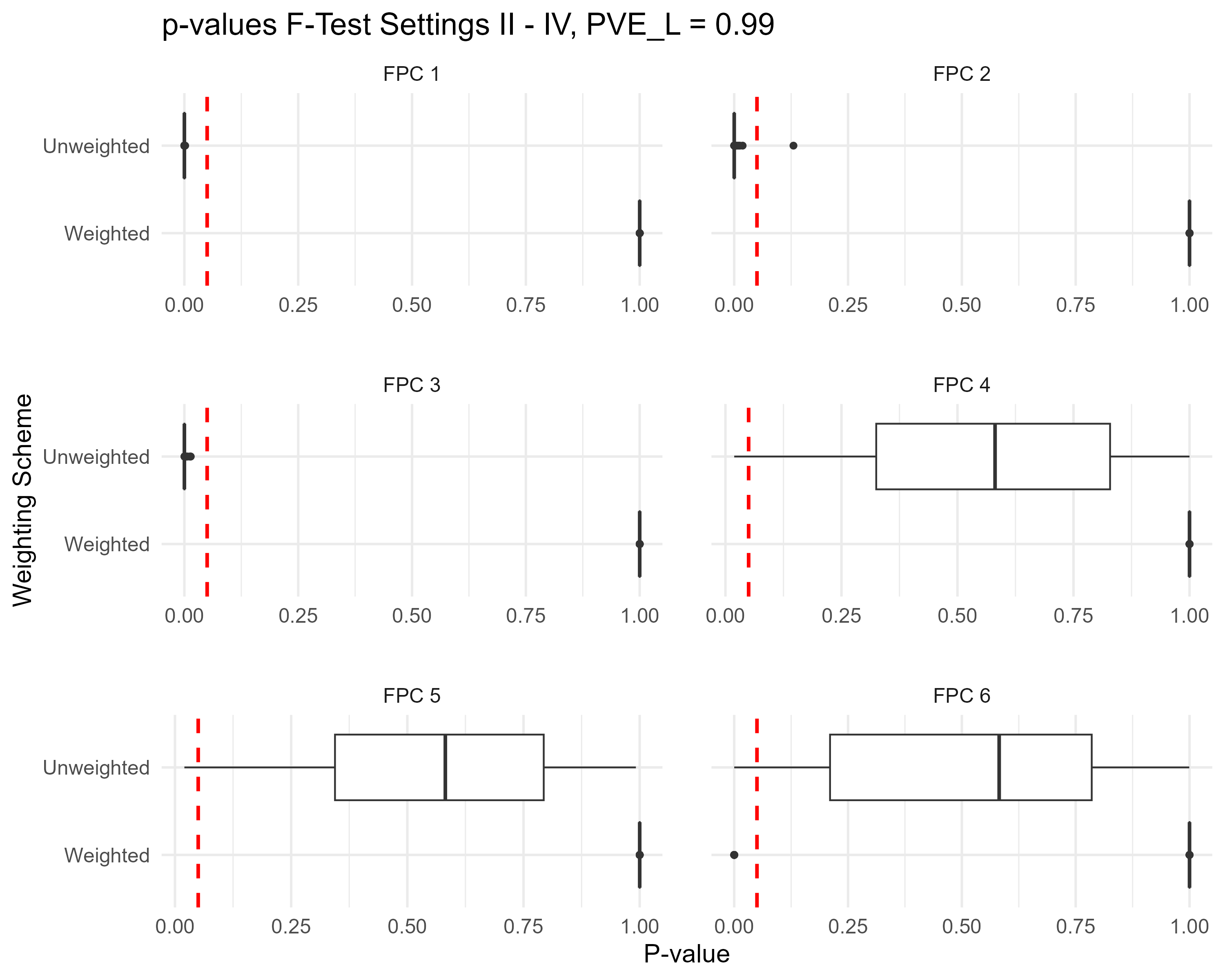}
 \subcaption{Settings~2 and~4, $\mathrm{PVE}_L = 0.99$}
 \end{minipage}

 \caption{Boxplots of $p$-values from the $F$-tests corresponding to
 Figure~\ref{fig:fstat}. The dashed horizontal line indicates the reference
 threshold $p = 0.05$.}
 \label{fig:pval}
\end{figure}

Figures~\ref{fig:fstat} and~\ref{fig:pval} summarize the distribution of $F$-statistics and corresponding $p$-values from weighted regressions of FPC scores on the covariates. The proposed method yields $F$-statistics concentrated near zero and $p$-values close to one across all settings, indicating weak residual dependence after weighting. In contrast, the unweighted case exhibits large $F$-statistics and small $p$-values, reflecting substantial dependence between FPC scores and covariates. Results are displayed separately for $\mathrm{PVE}_L = 0.95$ ($L = 4$ FPC scores) and $\mathrm{PVE}_L = 0.99$ ($L = 6$ FPC scores), and are grouped by the nature of the treatment-covariate relationship. The reference thresholds are $F = 10$ and $p = 0.05$. These results suggest that the proposed weighting scheme effectively reduces linear dependence between FPC scores and covariates, which is the form of confounding targeted by the balancing constraints.

% F-stats:

% P-vals: 

\begin{figure}[H]
 \centering
 \begin{minipage}{0.48\textwidth}
 \centering
 \includegraphics[width=\textwidth]{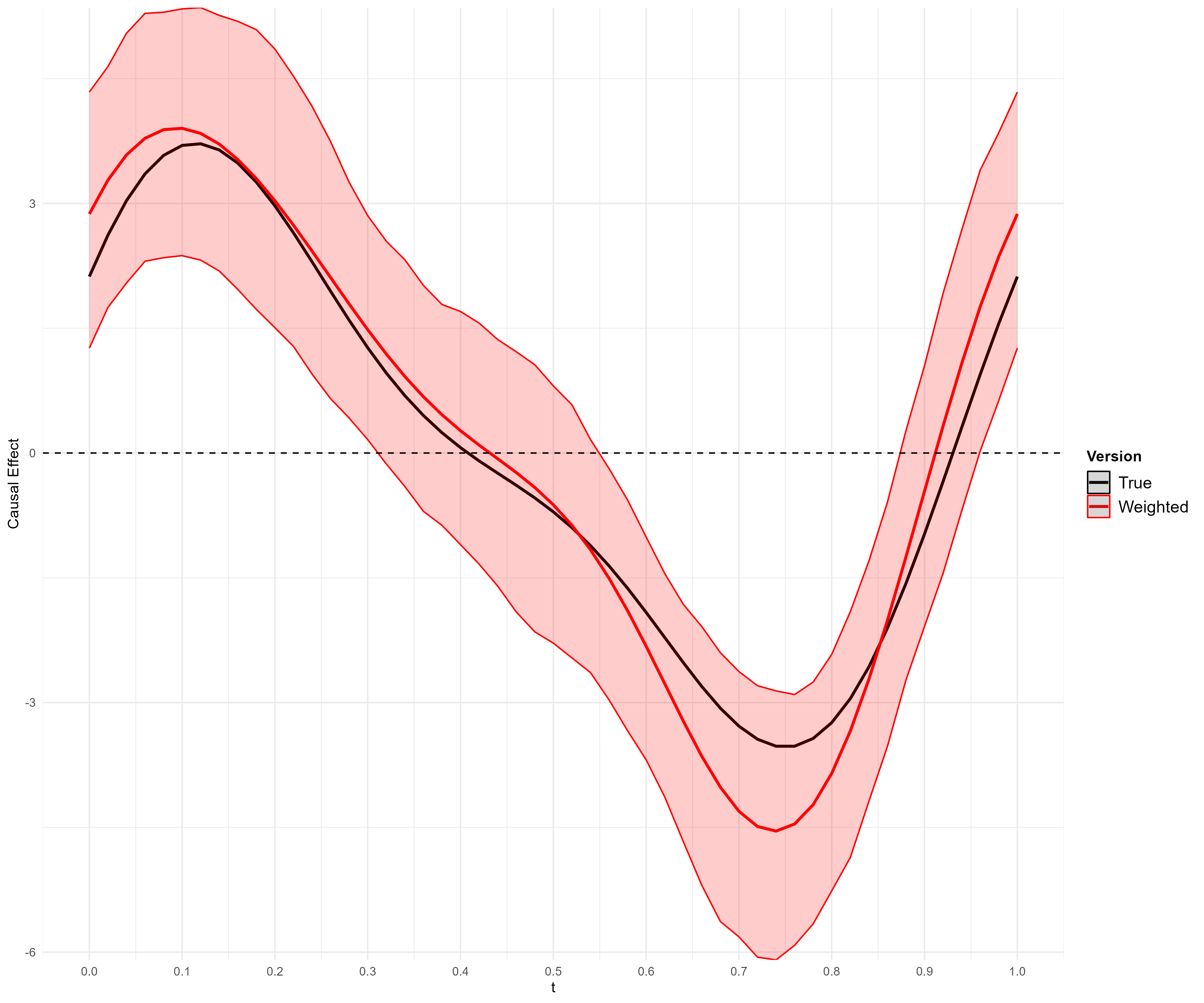}
 \subcaption{Run~34}
 \end{minipage}
 \hfill
 \begin{minipage}{0.48\textwidth}
 \centering
 \includegraphics[width=\textwidth]{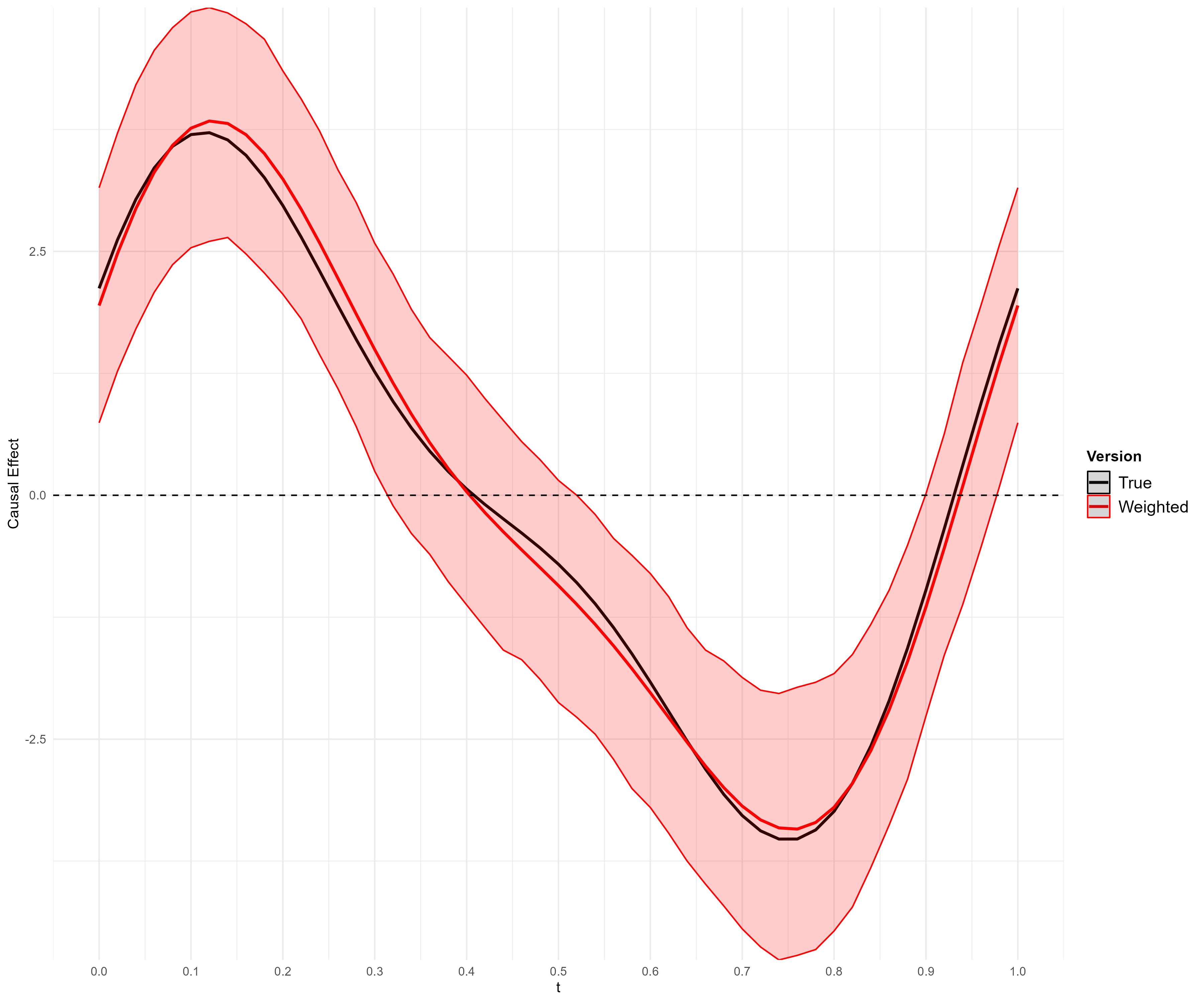}
 \subcaption{Run~126}
 \end{minipage}

 \caption{Estimated causal effect function $\hat{\mu}(t)$ and $99\%$ bootstrap reverse-percentile pointwise confidence bands for two representative simulation runs from setting~1 ($\mathrm{PVE}_L = \mathrm{PVE}_{L^*} = 0.95$). The true causal effect function $\mu(t)$ is shown as a solid black line. Confidence bands are based on $B = 1{,}000$ bootstrap samples.}
 \label{fig:bootstrap}
\end{figure}

To assess inferential performance, Figure~\ref{fig:bootstrap} presents $99\%$ bootstrap reverse-percentile confidence bands for the estimated causal effect function $\hat{\mu}(t)$ in two representative simulation runs (Runs~34 and~126) from setting~1, constructed using $B = 1{,}000$ bootstrap samples. In both cases, the bands cover the true effect function across the domain, suggesting adequate finite-sample performance.

\begin{figure}[H]
\centering
\includegraphics[width=0.8\textwidth]{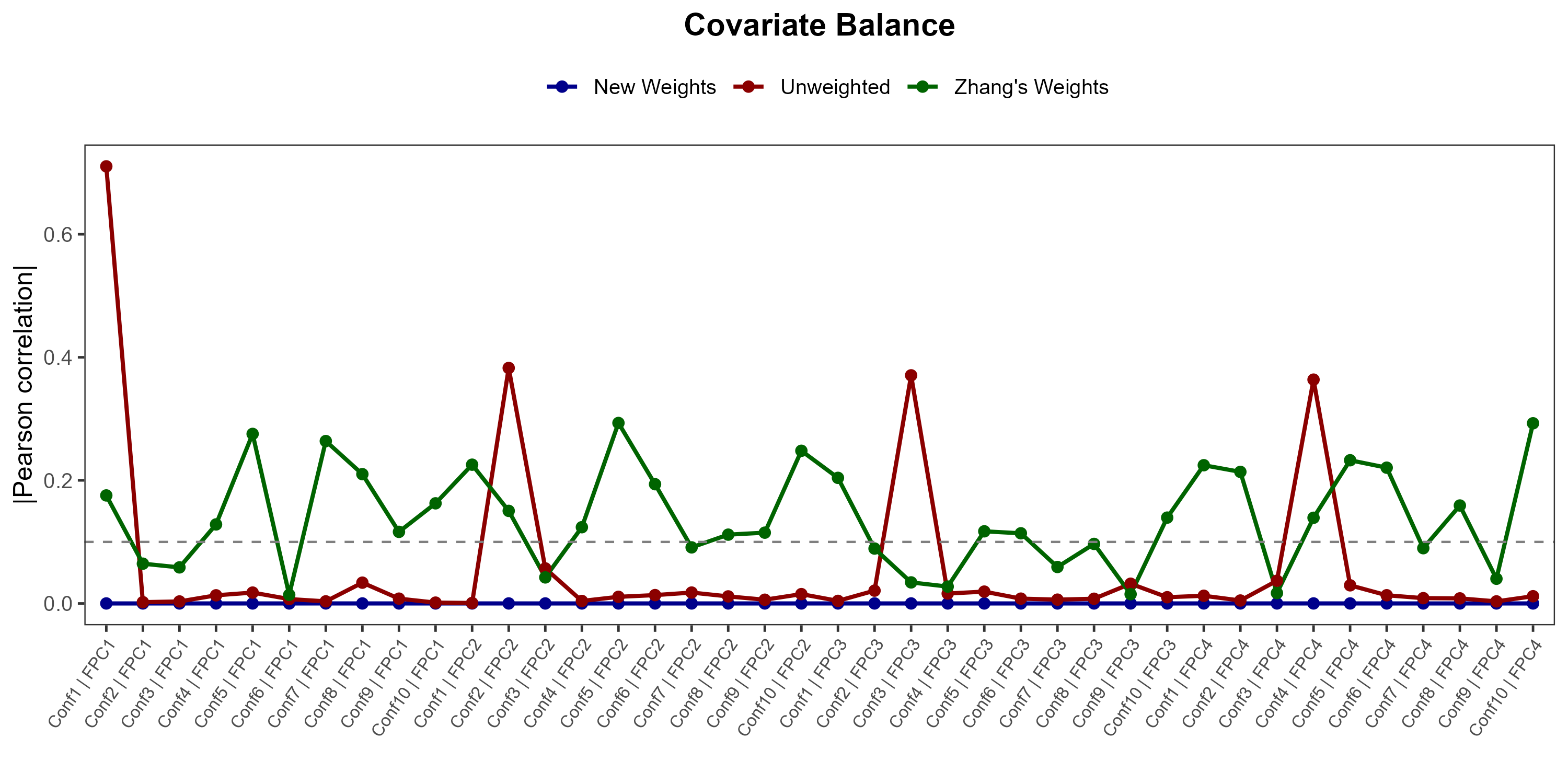}
\caption{Absolute Pearson correlations between confounder-FPC pairs under three weighting schemes: unweighted (green), proposed (red), and \citet{Zhang} (blue), for $n = 3000$ and $p = 10$ under Setting~1.}
\label{fig:balance_detail}
\end{figure}

For completeness, Figure~\ref{fig:balance_detail} reports the absolute Pearson correlations between confounder–FPC pairs under the unweighted estimator, \citet{Zhang}, and the proposed method for a representative configuration. The proposed method substantially reduces correlation magnitudes relative to the alternatives.

\newpage

\newpage
\section*{Supplementary Material D: Additional Analysis and Diagnostics for Functional Covariates and Outcome Simulation}
\label{sec:sm_balance_ext}

This section provides additional diagnostics supporting the simulation study reported in Section 3.2. We summarize the simulation settings and present detailed evidence on covariate balance and inferential behavior in the setting with a functional covariate and a functional outcome.

We consider four simulation settings that vary the treatment-covariate and covariate-outcome relationships:
\begin{itemize}
 \item Setting 1 (linear treatment--covariate and covariate--outcome relationships):\\
 $\lambda_{i1} = Z_{i1} + 4\mathcal{U}_{i1}$ and $g(D_i) = 2\lambda_{i1}$.

 \item Setting 2 (nonlinear treatment--covariate, linear outcome):\\
 $\lambda_{i1} = (Z_{i1} + 0.5)^2 + 4\mathcal{U}_{i1}$, with $g$ as in Setting~1.

 \item Setting 3 (linear treatment--covariate, nonlinear outcome):\\
 $\lambda_{i1}$ as in Setting~1 and $g(D_i) = 2\lambda_{i1} + \lambda_{i2}^2$.

 \item Setting 4 (nonlinear treatment--covariate and nonlinear outcome):\\
 $\lambda_{i1}$ as in Setting~2, with $g$ as in Setting~3.
\end{itemize}

Table~\ref{tab:ext_improvement} reports the percentage change of the weighted estimator relative to the unweighted estimator for MISE, AISE, and ISB across all settings and $\mathrm{PVE}_{L}, \mathrm{PVE}_{L^*_{\mathcal{T}}}$ and $\mathrm{PVE}_{L^*_{\mathcal{S}}}$ combinations.

\begin{table}[H]
\centering
\caption{Percentage improvement of the weighted estimator over the unweighted, functional covariate and outcome simulation. Negative values indicate deterioration.}
\label{tab:ext_improvement}
\begin{tabular}{lllcccccc}
\toprule
& & & \multicolumn{3}{c}{$\mathrm{PVE}_{L^*_{\mathcal{T}}} = \mathrm{PVE}_{L^*_{\mathcal{S}}}= 0.95$} & \multicolumn{3}{c}{$\mathrm{PVE}_{L^*_{\mathcal{T}}} = \mathrm{PVE}_{L^*_{\mathcal{S}}} = 0.99$} \\
\cmidrule(lr){4-6} \cmidrule(lr){7-9}
& & & MISE & AISE & ISB & MISE & AISE & ISB \\
\midrule
Setting 1
& $\mathrm{PVE}_{L_{\mathcal{T}}} = 0.95$ & & $20.92\%$ & $21.09\%$ & $19.41\%$ & $-39.04\%$ & $-84.07\%$ & $18.72\%$ \\
& $\mathrm{PVE}_{L_{\mathcal{T}}} = 0.99$ & & $20.86\%$ & $21.05\%$ & $19.40\%$ & $28.11\%$ & $29.48\%$ & $19.48\%$ \\
\midrule
Setting 2
& $\mathrm{PVE}_{L_{\mathcal{T}}} = 0.95$ & & $20.96\%$ & $21.13\%$ & $19.50\%$ & $-39.61\%$ & $-83.11\%$ & $19.22\%$ \\
& $\mathrm{PVE}_{L_{\mathcal{T}}} = 0.99$ & & $20.93\%$ & $21.09\%$ & $19.49\%$ & $28.91\%$ & $30.09\%$ & $19.65\%$ \\
\midrule
Setting 3
& $\mathrm{PVE}_{L_{\mathcal{T}}} = 0.95$ & & $1.31\%$ & $-2.30\%$ & $7.33\%$ & $-37.40\%$ & $-57.68\%$ & $7.08\%$ \\
& $\mathrm{PVE}_{L_{\mathcal{T}}} = 0.99$ & & $0.88\%$ & $-2.84\%$ & $7.36\%$ & $-11.09\%$ & $-15.52\%$ & $7.26\%$ \\
\midrule
Setting 4
& $\mathrm{PVE}_{L_{\mathcal{T}}} = 0.95$ & & $0.99\%$ & $-2.15\%$ & $7.37\%$ & $-36.45\%$ & $-55.94\%$ & $7.22\%$ \\
& $\mathrm{PVE}_{L_{\mathcal{T}}} = 0.99$ & & $0.65\%$ & $-2.92\%$ & $7.41\%$ & $-9.38\%$ & $-15.30\%$ & $7.35\%$ \\
\bottomrule
\end{tabular}
\end{table}

\begin{figure}[htbp]
    \centering
    \includegraphics[width=\textwidth]{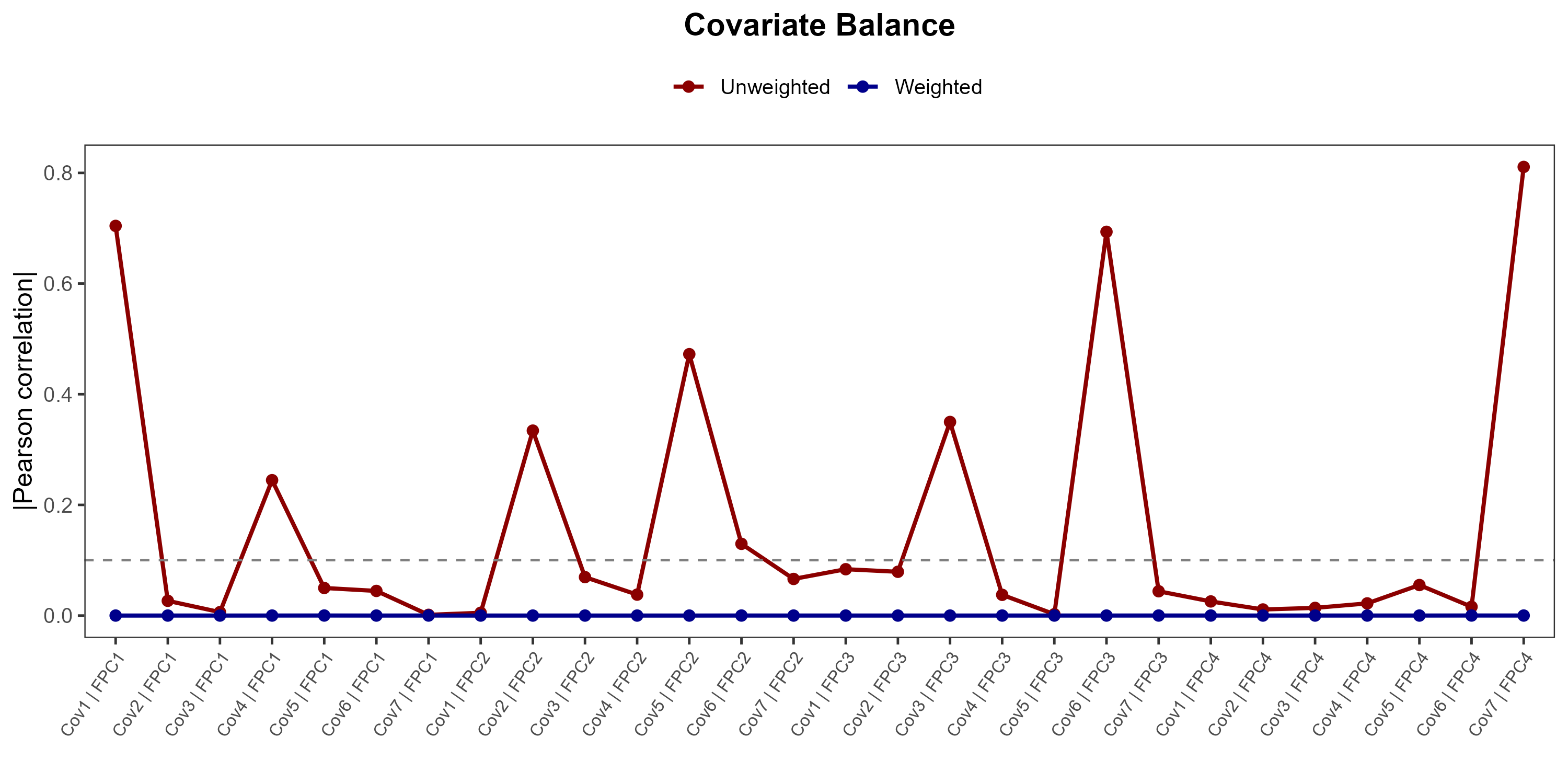}
    \caption{Correlation plot for settings I--III with $\mathrm{PVE}_L = 0.95$.}
    \label{fig:corrplot_ext}
\end{figure}

Figure~\ref{fig:corrplot_ext} assesses covariate balance through absolute Pearson correlations for settings I--III. 

\begin{figure}[H]
 \centering
 \includegraphics[width=0.75\textwidth]{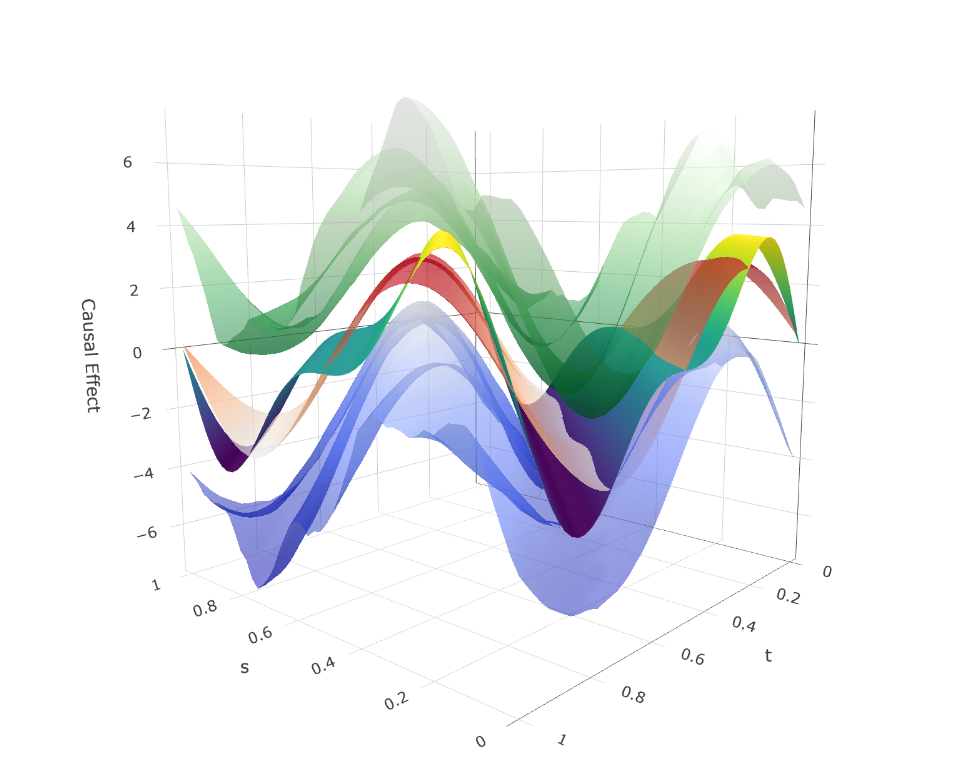}
 \caption{Three-dimensional visualization of the estimated causal effect surface
 $\hat{\mu}(s,t)$ and the true surface $\mu(s,t)$, with upper and lower $99\%$
 bootstrap reverse-percentile confidence surfaces. Run~126, setting~1,
 $\mathrm{PVE}_L = \mathrm{PVE}_{L^*} = 0.95$, $B = 1{,}000$ bootstrap
 samples.}
 \label{fig:boot3d}
\end{figure}

\begin{figure}[H]
 \centering
 \begin{minipage}{0.48\textwidth}
 \centering
 \includegraphics[width=\textwidth]{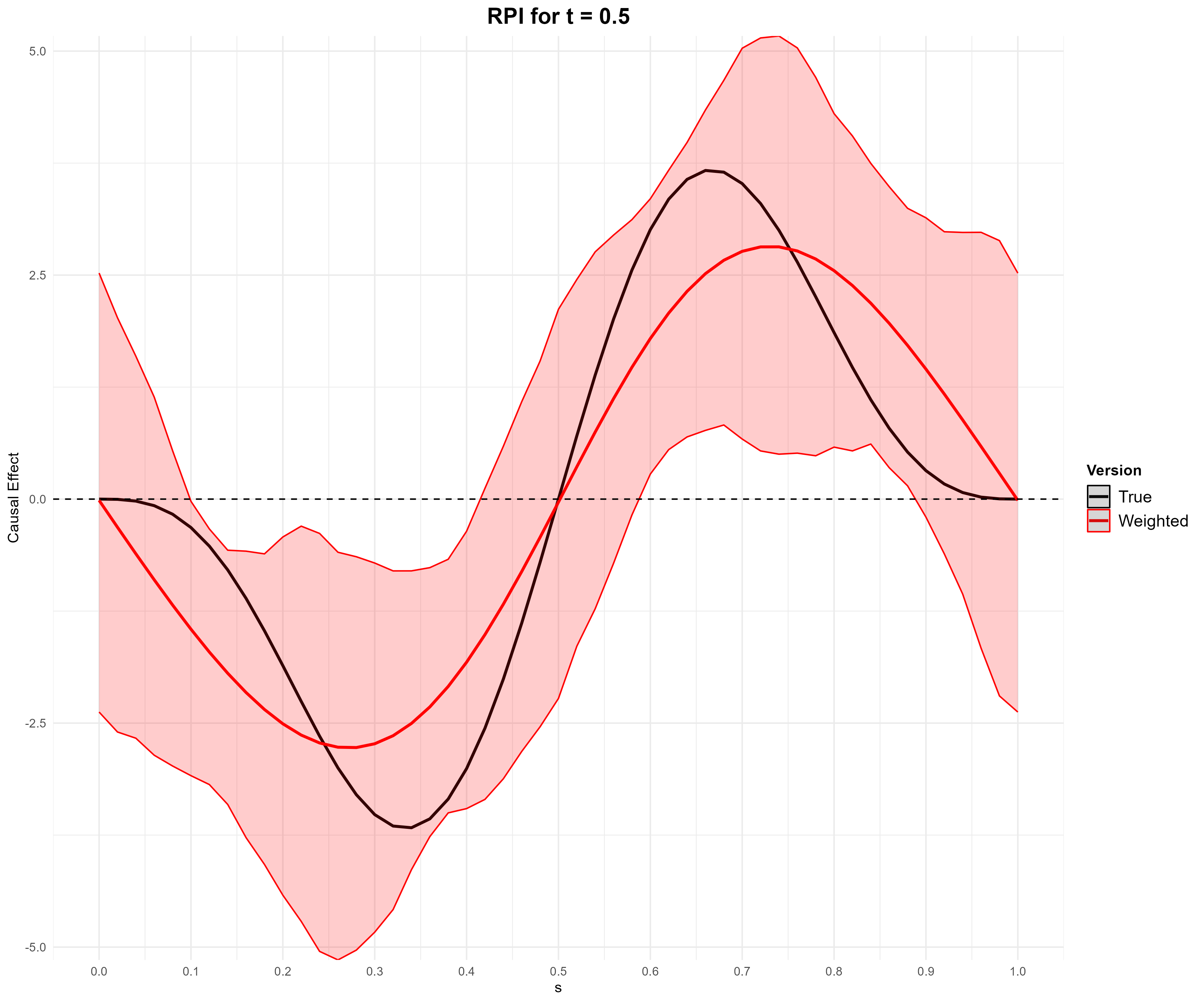}
 \subcaption{Cross-section at $t = 0.5$: $\mu(s, 0.5)$}
 \label{fig:boot2d_t}
 \end{minipage}
 \hfill
 \begin{minipage}{0.48\textwidth}
 \centering
 \includegraphics[width=\textwidth]{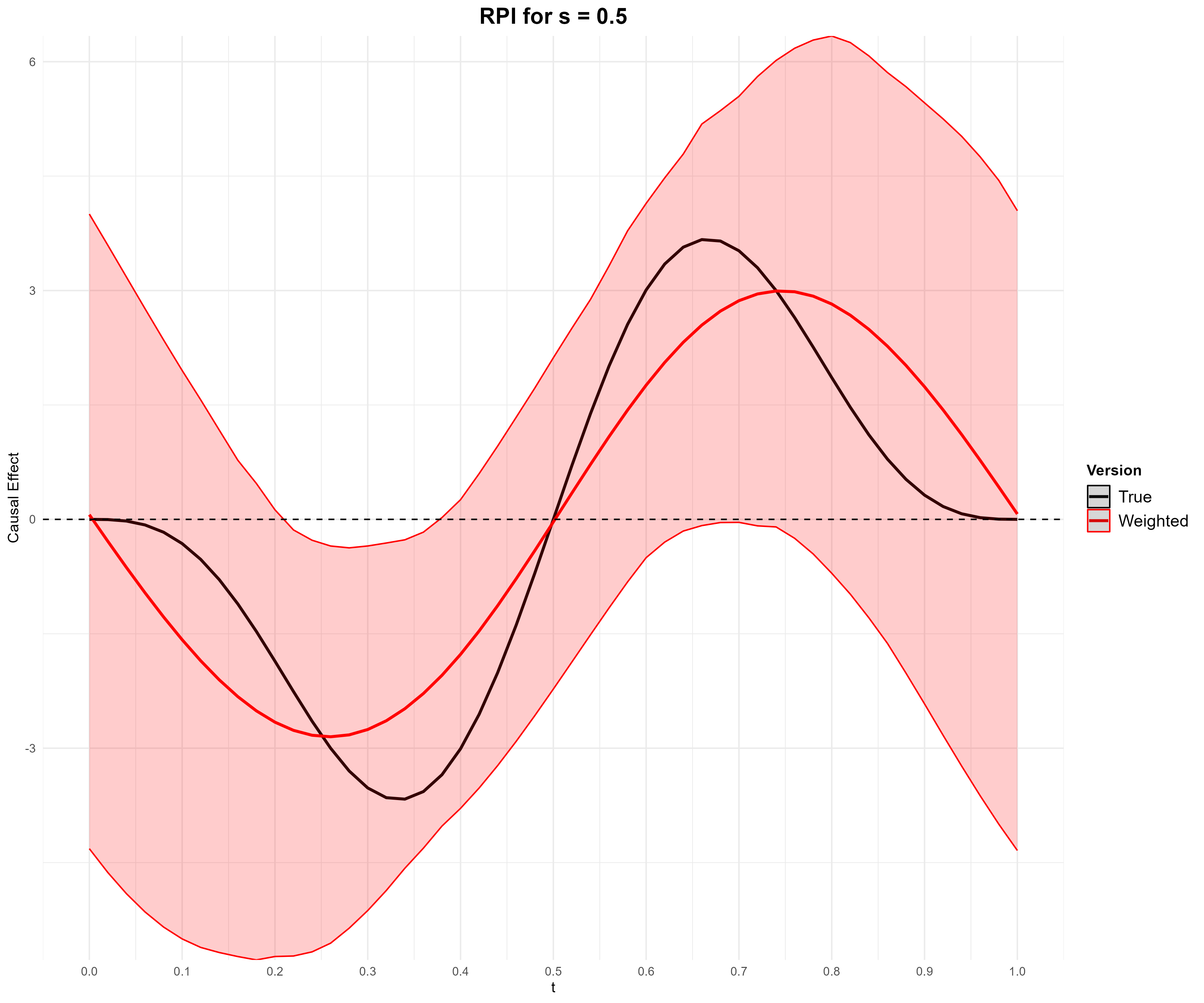}
 \subcaption{Cross-section at $s = 0.5$: $\mu(0.5, t)$}
 \label{fig:boot2d_s}
 \end{minipage}
 \caption{Two-dimensional cross-sections of the estimated causal effect surface,
 true surface, and $99\%$ bootstrap reverse-percentile confidence bands at
 fixed $t = 0.5$ (left) and fixed $s = 0.5$ (right). Run~126, setting~1.}
 \label{fig:bootstrap_ext}
\end{figure}

To assess inferential performance, we construct $99\%$ bootstrap reverse-percentile pointwise confidence surfaces for the estimated causal effect surface $\hat{\mu}(s,t)$ in a representative simulation run from Setting~1, using $B=1{,}000$ bootstrap samples. Because the estimand is bivariate, we report both three-dimensional visualizations and two-dimensional cross-sections. Figure~\ref{fig:boot3d} displays the estimated surface $\hat{\mu}(s,t)$, the true surface $\mu(s,t)$, and the upper and lower confidence surfaces. Figure~\ref{fig:bootstrap_ext} presents cross-sections at $t=0.5$ and $s=0.5$, providing a more detailed view of estimation accuracy and pointwise coverage along each axis. In these representative cross-sections, the true function lies within the confidence bands across the displayed domain.

\begin{figure}[H]
 \centering
 \includegraphics[width=\textwidth]{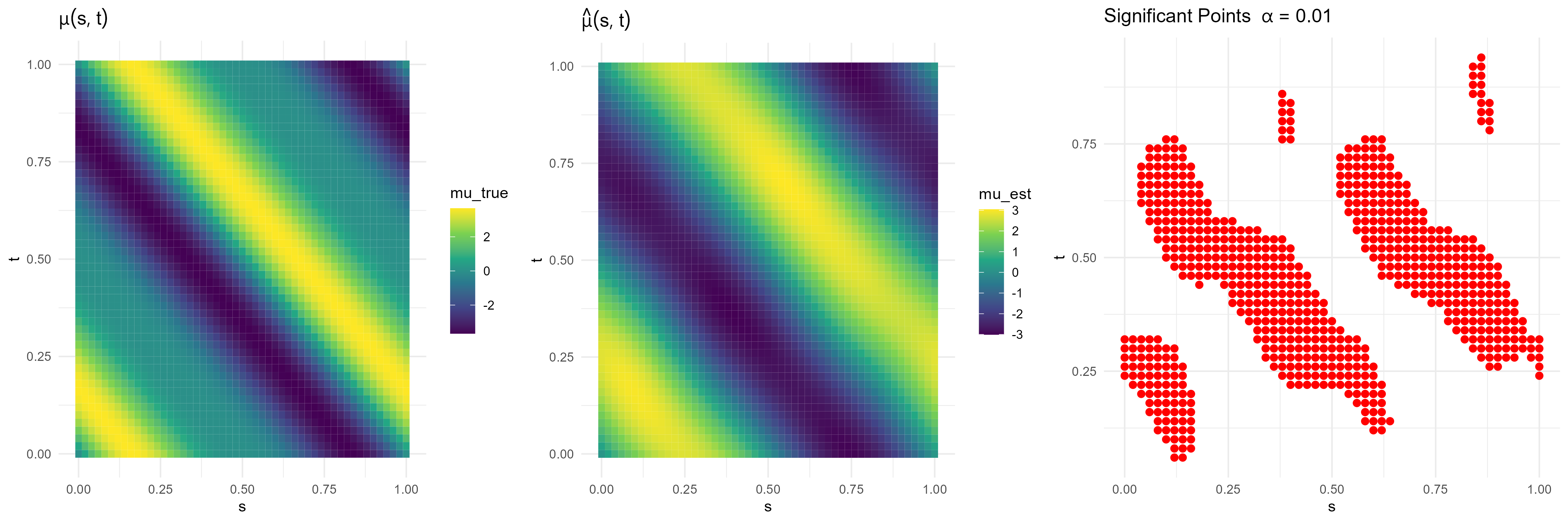}
 \caption{Significance map at $\alpha = 0.01$. Left: contour plot of the true
 surface $\mu(s,t)$. Middle: contour plot of the estimated surface
 $\hat{\mu}(s,t)$. Right: domain regions where zero is excluded from the
 $99\%$ confidence interval. Run~126, setting~1.}
 \label{fig:sig_points}
\end{figure}

Figure~\ref{fig:sig_points} displays regions of the domain where the $99\%$ confidence interval excludes zero, together with contour plots of the true and estimated causal effect surfaces. 

\newpage
\section*{Supplementary Material E: Additional descriptive and FPCA diagnostics for the UK Biobank Application}
\label{sec:sm_realdata}

This section provides additional descriptive summaries and FPCA diagnostics for the UK Biobank application in Section 4. %These results complement the covariate balance diagnostics and causal effect estimates reported in the main text.

\subsection*{Analysis 1: BMI trajectories and incident Type 2 Diabetes}

\begin{figure}[H]
    \centering
    \includegraphics[width=1\textwidth]{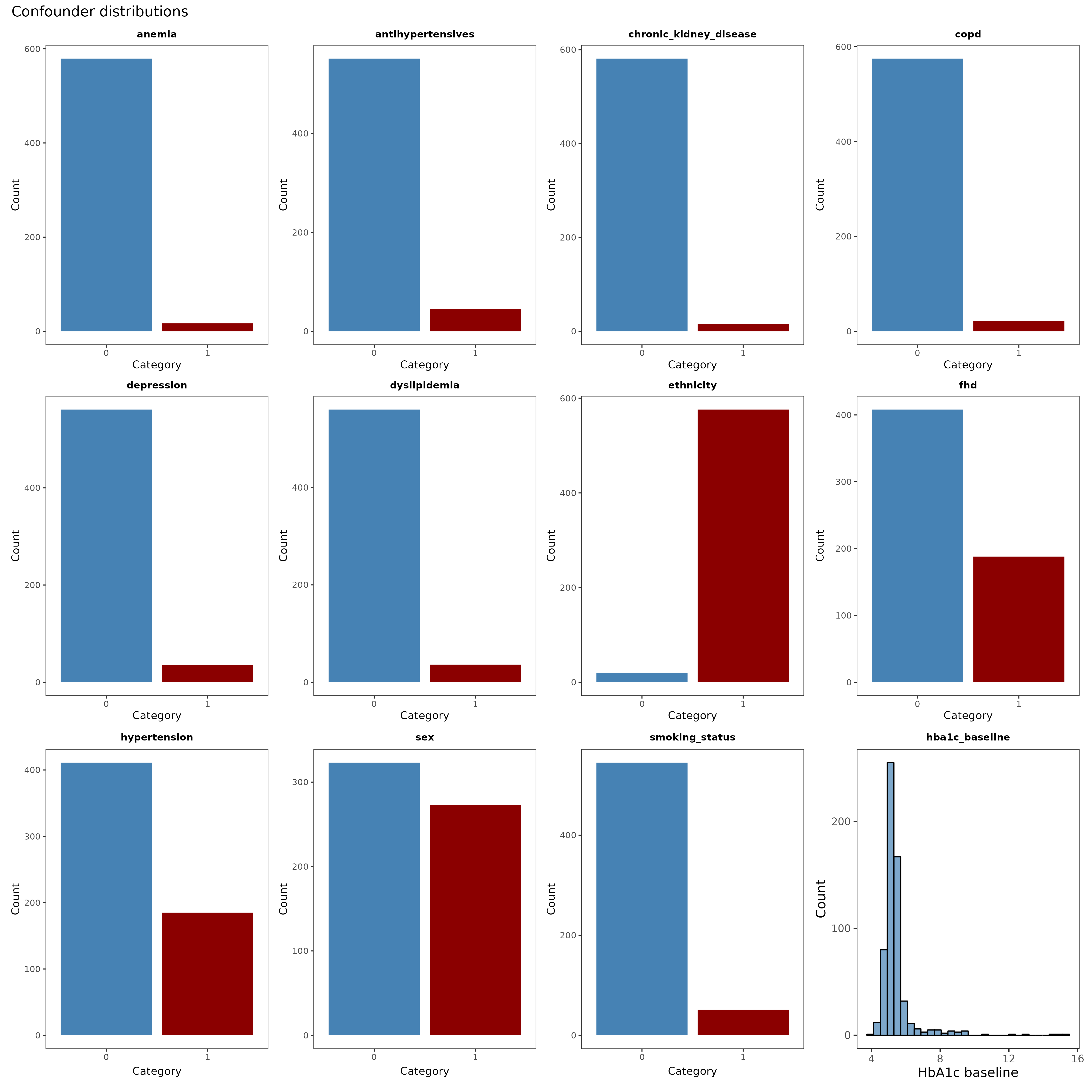}
    \caption{Distribution of categorical and continuous baseline confounders in Analysis 1. Baseline HbA1c is included as a continuous confounder to adjust for pre-existing metabolic status. fhd = family history of diabetes.}
    \label{fig:sm_a1_hba1c}
\end{figure}

Figure~\ref{fig:sm_a1_hba1c} summarizes the distribution of the categorical and continuous covariates used in Analysis~1. Baseline HbA1c is included as a continuous covariate to adjust for pre-existing metabolic status. Ischaemic heart disease is excluded from the analysis due to a severe imbalance between the exposure groups.

\begin{figure}[H]
    \centering
    \includegraphics[width=0.70\textwidth]{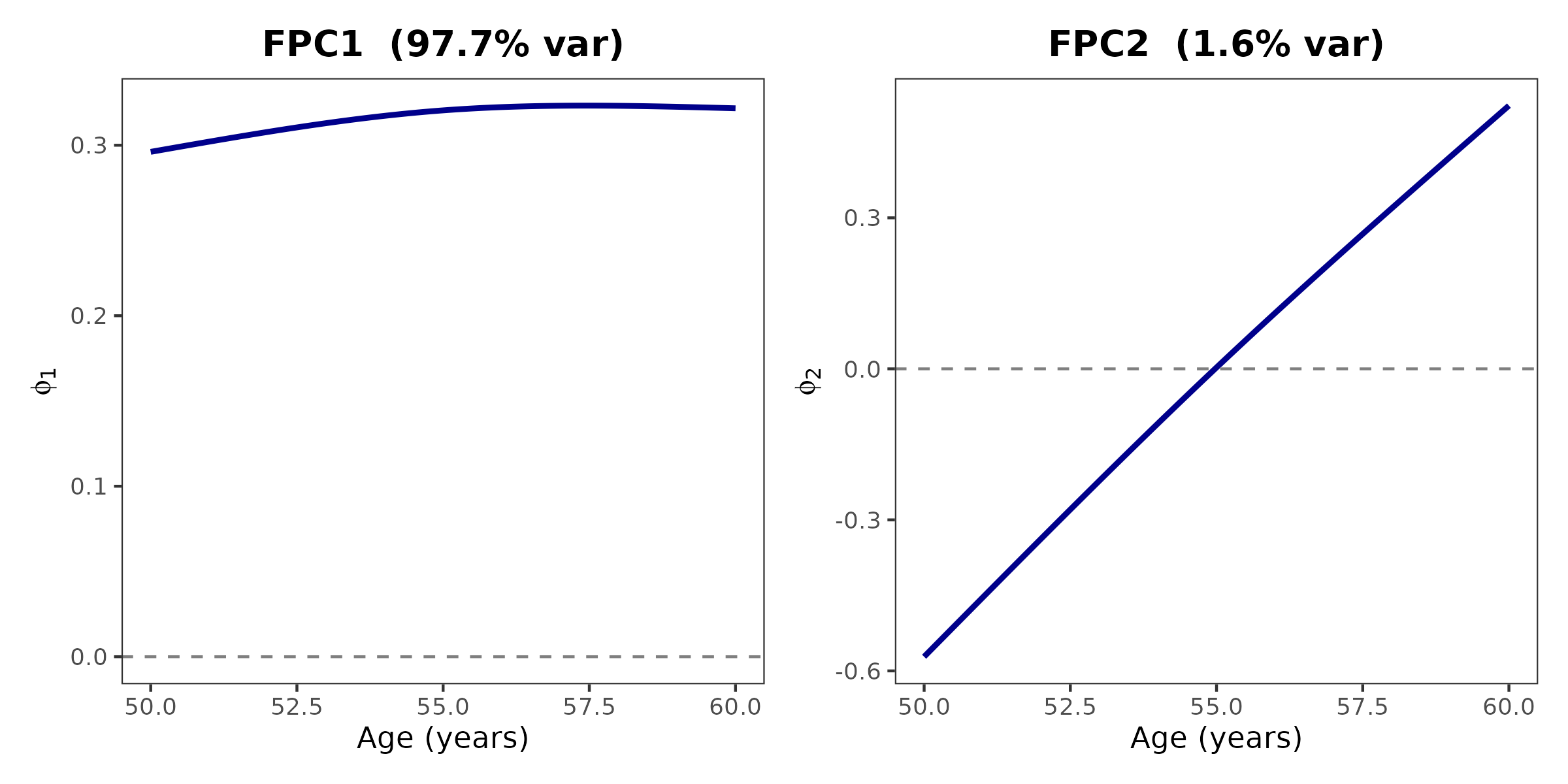}
    \caption{Selected functional principal components for the BMI exposure trajectory in Analysis~1.}
    \label{fig:sm_a1_fpc}
\end{figure}

Figure~\ref{fig:sm_a1_fpc} displays the selected FPCs for the BMI exposure trajectory in Analysis~1. These components provide the finite-dimensional representation used in the functional propensity score weighting procedure.

{\color{black}
To support the interpretation of the estimated effect functions, Figure~\ref{fig:sm_bmi_trajectories_a1} displays the smoothed BMI trajectories used in the analyses. 
The trajectories show gradual temporal variation over the age ranges considered, with limited evidence of rapid oscillations. This empirical pattern is consistent with the approximately linear leading functional principal components and with the smooth estimated effect functions reported in Section 4.2.

\begin{figure}[ht]
    \centering
    \includegraphics[width=0.7\textwidth]{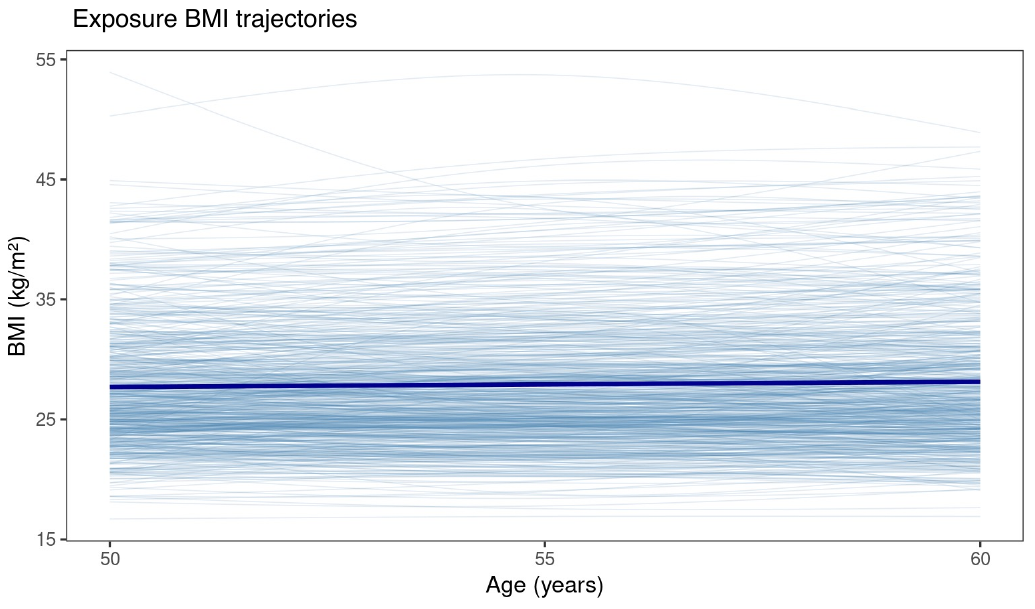}
    \caption{Smoothed BMI trajectories used in Analysis~1, where BMI over ages 50--65 is used as the functional treatment. Grey lines represent individual smoothed trajectories, and the red line represents the mean smoothed trajectory.}
    \label{fig:sm_bmi_trajectories_a1}
\end{figure}
}

{\color{black}
\textbf{Bootstrap confidence intervals.}
Pointwise confidence intervals were constructed using a subject-level bootstrap. At each bootstrap iteration, individuals were resampled with replacement, and the outcome, weights, and treatment FPC scores were resampled jointly. The weighted outcome model was then refitted on the bootstrap sample. We report reverse-percentile bootstrap intervals. Specifically, if $\hat{\mu}(t)$ denotes the estimate from the original sample and $q_{\alpha/2}(t)$ and $q_{1-\alpha/2}(t)$ denote the empirical quantiles of the bootstrap estimates at time $t$, the interval is given by:
$\left[2\hat{\mu}(t)-q_{1-\alpha/2}(t),\, 2\hat{\mu}(t)-q_{\alpha/2}(t) \right].$}
\subsection*{Analysis 2: BMI trajectories and HbA1c dynamics}

\begin{figure}[h]
    \centering
    \includegraphics[width=0.8\textwidth]{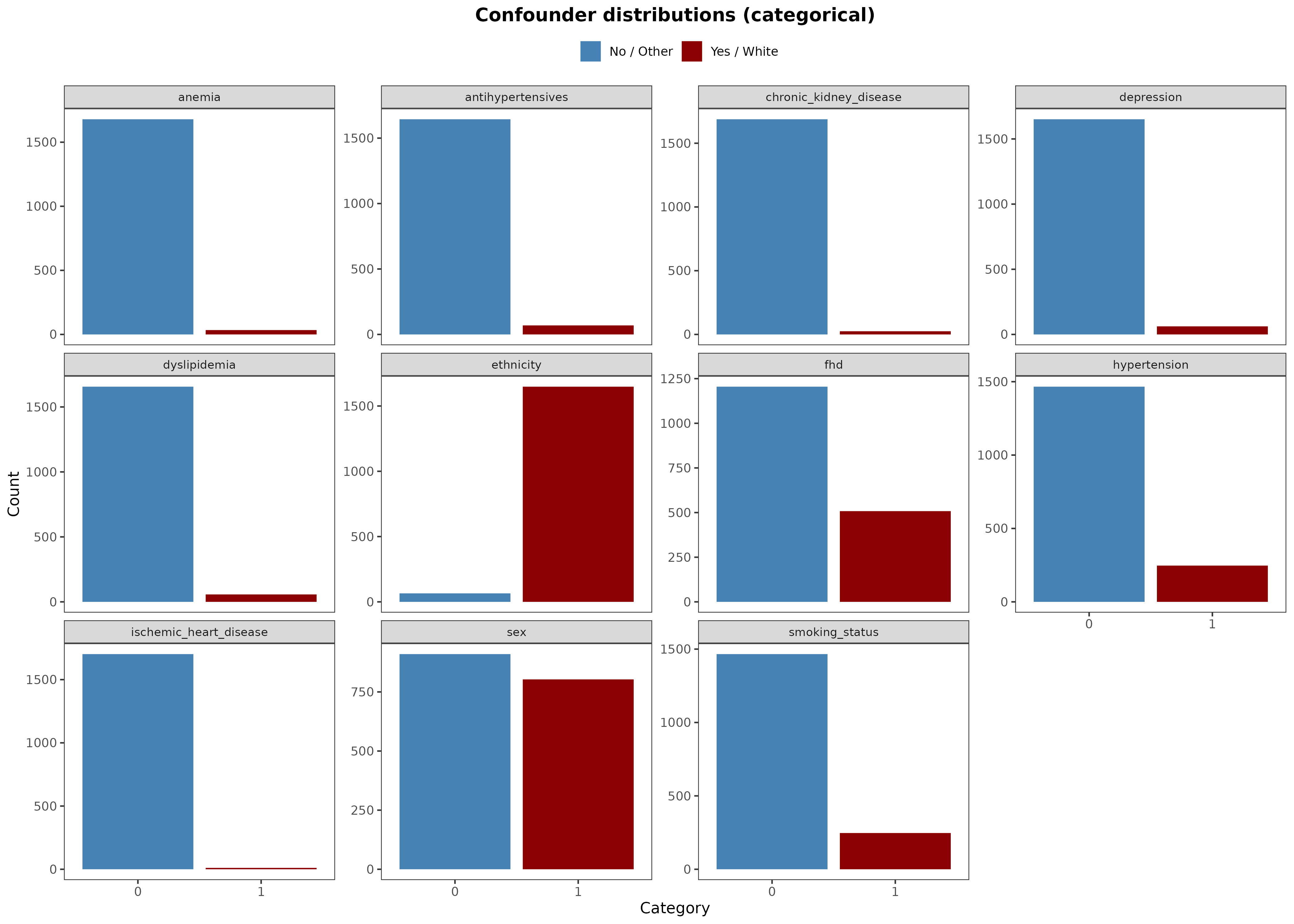}
    \caption{Distribution of categorical baseline confounders in Analysis 2.}
    \label{fig:sm_a2_cat}
\end{figure}

Figure~\ref{fig:sm_a2_cat} summarizes the distribution of categorical baseline covariates used in Analysis~2. In this analysis, baseline HbA1c is not included among the confounders because HbA1c is modeled as the functional outcome.

\begin{figure}[H]
    \centering
    \includegraphics[width=0.70\textwidth]{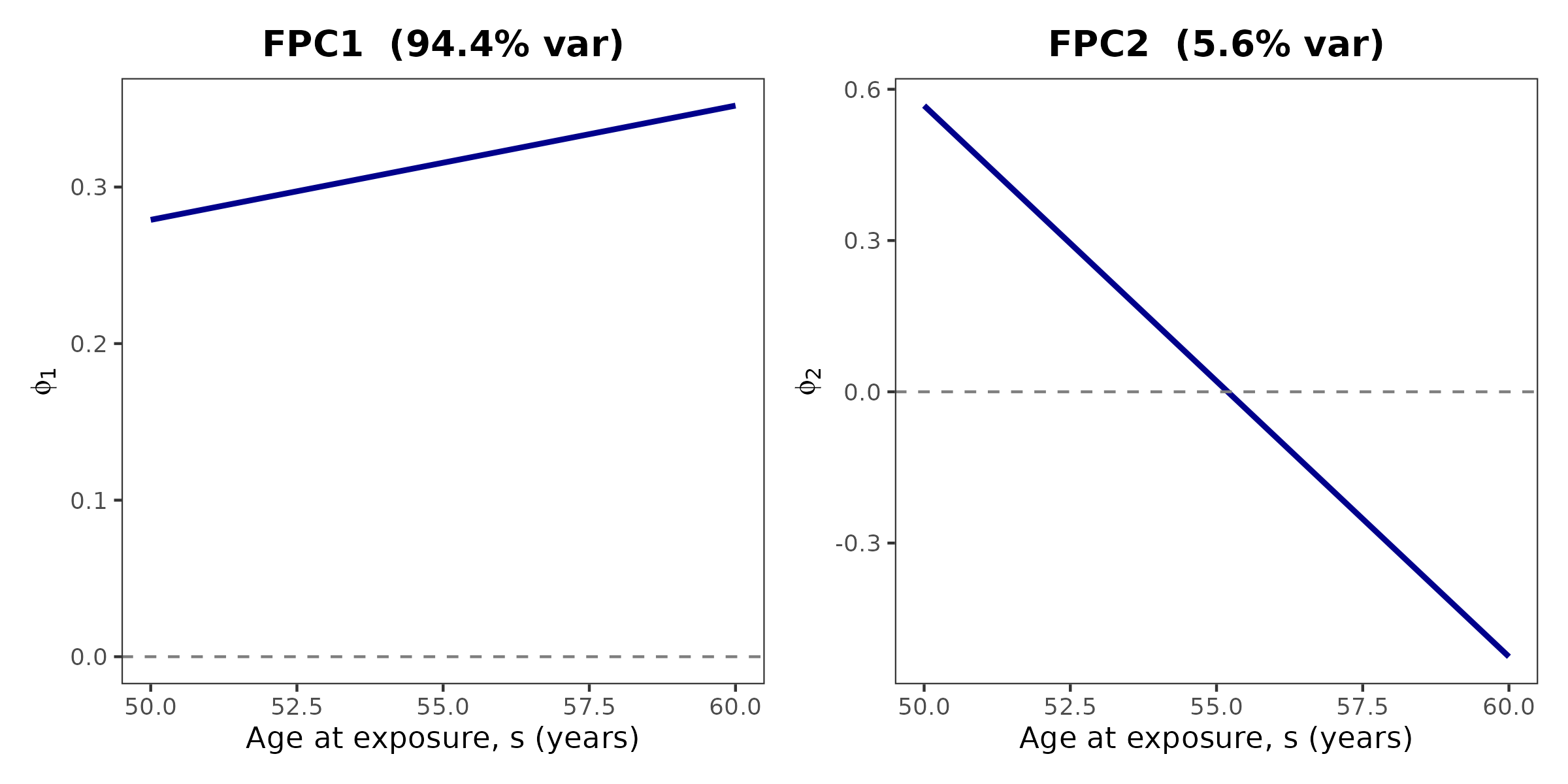}
    \caption{Selected functional principal components for the BMI exposure trajectory in Analysis~2.}
    \label{fig:sm_a2_fpc_exp}
\end{figure}

\begin{figure}[H]
    \centering
    \includegraphics[width=0.70\textwidth]{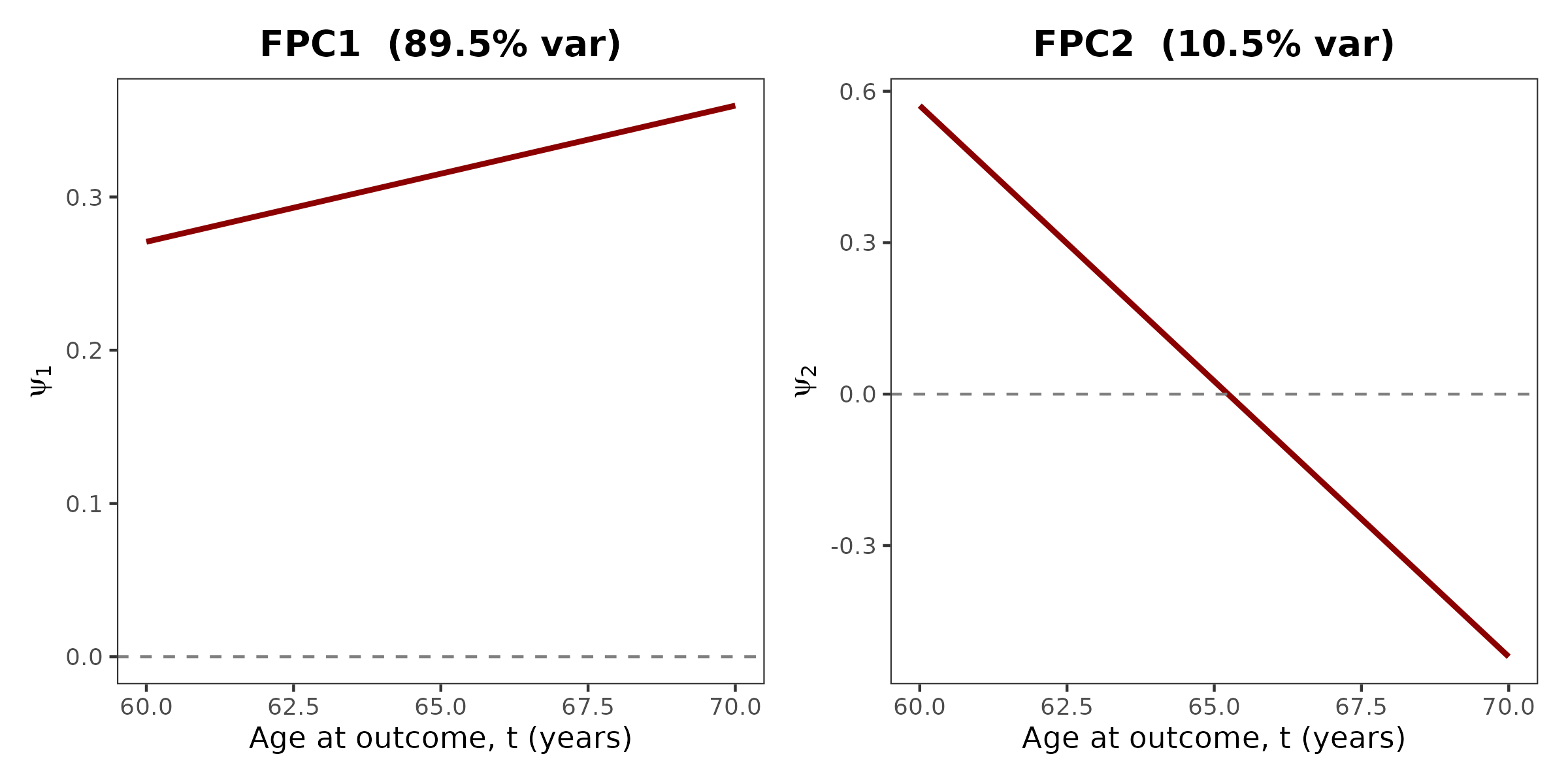}
    \caption{Selected functional principal components for the HbA1c outcome trajectory in Analysis~2.}
    \label{fig:sm_a2_fpc_out}
\end{figure}

Figures~\ref{fig:sm_a2_fpc_exp} and~\ref{fig:sm_a2_fpc_out} display the selected FPCs for the BMI exposure trajectory and HbA1c outcome trajectory, respectively. These components define the finite-dimensional representation used for the functional treatment and functional outcome models.

%MOVED TO THE MAIN MANUSCRIPT:
%Figure~\ref{fig:sm_surface} displays the estimated causal effect surface from Analysis~2. The surface provides a global visualization of how BMI at exposure age $s \in [50,60]$ affects HbA1c at outcome age $t \in [60,70]$. The one-dimensional slices and bootstrap confidence bands in the main text provide a more detailed inferential summary of this surface.

%\begin{figure}[H]
%    \centering
%    \includegraphics[width=0.50\textwidth]{Images/4_real_data_application/analisi2_heatmap_weighted.png}
%    \caption{Estimated causal effect surface $\hat{\mu}(s,t)$ in Analysis~2, where $s \in [50,60]$ denotes BMI exposure age and $t \in [60,70]$ denotes HbA1c outcome age. Warmer colors indicate larger positive estimated effects.}
%    \label{fig:sm_surface}
%\end{figure}

{\color{black}
Figure~\ref{fig:sm_surface_ci_a2} reports the lower and upper pointwise bootstrap confidence surfaces corresponding to the estimated weighted causal effect surface in Analysis~2. 

\begin{figure}[ht]
    \centering
    \includegraphics[width=0.9\textwidth]{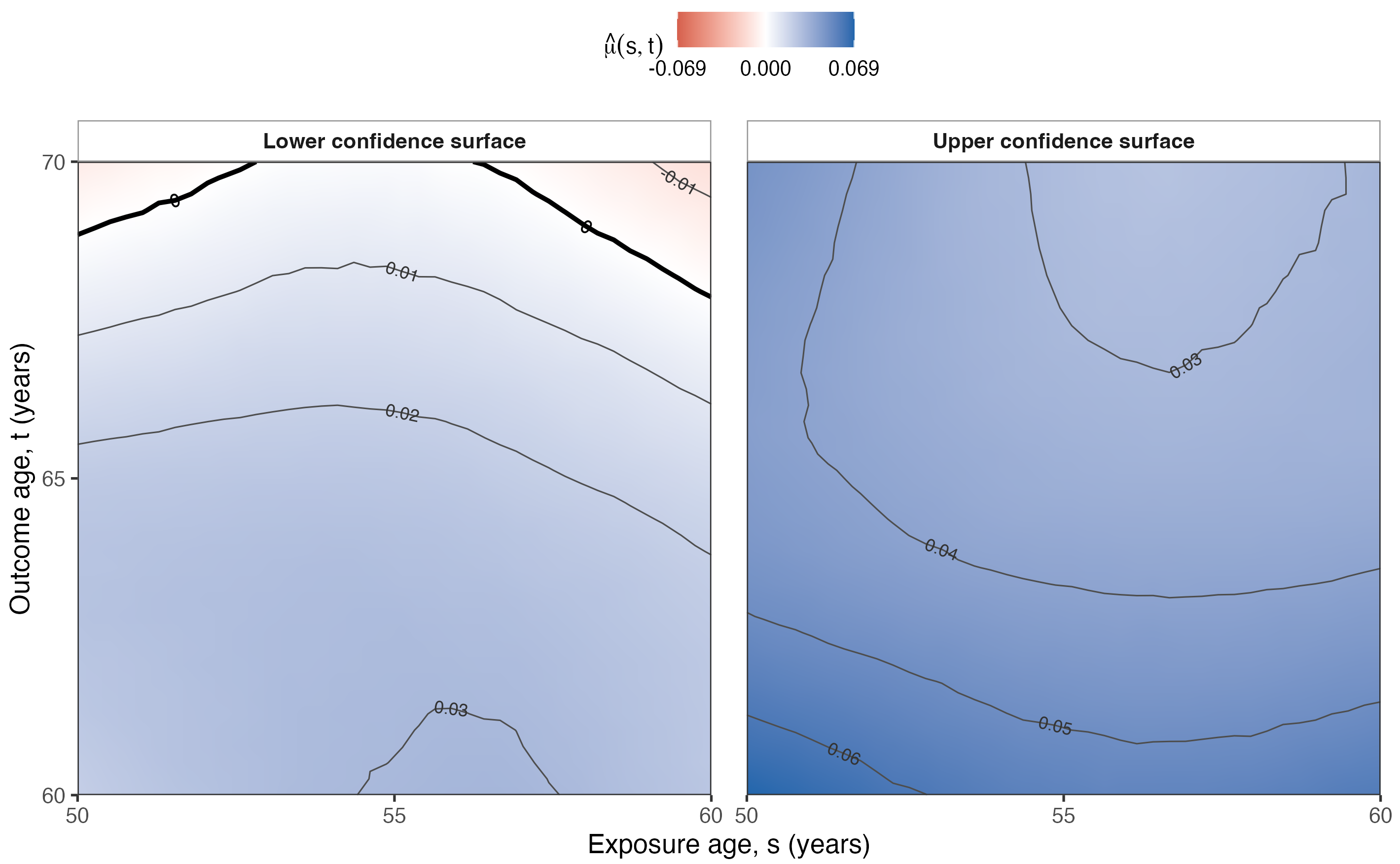}
    \caption{Lower and upper pointwise bootstrap confidence surfaces for the estimated weighted causal effect surface in Analysis~2. Contour lines indicate levels of equal estimated effect and are added to facilitate visual comparison.}
    \label{fig:sm_surface_ci_a2}
\end{figure}
}

{\color{black} 
Figure~\ref{fig:sm_unweighted_slices_a2} reports the unweighted estimates of the one-dimensional slices of the effect surface in Analysis~2, together with 95\% pointwise reverse-percentile bootstrap confidence intervals based on subject-level resampling ($B=1000$). The unweighted estimates show the same overall positive association between BMI and subsequent HbA1c as the weighted analysis, with larger effects at earlier exposure ages and attenuation at later exposure and outcome ages. %However, the direct comparisons with the weighted estimates indicate that adjustment through the proposed weighting scheme modifies both the magnitude and the uncertainty of the estimated effects.

\begin{figure}[ht]
    \centering
    \includegraphics[width=1\textwidth]{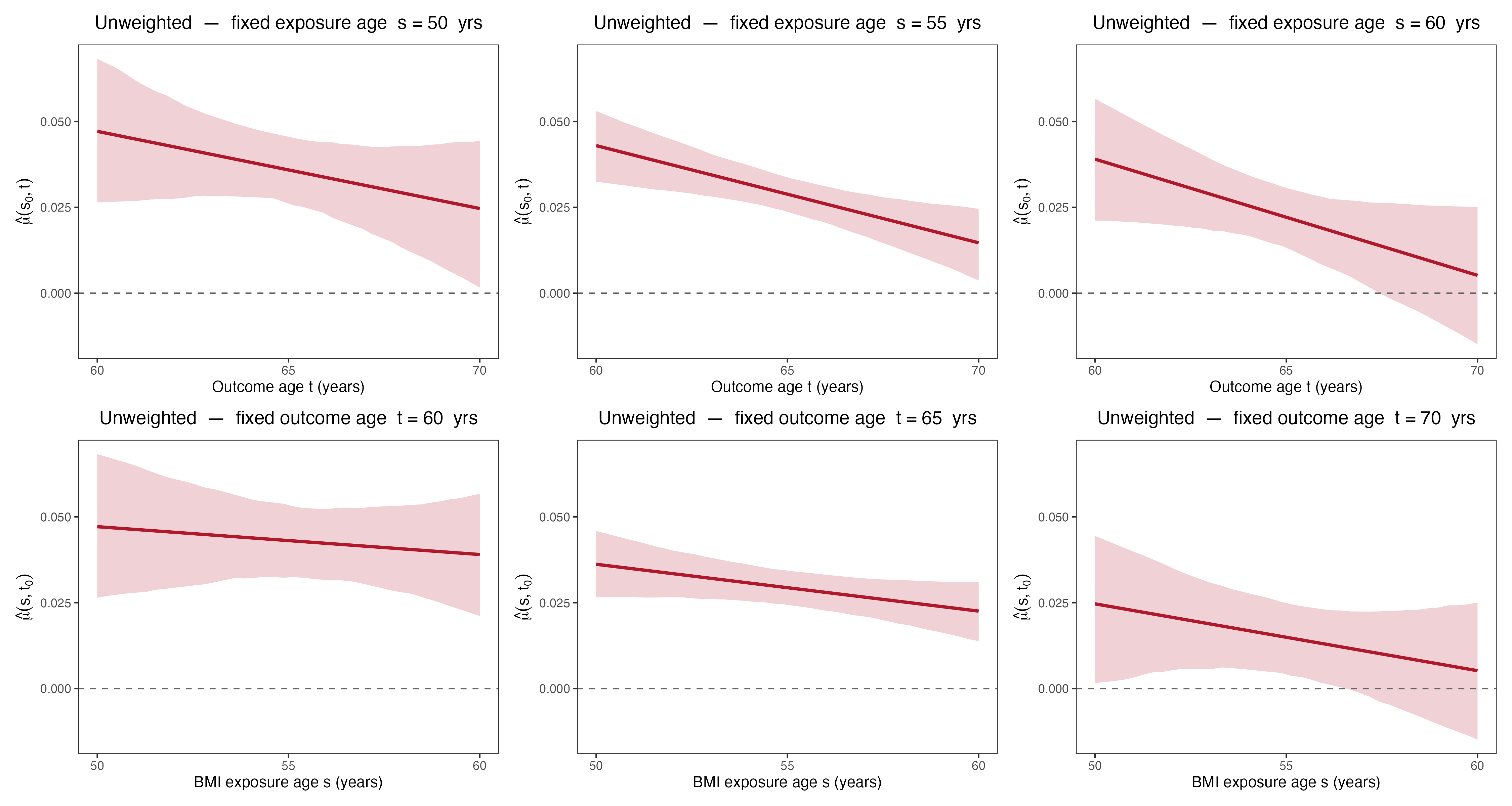}
    \caption{Unweighted slices of the estimated effect surface $\hat\mu(s, t)$ with 95\% pointwise bootstrap confidence bands. Stars mark pointwise significance ($\alpha = 0.05$). \textit{Top:} $\hat\mu(\cdot,\, t_0)$ as a function of exposure age $s$, for fixed outcome ages $t_0 \in \{60, 65, 70\}$. \textit{Bottom:} $\hat\mu(s_0,\, \cdot)$ as a function of outcome age $t$, for fixed exposure ages $s_0 \in \{50, 55, 60\}$.}
    \label{fig:sm_unweighted_slices_a2}
\end{figure}
}

\FloatBarrier
{\color{black}
To support the interpretation of the estimated effect surface, Figures~\ref{fig:sm_bmi_trajectories_a2} and~\ref{fig:sm_hba1c_trajectories_a2} display the smoothed treatment and outcome trajectories used in Analysis~2.
As for Analysis~1, the trajectories show gradual temporal variation over the age ranges considered, with limited evidence of rapid oscillations. This empirical pattern is consistent with the approximately linear leading functional principal components and with the smooth estimated effect surface reported in Section 4.3.

\begin{figure}[ht]
    \centering
    \includegraphics[width=0.7\textwidth]{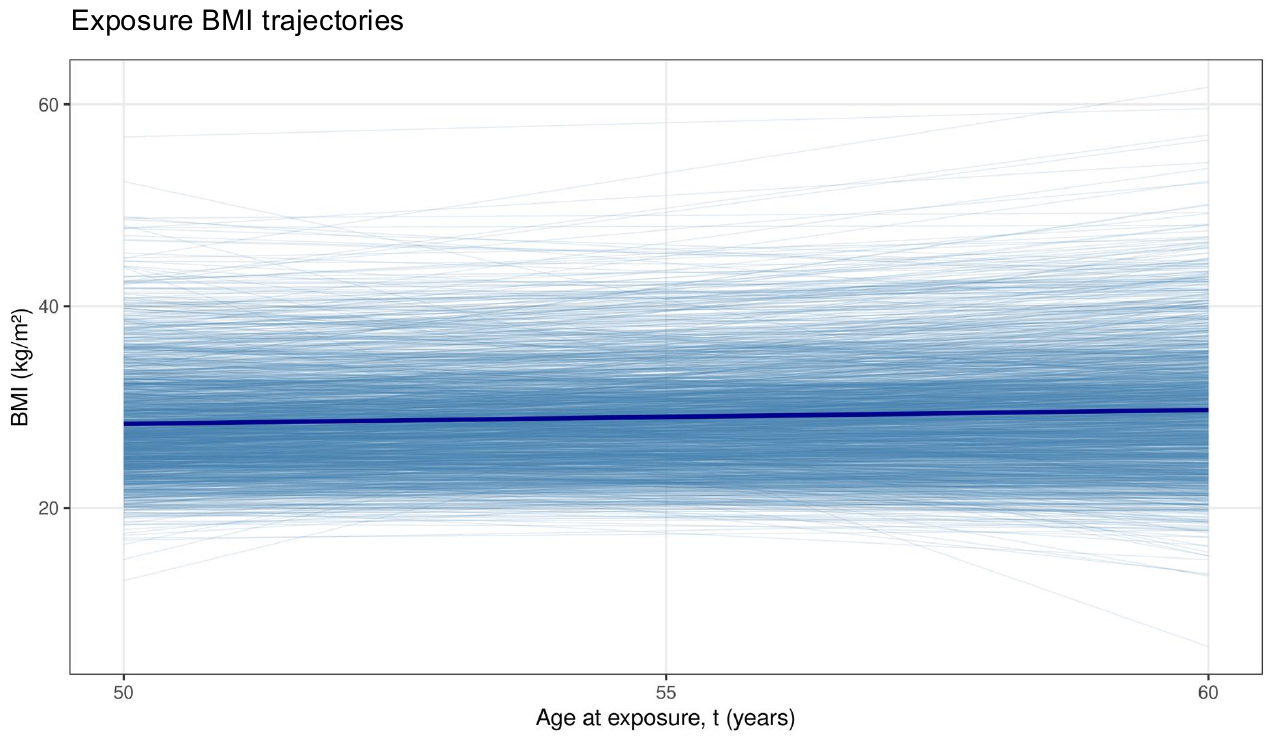}
    \caption{Smoothed BMI trajectories used in Analysis~2, where BMI over ages 50--60 is used as the functional treatment. Individual smoothed trajectories are shown in grey, with their pointwise mean shown in red.}
    \label{fig:sm_bmi_trajectories_a2}
\end{figure}

\begin{figure}[ht]
    \centering
    \includegraphics[width=0.7\textwidth]{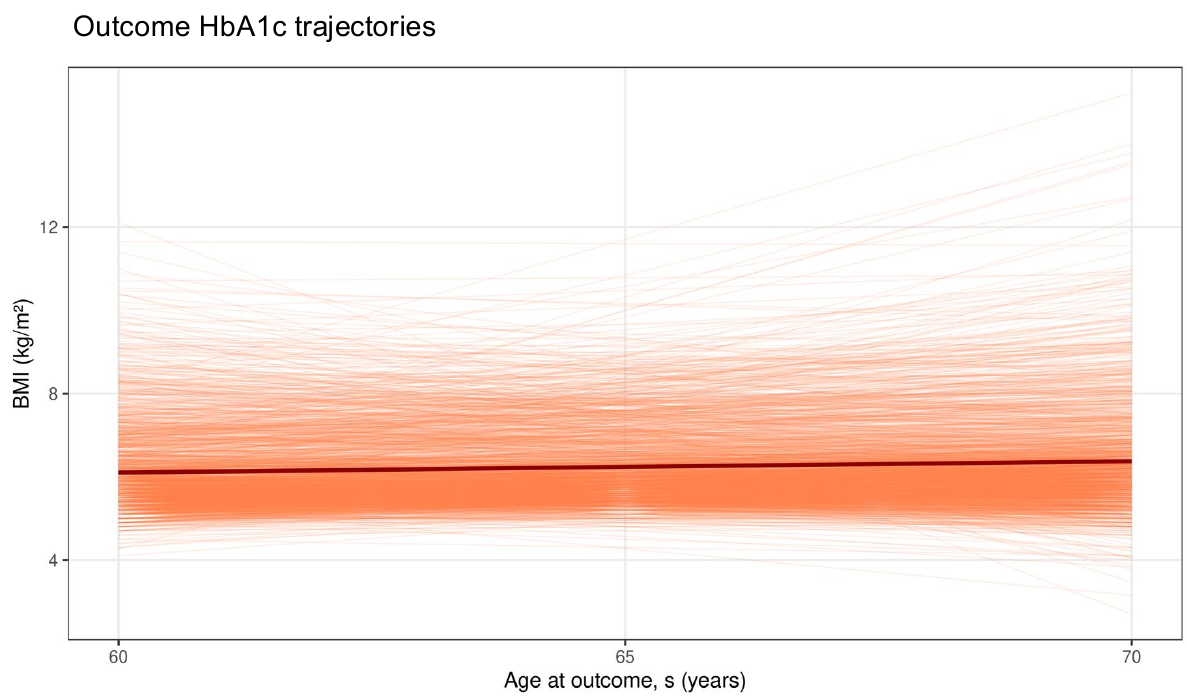}
    \caption{Smoothed HbA1c trajectories used in Analysis~2, where HbA1c over ages 60--70 is used as the functional outcome. Individual smoothed trajectories are shown in light blue, with their pointwise mean shown in blue.}
    \label{fig:sm_hba1c_trajectories_a2}
\end{figure}
}

%{\color{black}
%\textbf{Bootstrap confidence intervals.}
%Pointwise confidence intervals were constructed using a subject-level bootstrap. At each bootstrap iteration, individuals were resampled with replacement, and the outcome, weights, and treatment FPC scores were resampled jointly. The weighted outcome model was then refitted on the bootstrap sample. We report reverse-percentile bootstrap intervals.}

\end{document}